\documentclass[final]{elsarticle}
\usepackage[colorlinks=true,linkcolor=black,citecolor=black,urlcolor=black]{hyperref}
\pdfoutput=1
\usepackage{latexsym} 
\usepackage{graphics}
\usepackage{epsfig}
\usepackage{graphicx}
\usepackage{tikz-qtree}
\usepackage{tikz}
\usepackage{amsmath}
\usepackage{dirtytalk}
\usepackage{bbm}
\usepackage{algorithm}
\usepackage{lipsum}
\usepackage{pgfplots}
\usepackage{caption}
\usepgfplotslibrary{groupplots}
\usepackage[utf8]{inputenc}
\usepackage[english]{babel}
\usepackage{pgfplots}
\pgfplotsset{compat=1.17}
\usepackage{rotating}
\pgfplotsset{width=11cm,compat=newest}
\usepackage{algorithm}
\usepackage[noend]{algpseudocode}
\makeatletter
\def\BState{\State\hskip-\ALG@thistlm}
\makeatother
\algdef{SE}[DOWHILE]{Do}{doWhile}{\algorithmicdo}[1]{\algorithmicwhile\ #1}
\usepackage{graphicx}
\usepackage{subfig}
\usepackage{tikz}
\usepackage{pgfplots}

\usepackage{varwidth}
\usepackage{booktabs} 
\usepackage{epsfig}
\usepackage{amsfonts}
\usepackage{array}
\usepackage{rotating}
\usepackage{tabularx}
\usepackage{color}
\usepackage{multirow}
\usepackage{url}
\usepackage{lscape}
\usepackage{xargs} 
\newcolumntype{x}[1]{>{\centering\let\newline\\\arraybackslash\hspace{0pt}}p{#1}}
\usepackage{tabularx}
\usepackage{array}
\usepackage{colortbl}
\usepackage{multirow}
\usepackage{datetime}
\usepackage[mathscr]{eucal}
\usepackage{amssymb}
\usepackage{amsmath}
\usepackage{graphicx}
\usepackage{derivative}
\usepackage[utf8]{inputenc}
\usepackage[T1]{fontenc}
\usepackage{mathtools}
\usepackage[thinc]{esdiff}
\usepackage{wrapfig}
\usepackage{caption}
\usepackage{lscape}
\usepackage{rotating}
\usepackage{url}
\usepackage{babel}
\usepackage{translator}
\usepackage{pgfkeys,pgfcalendar}
\usepackage{graphicx}
\usepackage{multirow}
\usepackage{hhline}
\usepackage{diagbox}
\usepackage{eurosym}
\usepackage{amssymb}
\usepackage{pifont}
\usepackage{xcolor}
\usepackage{hyphenat}

\usepackage{acronym}
\usepackage{booktabs}
\def\preparecolorrefs#1{
  \setcounter{refindex}{0}
  \whiledo{\value{refindex}<#1}{
    \stepcounter{refindex}
    \expandafter\def\csname\therefindex color\endcsname{black}
  }
}

\makeatletter
\def\ps@pprintTitle{%
   \let\@oddhead\@empty
   \let\@evenhead\@empty
   \let\@oddfoot\@empty
   \let\@evenfoot\@empty}
\makeatother

\usepackage{color}
\usepackage{longtable}

\journal{}

\biboptions{numbers,sort}

\begin{document}

\begin{frontmatter}

\title{Machine Learning Applications to Diffuse Reflectance Spectroscopy in Optical Diagnosis; A Systematic Review}

\author{Nicola Rossberg$^1$}
\address{$^1$ Research Ireland Center for Research Training in Artificial Intelligence, School of Computer Science \& Information Technology, University College Cork, Ireland}
\ead{n.rossberg@cs.ucc.ie}

\author{Celina L. Li$^2$, Simone Innocente$^2$, Stefan Andersson-Engels$^{2, 3}$, Katarzyna Komolibus$^2$}
\address{$^2$Biophotonics@Tyndall, IPIC, Tyndall National Institute, Ireland}
\address{$^3$School of Physics, University College Cork, Ireland}
\ead{liyao.li@alumni.utoronto.ca, simoneinnocente98@gmail.com, stefan.andersson-engels@tyndall.ie, katarzyna.komolibus@tyndall.ie}

\author{Barry O'Sullivan$^4$, Andrea Visentin$^4$}
\address{$^4$Research Ireland Insight Centre for Data Analytics, School of Computer Science \& IT, University College Cork, Ireland}
\ead{osullivan.barry@ucc.ie, andrea.visentin@insight-centre.org}

\begin{abstract}

Diffuse Reflectance Spectroscopy has demonstrated a strong aptitude for identifying and differentiating biological tissues. However, the broadband and smooth nature of these signals require algorithmic processing, as they are often difficult for the human eye to distinguish. The implementation of machine learning models for this task has demonstrated high levels of diagnostic accuracies and led to a wide range of proposed methodologies for applications in various illnesses and conditions. In this systematic review, we summarise the state of the art of these applications, highlight current gaps in research and identify future directions. This review was conducted in accordance with the PRISMA guidelines. 77 studies were retrieved and in-depth analysis was conducted. It is concluded that diffuse reflectance spectroscopy and machine learning have strong potential for tissue differentiation in clinical applications, but more rigorous sample stratification in tandem with in-vivo validation and explainable algorithm development is required going forward.

\end{abstract}


\end{frontmatter}

\section{Introduction} \label{sec:Introduction}

Diffuse Reflectance Spectroscopy (DRS) is one of the most widely used optical techniques to analyse the properties of biological tissues \citep{vishwanath2010portable}. Its non-invasive nature and sensitivity to absorption related to tissue biomolecular content and scattering change, associated with subcellular morphology, make it an extremely powerful tool to analyse tissue composition, microstructure or oxygenation status, offering promising performance in applications such as cancer diagnostics and surgical guidance \citep{nogueira_evaluation_2021, veluponnar_diffuse_2023, akter2018medical, de_boer_towards_2018}. DRS signals are measured by delivering a typically white light source into the tissue and detecting diffusely reflected signals at a certain distance from the source, where the distance between the emitting and receiving fibres determines the tissue depth probed. Depending on the application and clinical objective, multiple illumination or detection fibres can be used to obtain more quantitative information and probe different depths. The light delivery and collection from tissue are often handled using optical fibres or fibre bundles. When incident on the tissue, the light undergoes scattering and absorption processes, which alter the light intensity across the measured spectrum \citep{mehta2023multimodal, veluponnar_diffuse_2023}. Since the optical properties of tissue and its inherent chromophores (e.g. oxygenated haemoglobin, deoxygenated haemoglobin, water, melanin) vary with wavelength, the wavelength range of DRS measurements limits the types of biomolecules that can be examined \citep{nogueira_evaluation_2021}. DRS has already shown significant potential to distinguish between cancerous and healthy tissues in multiple organs, including the breast, liver, colon, lung, and oral cavity \citep{akter2018medical}. For instance, recent studies have demonstrated its utility in breast cancer surgery, where the absorption and scattering information helped delineate tumour margins, potentially guiding surgeons toward more precise resection decisions \citep{de_boer_towards_2018}. Consequentially, DRS provides a valuable framework for understanding tissue morphology and pathology, making it a versatile tool for diagnosing a wide range of diseases and conditions.

Among other optical techniques (e.g. Raman spectroscopy, photoacoustics, optical coherence tomography) commonly used in medical diagnostics and surgical guidance, DRS offers unique benefits. Its ability to provide real-time, accurate information about both the underlying biological tissue composition and tissue structure makes it a versatile tool in clinical settings. Its simplicity, portability and relatively low cost of implementation allow for smooth integration into point-of-care devices, which can offer quicker access to necessary diagnostics, more frequent screening procedures and therefore improved patient outcomes \citep{kim2020optical}. Additionally, DRS probes can be integrated into small surgical devices, easing their medical deployment and enabling optically-guided surgical applications, for example, at the tip of a surgical drill \citep{duperron2019diffuse}. This ease of integration can help with a variety of applications, including low-visibility procedures such as revision knee arthroplasty or cancer resection procedures, which require clean resection margins to prevent re-occurrence \citep{li_frameworks_2023, brouwer_de_koning_toward_2018, baltussen_using_2020}. However, DRS remains limited by certain intrinsic and extrinsic factors. One of the main challenges is that DRS is predominantly surface-sensitive, and its probing depth is relatively shallow in highly absorbing and scattering tissues.  This limits its efficacy for certain diagnostic tasks,  particularly when deeper tissues need to be examined \citep{bess2019efficacy}. Additionally, inter and intra-patient variations in tissue heterogeneity can affect the accuracy and reliability of tissue characterisation. While inter-patient differences may be compensated through normalisation techniques  \citep{evers_diffuse_2012}, the intra-patient variations introduced by anatomical differences in tissue structure are more difficult to address and may require multiple measurements across different areas to fully capture tissue characteristics \citep{de2016using}. Moreover, being a contact measurement technique, DRS suffers from variations introduced by probe positioning, angle and changes in the pressure of the probe against tissue, which may influence the optical properties of the underlying tissue \citep{nogueira_evaluation_2021, blondel2021spatially}. Finally, the wide range of available instrumentation, different set-up geometries and data preprocessing and analysis techniques make it challenging to compare and synthesize conclusive findings across multiple studies.

Processing DRS signals commonly involves supervised machine learning (ML) algorithms. ML is a branch of artificial intelligence (AI) that refers to the design and implementation of algorithms which can learn from data and improve without necessitating the definition of domain-specified rules. By training algorithms to recognise patterns and make predictions, ML has achieved remarkable results in medicine, including applications in drug discovery, diagnostics, and remote patient monitoring \citep{vamathevan2019applications, shaik2023remote, kononenko2001machine}. ML is a common choice to process DRS signals due to the broadband nature of the DRS spectral data, with only subtle changes observed between different classes. Consequentially, the inter-class differences become obscured and difficult to differentiate for human observers. The ability of ML models to discern complex patterns simplifies this task, as they can decrease data complexity through feature reduction and identify otherwise invisible relations. Before the training process, ground-truth labels are derived through gold-standard medical procedures (e.g. for cancer diagnosis through histopathology), and models are subsequently trained with a supervised paradigm, using these labels. Models learn the patterns associated with each specified class and subsequently can classify new data with high accuracy \citep{sidey2019machine}. Some examples of previous implementations of ML for DRS processing include the classification of different ovine joints, the diagnosis of breast cancer and the estimation of free haemoglobin concentrations in blood bags \citep{gunaratne_wavelength_2020, chaudhry_breast_2023, can_estimation_2018}. Several previous reviews have considered the implementations of machine learning for spectroscopy techniques \citep{meza2021applications, zhang2022brief}. 

However, no reviews analysing the integration of DRS and ML in the field of medical diagnostics have been retrieved. This systematic review aims to bridge this gap by searching a wide range of databases to identify studies applying ML techniques to classify DRS signals in medical contexts and analysing these studies by their specific applications. It is vital to synthesize the outcomes of the existing studies, as DRS has repeatedly demonstrated strong potential for diagnosis in clinical applications. However, the generalisability and consequently, the reliability and integration, of the proposed methodology is limited due to the data scarcity and small sample size associated with medical research. To address the challenges posed by limited data, this review thus has three aims. First, to summarise and analyse the existing state of knowledge and permit drawing more robust conclusions regarding the feasibility of the proposed models and methodology. Second, to identify research gaps and areas for improvement that remain to be addressed. Finally, to identify key considerations for future research to improve the quality and consistency of studies in this field. These three aims are summarised in the research questions listed below. 

\begin{enumerate}
    \item \textbf{RQ 1:} What are the main research areas and applications of DRS and ML in current studies?
    \item \textbf{RQ 2:} What gaps and areas for improvement exist in the current research?
    \item \textbf{RQ 3:} What key considerations should future research address?
\end{enumerate}

The remainder of this review is organised as follows: Section \ref{sec:Methodology} describes the search strategies, inclusion criteria, and databases queried. Section \ref{sec:Results} provides an overarching analysis of the included studies, detailing the models, tissue types, and distribution of applications. A more detailed breakdown structured for the proposed application of the papers follows, covering the aims, methodologies, and results of identified studies. Finally, Section \ref{sec:Discussion} synthesises key findings, identifies gaps in the analysed literature and explores potential future directions for implementing ML in DRS analysis across various medical applications.

\section{Methodology} \label{sec:Methodology}

\subsection{Literature Search}

\begin{figure}[!h]
    \centering    \includegraphics[width=\textwidth,height=\textheight,keepaspectratio]{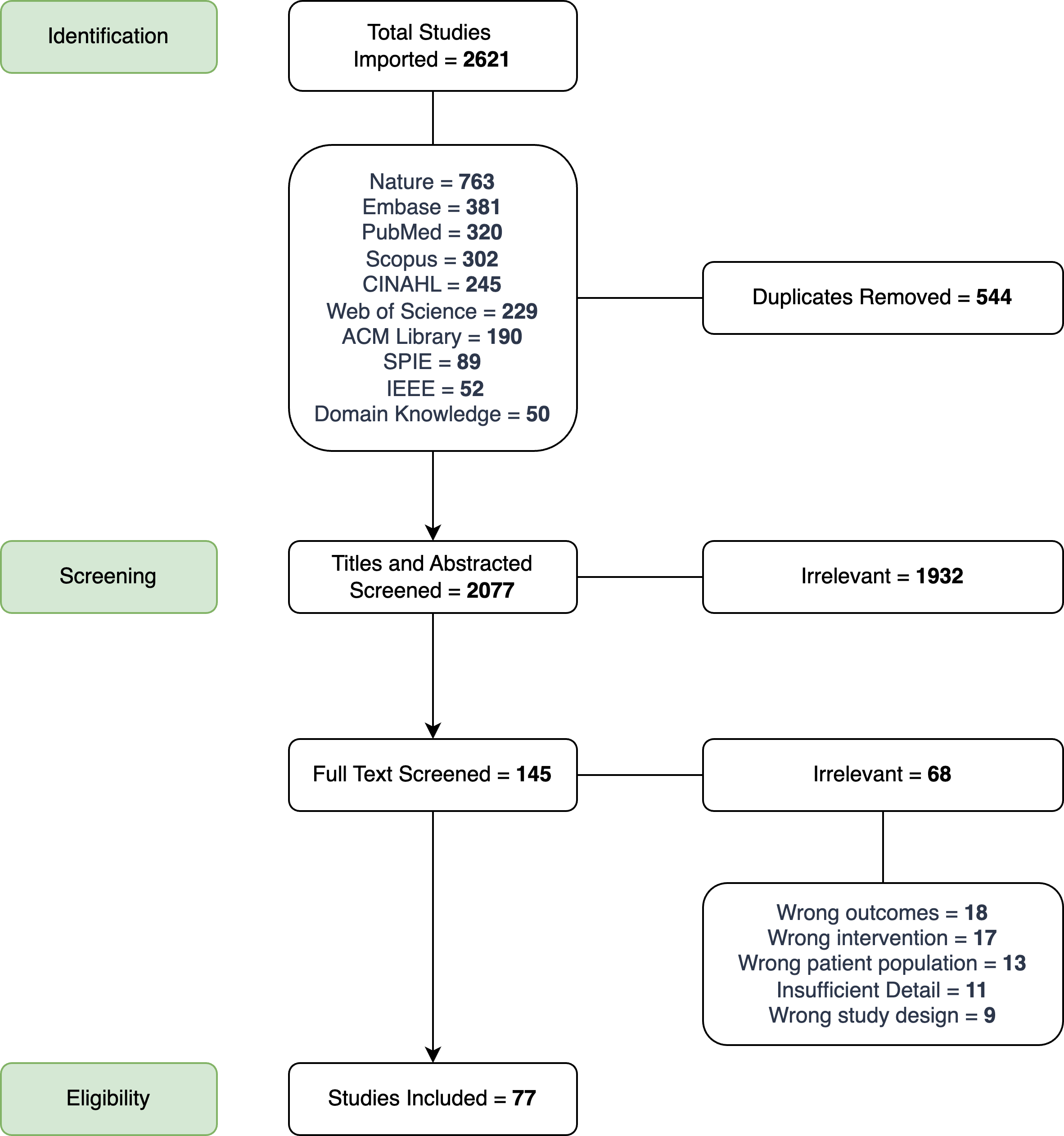}
    \caption{PRISMA diagram of literature search.}
    \label{fig:PRISMA_Diagram}
\end{figure}

This systematic review followed the principles of the PRISMA (Preferred Reporting Items for Systematic Reviews and Meta-analyses) guidelines \citep{page2021prisma}. The selected databases were PubMed, IEEEexplore, Nature, ACM Library, Scopus, Web of Science, CINAHL, and Embase and the last search was conducted on the 7th of May 2024. Databases were identified in collaboration with a library information specialist, and searches were restricted to studies published after 2004 due to the rapid increase in ML and deep learning potential during this time. Table \ref{tab:search phrases} shows the chosen search phrases for each database and the number of identified records. For some databases, more than one query was searched to broaden the scope of the identified results. An additional 89 and 50 studies were identified based on the SPIE library and the authors' domain knowledge, respectively. A series of inclusion and exclusion criteria were defined and are presented in Table \ref{tab:inclusion_exclusion}. PICOS (Population, Intervention, Comparison, Outcome, Study Population) principles were followed during this definition. Two screening phases were completed. The former considered only the title and abstract of articles. The latter screened the full text of previously included articles. A total of 77 articles were retained after full-text screening, and data and quality extraction were completed for these articles. Data extraction distilled the pertinent information from each study and denoted it in a standardised format to ease later analysis. Quality extraction assessed the risk of bias in each study across several criteria, including selective outcome reporting and improper preprocessing. This ensured the identification of potential bias in the included studies. Two authors completed article screening and extraction, and conflicts were resolved via discussion. Information about all retained studies alongside their key attributes is available on GitHub \footnote{\href{https://github.com/ncrossberg/SystematicReview/tree/main}{Link to GitHub Repo with all article information.}}. The PRISMA flowchart resulting from the identification, screening and extraction of articles for this review is shown in Figure \ref{fig:PRISMA_Diagram}. 

\renewcommand{\arraystretch}{1.7}
\begin{table}[!ht]
\centering
\caption{Searched databases, boolean phrases and number of retrieved papers.}
\label{tab:search phrases}
\resizebox{\textwidth}{!}{%
\begin{tabular}{|l|l|l|}
\hline
Venue                           & Search Phrase                                                                                                                                                                                                                                                                                                                                                                                                                             & Nr. Results \\ \hline
\multirow{2}{*}{Nature}         & \begin{tabular}[c]{@{}l@{}}("diffuse reflectance spectroscopy" OR "DRS") AND \\ ( "medicine") AND ( "artificial intelligence" OR "AI")\end{tabular}                                                                                                                                                                                                                                                                                       & 748         \\ \cline{2-3} 
                                & \begin{tabular}[c]{@{}l@{}}"Diffuse Reflectance Spectroscopy" AND \\ ("Artificial Intelligence" OR "Machine Learning")\end{tabular}                                                                                                                                                                                                                                                                                                       & 17          \\ \hline
\multirow{2}{*}{Embase}         & \begin{tabular}[c]{@{}l@{}}"artificial intelligence" OR "machine learning" OR "deep learning" OR "AI" OR "ML":ab,ti AND \\ "medicine" OR "medical" OR "healthcare" OR "medical informatics" OR "diagnosis" OR \\ "diagnostics" OR "surgery":ab,ti AND "diffuse reflectance spectroscopy" OR "DRS":ab,ti\end{tabular}                                                                                                                      & 361         \\ \cline{2-3} 
                                & \begin{tabular}[c]{@{}l@{}}'diffuse reflectance spectroscopy':ab,ti AND \\ ('artificial intelligence':ab,ti OR 'machine learning':ab,ti)\end{tabular}                                                                                                                                                                                                                                                                                     & 31          \\ \hline
PubMed                          & \begin{tabular}[c]{@{}l@{}}("Spectrum Analysis"{[}Mesh{]} AND "Artificial Intelligence"{[}Mesh{]}) \\ AND ("Medical Informatics"{[}Mesh{]})\end{tabular}                                                                                                                                                                                                                                                                                  & 320         \\ \hline
\multirow{2}{*}{Scopus}         & \begin{tabular}[c]{@{}l@{}}( TITLE-ABS-KEY ( "medicine" OR "medical" OR "healthcare" OR "medical informatics" OR \\ "diagnosis" OR "diagnostics" OR "surgery" ) ) AND ( TITLE-ABS-KEY \\ ( "diffuse reflectance spectroscopy" OR drs ) ) AND ( TITLE-ABS-KEY \\ ( "artificial intelligence" OR "machine learning" OR "deep learning" OR "AI" OR "ML" ) )\end{tabular}                                                                     & 196         \\ \cline{2-3} 
                                & \begin{tabular}[c]{@{}l@{}}TITLE-ABS-KEY "Diffuse Reflectance Spectroscopy" AND \\ TITLE-ABS-KEY ( "artificial intelligence" OR "machine learning")\end{tabular}                                                                                                                                                                                                                                                                          & 106         \\ \hline
\multirow{2}{*}{Cinhal}         & \begin{tabular}[c]{@{}l@{}}TI ( (MH "Medicine+") OR (MH "Medical Informatics") OR (MH "Diagnosis+") ) AND \\ AB ( (MH "Medicine+") OR (MH "Medical Informatics") OR (MH "Diagnosis+") ) AND \\ TI ( (MH "Artificial Intelligence+") OR (MH "Machine Learning+") OR \\ (MH "Deep Learning") ) AND AB ( (MH "Artificial Intelligence+") OR \\ (MH "Machine Learning+") OR (MH "Deep Learning") ) AND (MH "Spectrum Analysis+")\end{tabular} & 112         \\ \cline{2-3} 
                                & \begin{tabular}[c]{@{}l@{}}MH "Spectrum Analysis+" AND ((MH "Artificial Intelligence+") OR \\ (MH "Machine Learning+"))\end{tabular}                                                                                                                                                                                                                                                                                                      & 133         \\ \hline
\multirow{2}{*}{Web of Science} & \begin{tabular}[c]{@{}l@{}}(Topic) "diffuse reflectance spectroscopy" OR "DRS" AND \\ (Topic) "artificial intelligence" OR "machine learning" OR "deep learning" OR \\ "AI" OR "ML" AND (Topic) "medic*" OR "healthcare" OR \\ "medical informatics" OR "diagnos*" OR "surgery"\end{tabular}                                                                                                                                              & 124         \\ \cline{2-3} 
                                & \begin{tabular}[c]{@{}l@{}}(Topic) "diffuse reflectance spectroscopy" AND \\ (Topic) "artificial intelligence" OR "machine learning"\end{tabular}                                                                                                                                                                                                                                                                                         & 105         \\ \hline
\multirow{2}{*}{ACM Library}    & \begin{tabular}[c]{@{}l@{}}( "artificial intelligence" OR "machine learning" OR "deep learning" OR \\ "AI") AND ("medicine" OR "medical" OR "healthcare" OR \\ "medical informatics" OR "diagnosis" OR "diagnostics" OR "surgery") AND \\ (""diffuse reflectance spectroscopy" OR "DRS")\end{tabular}                                                                                                                                     & 277         \\ \cline{2-3} 
                                & "Diffuse Reflectance Spectroscopy" AND ("Artificial Intelligence" OR "Machine Learning")                                                                                                                                                                                                                                                                                                                                                  & 1           \\ \hline
\multirow{2}{*}{IEEE}           & \begin{tabular}[c]{@{}l@{}}"artificial intelligence" OR "machine learning" OR "deep learning" OR "AI" OR "ML" AND \\ "medicine" OR "medical" OR "healthcare" OR "medical informatics" OR "diagnosis" OR \\ "diagnostics" OR "surgery" AND "diffuse reflectance spectroscopy" OR "DRS"\end{tabular}                                                                                                                                        & 27          \\ \cline{2-3} 
                                & "Diffuse Reflectance Spectroscopy" AND ("Artificial Intelligence" OR "Machine Learning")                                                                                                                                                                                                                                                                                                                                                  & 25          \\ \hline
\end{tabular}%
}
\end{table}

\begin{table}[!h]
\caption{Inclusion and exclusion criteria in PICOS format.}
\label{tab:inclusion_exclusion}
\resizebox{\textwidth}{!}{%
\begin{tabular}{|l|l|l|}
\hline
 & Inclusion & Exclusion \\ \hline
Population & \begin{tabular}[c]{@{}l@{}}DRS Signals,\\ Human or animal tissue,\\ Less than 20\% simulated or missing data,\\ Mixed methods if the results are \\ reported separately for each method.\end{tabular} & \begin{tabular}[c]{@{}l@{}}Non-human or non-animal tissue,\\ Phantom studies,\\ Studies with Monte-Carlo \\ simulated data.\end{tabular} \\ \hline
Intervention & \begin{tabular}[c]{@{}l@{}}Machine Learning for \\ classification or regression.\end{tabular} & \begin{tabular}[c]{@{}l@{}}No Machine Learning,\\ Machine Learning for something \\ other than classification or regression.\end{tabular} \\ \hline
Comparison, & \begin{tabular}[c]{@{}l@{}}No comparison,\\ Other spectroscopic methods.\end{tabular} &  \\ \hline
Outcome & \begin{tabular}[c]{@{}l@{}}Objectives Measures of classification, \\ Objectives Measures of regression.\end{tabular} & \begin{tabular}[c]{@{}l@{}}No reported measures,\\ Biased measures,\\ Qualitative measures.\end{tabular} \\ \hline
\begin{tabular}[c]{@{}l@{}}Study \\ Population\end{tabular} & \begin{tabular}[c]{@{}l@{}}English Language,\\ Analytical Study,\\ Report Quantitative Outcomes,\\ Primary Studies,\\ Research studies.\end{tabular} & \begin{tabular}[c]{@{}l@{}}Non-English,\\ Systematic Review,\\ Meta Analysis,\\ Literature Review, \\ Before 2005.\end{tabular} \\ \hline
\end{tabular}%
}
\end{table}

\subsection{Data Collection and Extraction}

Data extraction and a quality assessment were conducted for each study included after full-text screening. The following information was extracted for each study:

\begin{enumerate}
    \item Authors, institutions, sponsors and publication type;
    \item Study design, study aims, light source and detector type, the number of classes, features and instances, the range of DRS signals recorded, and the type of illness treated;
    \item Group sizes, method of establishing ground truth, method of participant recruitment;
    \item Preprocessing, total sample size, group differences, and age, percentage missing and simulated cases, in-vivo or ex-vivo data and type of tissue (human or animal);
    \item Model hyperparameters and methods and type of test (regression or classification);
    \item Reported outcome and performance measures.
\end{enumerate}

A quality assessment was conducted for all studies, inquiring about five aspects of the study's quality: Selective outcome reporting, unsuitable outcome metric, improper preprocessing, unsuitable model choice and no data stratification. For each aspect, the probability of resulting biases was assessed as 'high', 'low' or 'unclear'. None of the retained studies were found to contain excessive bias, and hence, all were included in the analysis. 

\section{Results} \label{sec:Results}

The results section is organised as follows. First, an overview of all identified papers is presented, including analysis of tissue and measurement type, country of origin and implemented methodologies. Subsequently, papers are divided by their proposed application, and their aims, methodology and results are reviewed. Extensive research is conducted on the topic of cancer diagnosis, and papers are divided by whether they address cancer-related ailments and then further split by the specific application.

\subsection{Research Origin}
\begin{figure}[!ht]
    \centering    \includegraphics[width=\textwidth,height=\textheight,keepaspectratio]{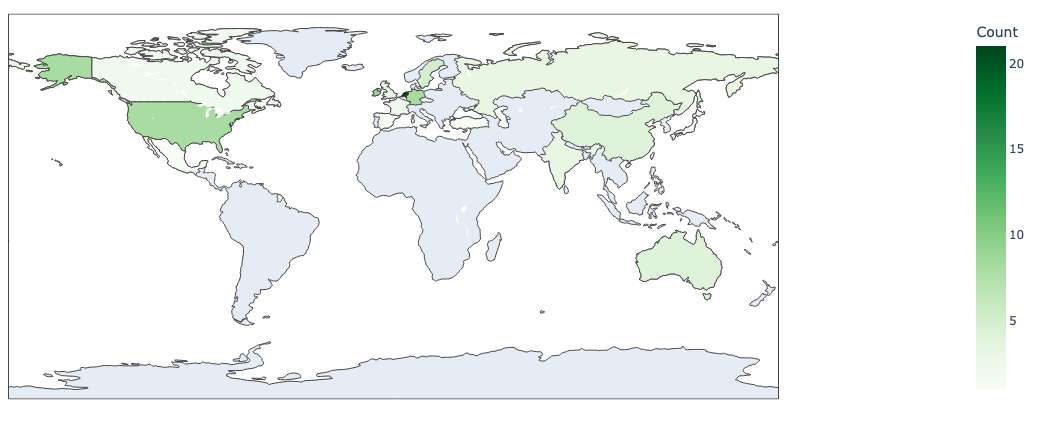}
    \caption{Geographical Representation of the country of origin of identified records.}
    \label{fig:CountryOfOrigin}
\end{figure}

The analysis of research origin shows that twenty-one of the identified studies were published in the Netherlands, making it the most frequent country of origin in the review. After the Netherlands, Ireland, Germany, and the United States are the most frequent publishers with nine, eight and eight records, respectively. Figure \ref{fig:CountryOfOrigin} shows a geographical representation of the countries of origin of the identified papers, with darker colours representing a larger number of papers published. One interesting trend is the considerable number of studies published by Western countries. Given the differences in skin colour, diet, sun exposure and other environmental factors, it is hence important to consider that the findings of these studies may not generalise to the rest of the world. Since DRS is sensitive to superficial layers, factors such as skin colour and sun exposure may be reasonably expected to introduce systematic bias into the measurement process. As such, future research should either aim to collect more diverse datasets during initial training or ensure the validation of developed systems on diverse data before implementation to ensure the safety of all patients. 

\subsection{Tissue and Examination Type}
\begin{figure}[!ht]
    \centering    \includegraphics[width=\textwidth,height=\textheight,keepaspectratio]{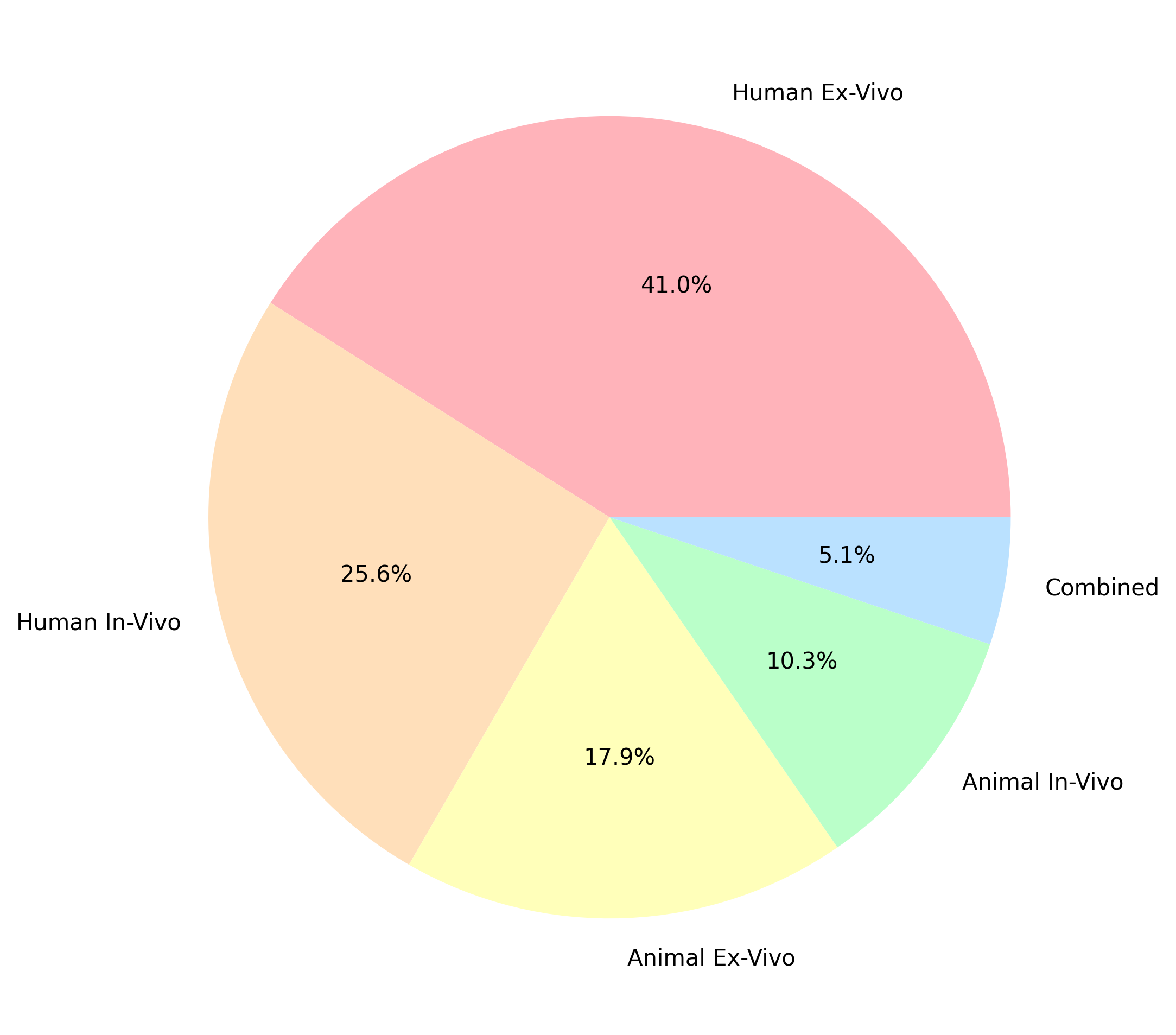}
    \caption{Frequency of the tissue and examination type.}
    \label{fig:VennTissueType}
\end{figure}

In addition to the origin of the included studies, the type of tissue and examination type of each study are analysed. The tissues can be differentiated by the origin (human or animal) and the place of measurement (in-vivo vs. ex-vivo). Figure \ref{fig:VennTissueType} shows the distribution of tissue and sampling type across the studies. A small subset of four studies considered both in and ex-vivo signals and is denoted in the figure under the caption of 'Combined'. The majority of studies rely on ex-vivo signals. This is attributed to the relative ease of attaining this data both physiologically and ethically. In addition, the classification of ex-vivo signals is comparably more straightforward due to fewer contaminants, more consistent illumination and consequentially less noise. Notably, human tissue is predominantly used for analysis despite the ethical and access barriers in place for patient protection. Human data is preferable due to the generalisability of findings and proximity to the proposed application's target group. As such, the large proportion of human, and especially in-vivo human data, is promising for the successful integration of the methodologies proposed by these studies. 

\subsection{Implemented Models} \label{subsec: Implemented  Models}
\begin{figure}[!ht]
    \centering    \includegraphics[width=\textwidth,height=\textheight,keepaspectratio]{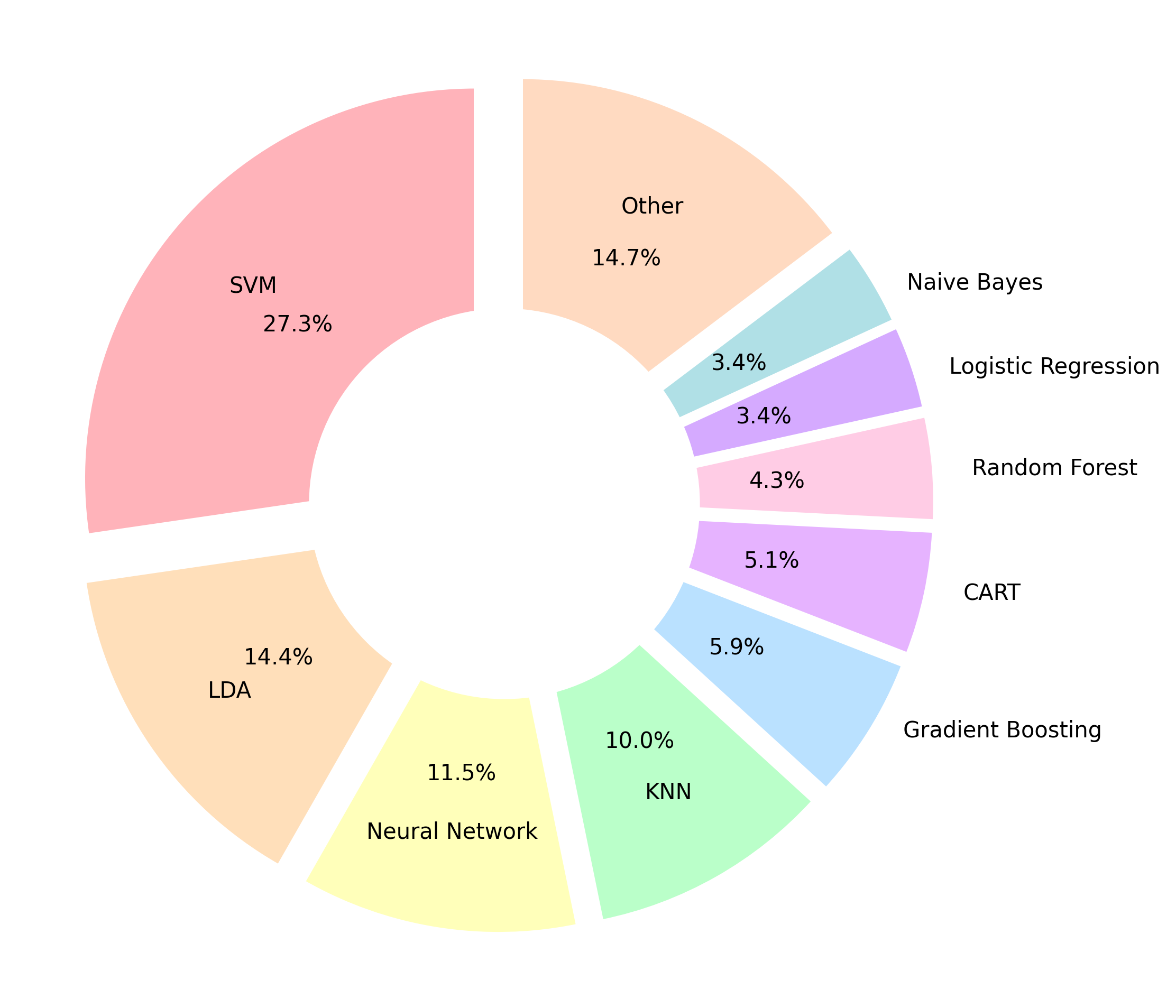}
    \caption{Frequency of the implemented machine learning models.}
    \label{fig:ModelType}
\end{figure}

Figure \ref{fig:ModelType} shows the distribution of ML models used in the retrieved studies. Note that several studies implemented more than one model, in which case all models are included in the chart. The most commonly implemented methods, alongside their advantages and limitations concerning the processing of DRS data, are discussed below. The most popular model, support vector machine (SVM), constituted 29.3\% of all implemented algorithms. In the majority of cases, SVM was implemented in combination with a feature extraction technique to reduce the data dimensionality before processing. The popularity of SVM may be linked to its ability to process non-linear relationships commonly found in DRS data due to variations in tissue structure and scattering \citep{ghosh2019study, begum2021diffuse}. Furthermore, given suitable preprocessing, SVM models are robust to the small sample sizes ubiquitous with medical research, hence decreasing the likelihood of overfitting. Finally, the models are capable of identifying an efficient decision boundary, which is vital for DRS data, where subtle differences between samples may be diagnostically significant \citep{valkenborg2023support, rossel2010using}.

Linear Discriminant Analysis (LDA) was the second most frequently implemented model, comprising over 15\% of the implemented algorithms. LDA offers advantages similar to SVM, including a clear decision boundary and low computational cost. However, it assumes Gaussian distributions within the data and may struggle with non-linear boundaries between classes, which makes it somewhat less suitable for the processing of DRS signals compared to SVM \citep{begum2021diffuse}. Furthermore, both SVM and LDA models rely greatly on successful preprocessing, which is both time-intensive and lacks a gold standard methodology, introducing a further drawback \citep{bona2007coal}. As such, while both SVM and LDA are capable of separating DRS data with low computational cost, their implementation is reliant on efficient preprocessing and the user's understanding of the data structure. 

These limitations are bridged by neural networks, which do not require prior data engineering and constitute over 12\% of the deployed models. Neural networks offer the advantage of automatic feature extraction and efficient pattern recognition during the training process, hence reducing reliance on domain knowledge and data engineering \citep{dara2018feature}. Furthermore, their pattern recognition may allow them to differentiate interlinked classes, which recommends them for the processing of noisy DRS data \citep{kretzschmar2005feedforward, chatterjee2021performance}. However, they are also more computationally expensive and less transparent than SVM or LDA models. Additionally, neural networks are generally large, data-hungry models which are prone to overfitting in smaller datasets \citep{lawrence1997lessons}. As such, while they have the potential to yield high classification accuracies, these networks require extensive datasets and should not be deployed liberally due to their high computational costs. 

Another frequently deployed model is K nearest neighbours (KNN), which functions based on the distance between a new data point and the K closest neighbours, as computed by a chosen distance metric. In the retrieved studies, over 10\% of deployed models were KNNs. The non-parametric nature of KNN recommends it for DRS signals, as it can capture non-linear and complex relationships. Additionally, KNNs are lazy learners and do not require explicit training, hence decreasing computational cost and time \citep{garcia2009completely}. However, similar to neural networks, these algorithms are computationally expensive as the distance computations between the test sample and training set are inefficient \citep{dhanabal2011review}. Furthermore, they are sensitive to noisy data, making effective preprocessing vital to their deployment for DRS data. One vital consideration for the implementation of KNNs for medical data is their associated lack of privacy. As new points are classified based on the existing dataset, sharing models inevitably include the sharing of collected data, which is not feasible in the field of medicine due to the associated privacy concerns. While privacy-preserving KKNs have been designed, involving separate cloud servers with high computational and communication costs, these systems are highly complex and may not be feasible for implementation by non-specialists \citep{song2022privacy}. Furthermore, the implementation of privacy-preserving measures is found to lower accuracy compared to traditional KNNs. As such, while these models are straightforward to deploy, the privacy concerns alongside their computational costs may not recommend them for clinical usage. 

\subsection{Publication Years of Studies}
\begin{figure}[!ht]
    \centering    \includegraphics[width=\textwidth,height=\textheight,keepaspectratio]{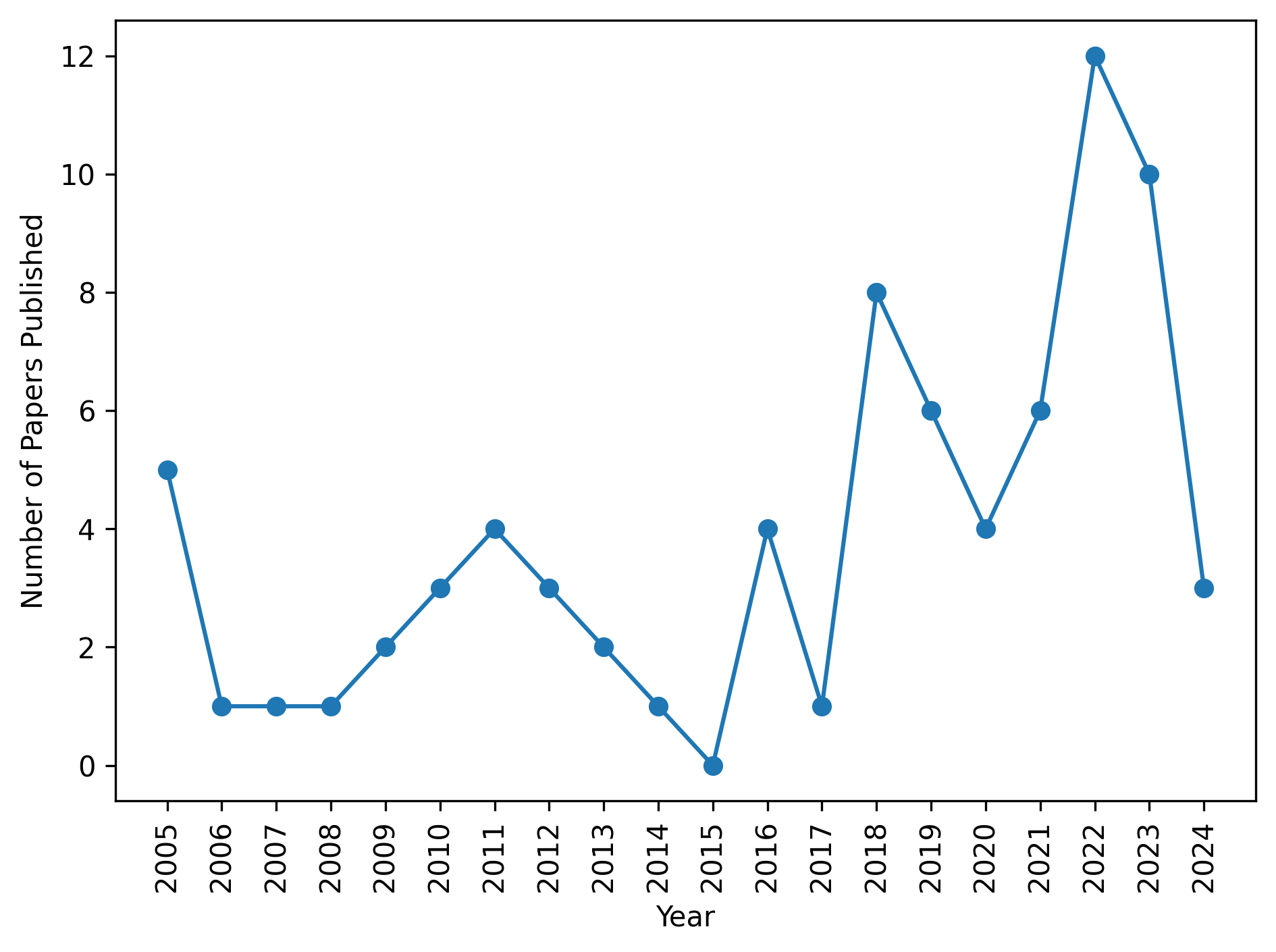}
    \caption{Trends in the number of publications per year.}
    \label{fig:PublicationYears}
\end{figure}

To assess the distribution of publications across time, the number of papers published per year is shown in Figure \ref{fig:PublicationYears}. Interestingly, the number of papers has increased in recent years, with the largest number of papers being published in 2022. This may be partially attributed to the rapid traction AI gained during this period due to the popularity of Generative AI models such as ChatGPT. It is worthwhile noting that the last search was conducted on the 7th of May 2024, which is the likely cause for the comparatively low number of records for the year 2024.

\subsection{Recorded Wavelengths}
\begin{figure}[!ht]
    \centering    \includegraphics[width=\textwidth,height=\textheight,keepaspectratio]{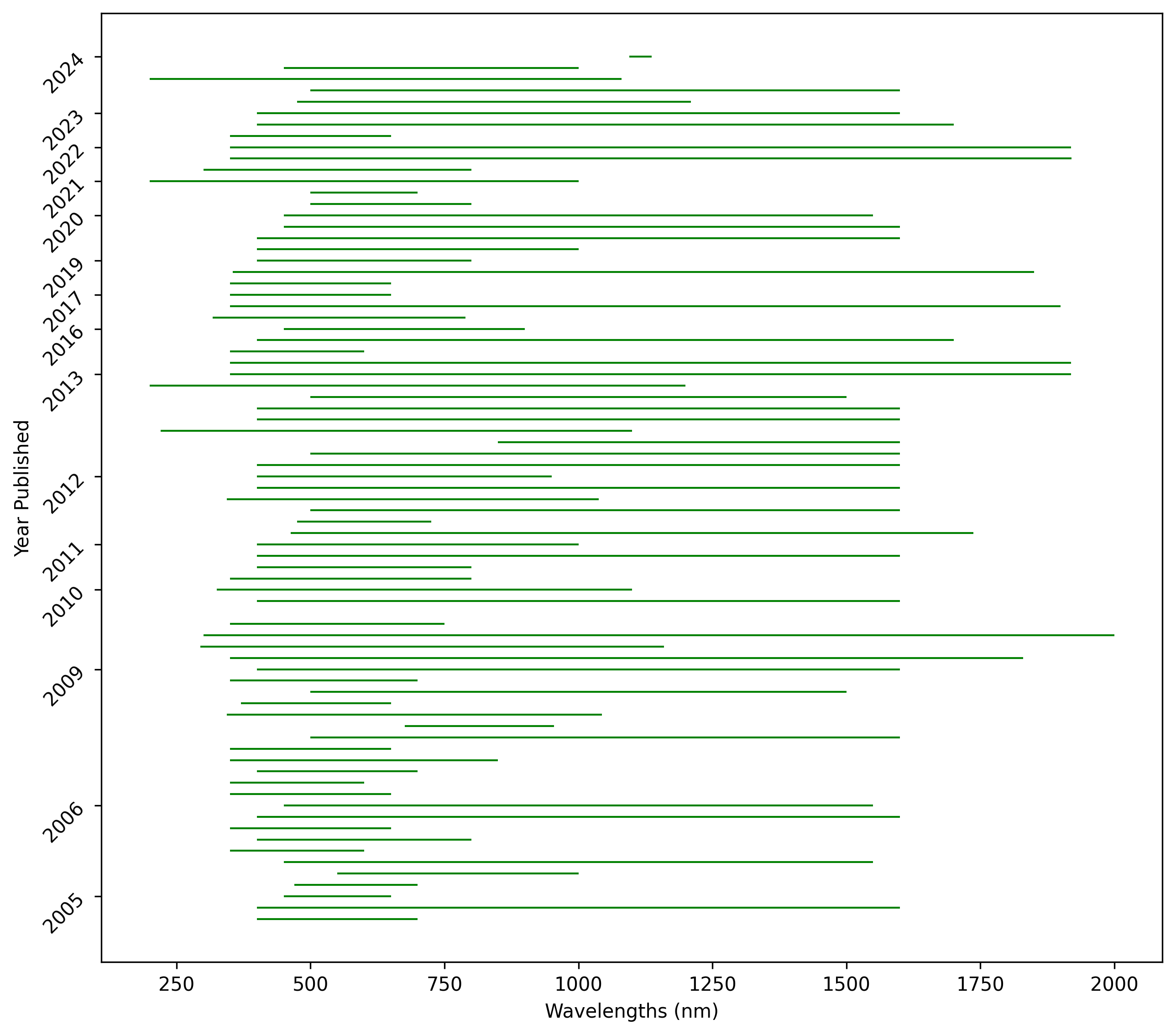}
    \caption{Ranges of recorded wavelengths over time.}
    \label{fig:Wavelength Distribution}
\end{figure}

The optical properties identified in a given tissue vary with wavelength. Consequentially, the recorded wavelength range defines the analysed biomolecules within the tissue \citep{nogueira_evaluation_2021}. To inspect the distribution of analysed wavelength over time and gain insight into developments in engineering developments and recording choices, the recorded ranges of all studies are plotted in Figure \ref{fig:Wavelength Distribution}. Most studies focus on the region between 400 - 1100 nm, with some extending the wavelength range to 1600 nm and several reaching over 1800 nm. The extension of the measurement range beyond 1100 nm requires an additional spectrometer and is referred to as Extended Wavelength DRS (EWDRS) in the literature. EWDRS offers some unique advantages, including the ability to capture reflectance in a wider range and, hence, capture tissue variations not visible in smaller measurement ranges. However, the measurement of EWDRS requires an additional spectrometer based on indium gallium arsenide to allow sensitivity to longer wavelengths. These spectrometers are expensive and less sensitive than traditional silicone ones, making them inaccessible to some labs and research settings. 

\subsection{Bias Analysis} \label{sec: Bias Analysis}
\begin{figure}[!ht]
    \centering    \includegraphics[width=\textwidth,height=\textheight,keepaspectratio]{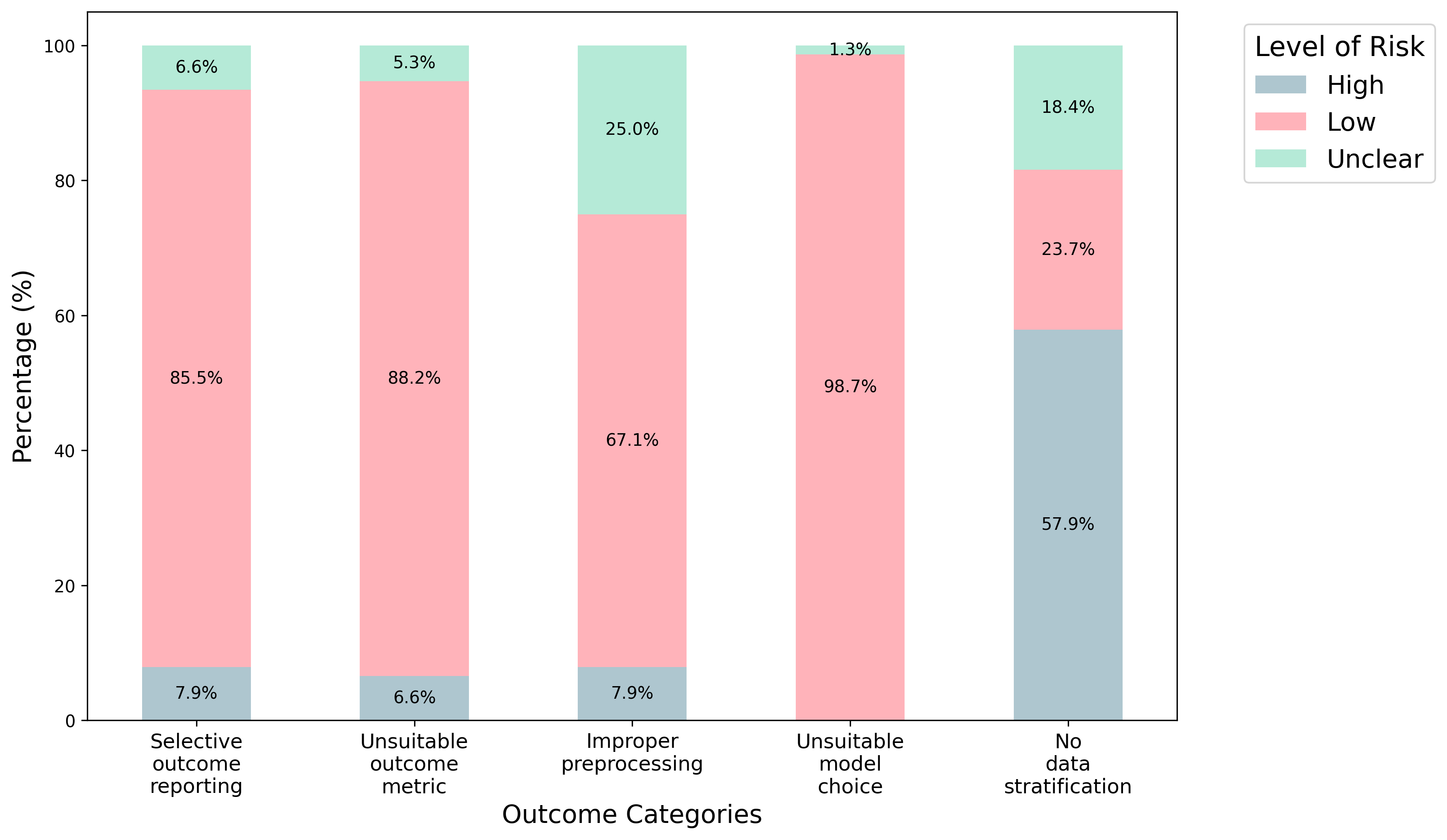}
    \caption{Frequency of risk level across the analysed sources of bias.}
    \label{fig:Bias Analysis}
\end{figure}

In line with PRISMA recommendations, a bias analysis of all studies was conducted, and results are presented in Figure \ref{fig:Bias Analysis}. The highest risk of bias was found regarding a lack of patient-wise data stratification, with 57.9\% of studies being considered at high risk. This is discussed in detail in Section \ref{subsec: patient stratification}. The stratification of patients between training and testing sets is a relatively novel concept in the field of photonics and hence, studies were not excluded on its basis. Interestingly, 7.9\% of studies were found at high risk of selectively reporting outcome data. This is most likely attributed to the sheer number of experiments conducted throughout the studies as well as the page limit of many conferences. It is, hence, likely that incomplete or selective outcome reporting is not completed in bad faith but rather a consequence of focussing studies on the most important findings. The high proportion of uncertainty regarding preprocessing bias stems from the lack of detailed reporting of implemented preprocessing techniques and information regarding data structures. Future research should include information regarding both the number of wavelengths and spectra recorded, as well as provide a detailed description of preprocessing steps to allow for the fair assessment of potential bias during these procedures. No single study was found to present an excessive risk of bias, and hence, no studies were excluded based on this.

The remainder of this results section is dedicated to the in-depth analysis of the retained studies. The retrieved articles are divided by the condition or illness they target. This division is chosen to permit a cross-study comparison of outcomes and a general synthesis of the feasibility of DRS and ML for each condition. It furthermore permits insight into which methodologies have been implemented for a given application. As such, specific recommendations for future research can be made for each domain. Going forward, section \ref{sec:Cancer-related Applications} presents research related to the identification and classification of cancerous tissue. Section \ref{sec: Other Applications} details the outcomes of studies targetting non-cancerous conditions and illnesses. 

\subsection{Cancer-related applications} \label{sec:Cancer-related Applications}

Cancer is one of the world's leading health concerns, with an estimated 20 million new cases and close to 10 million cancer deaths in 2022 \citep{bray2024global}. While mortality rates are found to decrease in the United States of America, partially due to decreased smoking and earlier detection, disparities between economic classes remain high, and incidences of cancer, including breast, prostate and cervical cancer, continue to increase \citep{siegel2024cancer}. 

The main application of DRS in the field of oncology is for the differentiation of cancerous and non-cancerous tissue. The implementation of DRS for such a task has demonstrated consistently high accuracy rates, identifying cancerous tissue across various settings, with the most popular applications including colorectal, breast and skin cancer. Currently, most applications complete classification on ex-vivo tissue due to the comparably easier access and decreased ethical constraints in this setting. Additionally, elucidating the capacity of DRS ex-vivo is a vital predecessor to in-vivo applications, as it provides a proof-of-concept and helps in optimisation of the device design. Once verified, DRS systems are generally implemented in-vivo to verify their efficacy in the targetted setting. Ultimately, most studies aim to implement DRS as a surgical support system to ease tissue identification during surgical procedures and improve positive resection margins, among other applications \citep{dahlstrand_extended-wavelength_2019, baltussen_diffuse_2017, gkouzionis_real-time_2022}. However, in-vivo measurements are associated with a range of difficulties, including interference from lights in the operating theatre, presence of body fluids (e.g. blood) affecting the measured spectrum, calibration or motion artifacts. As such, it is vital to optimise ex-vivo measurement beforehand to ensure the best initial results in this more challenging setting. Cancer-related illnesses were found to be the most prominent target in the majority of identified studies, with forty-six of the included research papers targeting the identification or resection of cancerous tissues. The analysis of the identified studies is divided by the types of cancer treated, and the following section provides a more in-depth analysis of the retained studies. Table \ref{tab:Cancer_Studies} shows the types of cancers as well as the included studies and their associated sections.

\begin{table}[!ht]
    \centering
    \resizebox{\textwidth}{!}{%
    \begin{tabular}{|c|c|c|}
    \hline
         Section&  Type of Cancer& Studies Included\\ \hline
         \ref{sec: Colorectal Cancer} & Colorectal Cancer & \citep{nazarian_real-time_2024, baltussen_diffuse_2017, baltussen_optimizing_2019, baltussen_tissue_2019, baltussen_using_2020, nogueira_benefit_2021, nogueira_evaluation_2021, nogueira_intestinal_2024, saito_nogueira_improving_2022, geldof_diffuse_2023, saito_nogueira_insights_2022, saito_nogueira_diffuse_2023, nogueira_tissue_2021}\\
         \ref{sec: Breast Cancer}& Breast Cancer & \citep{soares_diagnostic_2013, veluponnar_margin_2024, de_boer_towards_2018, zhu_diagnosis_2006, zhu_use_2005, chaudhry_breast_2023, de_boer_optical_2021, evers_diffuse_2013, nachabe_diagnosis_2011, majumder_comparison_2008, veluponnar_diffuse_2023} \\
         \ref{sec: Skin Cancer} & Skin Cancer & \citep{amouroux_classification_2009, murphy_toward_2005, kupriyanov_assessing_2023, kupriyanov_evaluating_2023, devpura_critical_2011} \\
         \ref{sec: Oral Cancer} & Oral Cancer & \citep{de_veld_autofluorescence_2005, jayanthi_diffuse_2011, brouwer_de_koning_toward_2018, skala_comparison_2007, a_v_kolpakov_decision_2023} \\ 
        \ref{sec: Brain Cancer} & Brain Cancer &  \citep{li_situ_2023, lai_automated_2020, skyrman_diffuse_2022} \\
        \ref{sec: Cervical Cancer} & Cervical Cancer & \citep{liu_study_2021, shaikh_comparative_2017, marin_diffuse_2005} \\
         \ref{sec: Liver Cancer} & Liver Cancer & \citep{tanis_vivo_2016, reistad_distinguishing_2022, keller_diffuse_2018} \\
        \ref{sec: Lung Cancer} & Lung Cancer & \citep{spliethoff_improved_2013, evers_diffuse_2012} \\
         \ref{sec: Gastrointestinal Cancer} & Oesophageal Cancer & \citep{gkouzionis_real-time_2022, nazarian_real-time_2022} \\
         \ref{sec: Other Cancers} & Other Cancers & \citep{geldof_layer_2022, geldof_toward_2024, werahera_diffuse_2016} \\
         \hline
    \end{tabular}
    }
    \caption{Included cancer-related studies}
    \label{tab:Cancer_Studies}
\end{table}

\subsubsection{Colorectal cancer} \label{sec: Colorectal Cancer}

The most common application of DRS and ML for cancer diagnostics is the identification of colorectal and bowel cancer. Colorectal cancer or colorectal carcinoma is the third most common and second most deadly cancer, with over 1.9 million new cases diagnosed and over 900,000 deaths in 2022 \citep{bray2024global}. In this systematic search, thirteen papers investigating the use of DRS in combination with ML to identify colorectal cancer were retrieved. All identified studies used human tissue for the analysis, with the number of DRS measurements varying from 117 \citep{baltussen_using_2020} to 6634 \citep{baltussen_optimizing_2019}. A variety of ML models were implemented for the diagnosis of colorectal cancer, with the most common one being SVM \citep{nazarian_real-time_2024, baltussen_diffuse_2017, baltussen_optimizing_2019, baltussen_tissue_2019, baltussen_using_2020, nogueira_benefit_2021, nogueira_evaluation_2021, nogueira_intestinal_2024, saito_nogueira_improving_2022, geldof_diffuse_2023}. The remaining studies used KNN \citep{saito_nogueira_insights_2022}, Classification and Regression Tree (CART) \citep{saito_nogueira_diffuse_2023}, Decision Tree \citep{nogueira_tissue_2021} or multiple models including SVM \citep{nazarian_real-time_2024, baltussen_optimizing_2019}. All models were implemented for the classification of cancerous and healthy tissue, with nine studies conducting binary classification between healthy and cancerous tissue \citep{saito_nogueira_insights_2022, nazarian_real-time_2024, baltussen_optimizing_2019, nogueira_benefit_2021, saito_nogueira_improving_2022, geldof_diffuse_2023, nogueira_tissue_2021, baltussen_using_2020, baltussen_diffuse_2017} and the remaining four studies including fat in addition to the other two classes for classification \citep{baltussen_tissue_2019, nogueira_evaluation_2021, nogueira_intestinal_2024, saito_nogueira_diffuse_2023}. All studies achieved reasonably high classification results with \citet{baltussen_diffuse_2017} reporting 95\% classification accuracy in the binary case, albeit the dataset was somewhat imbalanced with 63.4\% of cases belonging to the healthy class. However, similarly high results are reported by \citet{saito_nogueira_diffuse_2023}, who achieved 90.2\% accuracy (95.9\% sensitivity, 98.9\% specificity and 95.5\% Area Under the Curve (AUC)) and 94\% accuracy (96.9\% sensitivity, 98.9\% specificity and 96.7\% AUC) using probes with shorter (630 $\mu m$) and longer (2500 $\mu m$) source-detector separation distances, respectively. The majority of studies collected data ex-vivo, likely due to the difficulty of accessing the relevant tissue. Interestingly, the study utilising in-vivo data retained relatively high classification results, reporting 91\% accuracy \citep{baltussen_tissue_2019}. Given the comparative difficulty of in-vivo classification, these findings are promising of the potential for DRS integration into a surgical setting. Future research should aim at validating these accuracy values in a different dataset, testing the effects of diverse data and alterations in lighting and probe architecture. Given the difficulty of attaining data in-vivo, the construction of a transfer learning algorithm may be recommendable. Transfer learning is a deep learning algorithm which is capable of learning patterns from diverse datasets and applying these to a new domain, hence reducing the need for large dataset collection in a new task. Several studies by the group of Nogueira et al. evaluated the use of different probing depths for the differentiation of cancerous and healthy tissue. The majority of these works supported the use of a 2500 $\mu m$ source-detection distance and suggested that it may lead to better classification results. One possible explanation for these findings is that the biomolecular differences important for colorectal cancer detection occur at deeper tissue layers and are hence more effectively detected by the 2500 $\mu m$ probe. However, one study by the same team found the shorter probe to perform better for the differentiation of tumours and mucosal tissue \citep{nogueira_tissue_2021}. While most considered studies thus imply that deeper probing depth is beneficial for the discrimination of colorectal cancer tissue, some contrary findings exist, and future research may wish to replicate the current studies to ascertain the most suitable source-detector distance for the detection of colorectal cancer. Overall, these studies demonstrate high classification accuracies in both the binary case, differentiating healthy and cancerous tissue, as well as in the multiclass scenario, differentiating cancerous tissue from healthy colon tissue and fat.

\subsubsection{Breast cancer} \label{sec: Breast Cancer}

Breast cancer is the second most commonly diagnosed cancer and second leading cause of death among women, with early diagnosis being vital for prevention and successful treatment \citep{bray2024global}. Eleven papers considering the use of DRS for breast cancer identification were retrieved. Seven of the identified studies employed SVM for classification \citep{soares_diagnostic_2013, veluponnar_margin_2024, de_boer_towards_2018, zhu_diagnosis_2006, zhu_use_2005, chaudhry_breast_2023, de_boer_optical_2021}. The remaining studies employed CART \citep{evers_diffuse_2013, nachabe_diagnosis_2011}, Logistic Regression \citep{majumder_comparison_2008} or multiple methods including KNN, LDA and Gradient Boosting \citep{veluponnar_diffuse_2023, chaudhry_breast_2023, veluponnar_margin_2024}. The majority of studies focusing on the binary differentiation of healthy and malignant breast tissue \citep{de_boer_optical_2021, de_boer_towards_2018, veluponnar_diffuse_2023, zhu_use_2005, chaudhry_breast_2023, evers_diffuse_2013, zhu_diagnosis_2006}, while the remaining research conducted multiclass classification with three classes \citep{soares_diagnostic_2013}, four classes \citep{majumder_comparison_2008}, and five classes \citep{nachabe_diagnosis_2011}. The binary classification achieved high accuracies, with classification quality varying as a function of set-up and preprocessing. One interesting finding by \citet{chaudhry_breast_2023} showed that an extended wavelength range may improve classification accuracy. Where a standard DRS range of 450 - 900 nm achieved a sensitivity of 40\% and a specificity of 71\%, an extended wavelength range improved results to a sensitivity of 94\% and a specificity of 91\%, emphasising the potential of larger collection ranges. This may be attributed to two potential causes. From a biological perspective, the differences in biomolecular and microstructures relevant to the distinction of the classes may reflect light in the extended wavelength region, and the range extension, therefore, collects more pertinent information. From an ML perspective, an increased wavelength range provides more data, which may increase model precision and, hence, improve discrimination. However, this is dependent on efficient model training as an extended feature range may also lead to overfitting. As such, an increased wavelength range is not a guarantee for increased model performance. Additionally, the increase in expenses associated with recording an extended wavelength range should be taken into consideration, as it may bar lower-income settings from the use of this technology. In the multiclass case, classification performance varied widely with \citet{majumder_comparison_2008} reporting accuracies between 28\% and 86\% for the differentiation of four classes, including cancerous, benign and healthy tissue. On the other hand, \citet{nachabe_diagnosis_2011} reached accuracies as high as 94\% for the classification of five different types of breast tissue. Overall, the studies concluded that DRS is a viable method for differentiating healthy and cancerous breast tissue due to consistently high classification accuracies. However, the majority of the included studies considered only ex-vivo tissues. \citet{veluponnar_margin_2024} is the only included study using in-vivo tissues, and the authors found encouraging results with a RUSboost model achieving a Mathews correlation coefficient (MCC) of 0.76 when trained on ex- and in-vivo data and tested on in-vivo data only. While the high classification accuracies of the ex-vivo case are hence not directly transferable to in-vivo data, initial findings are promising and future research may wish to further validate implemented models on in-vivo data. One potential issue discussed in previous literature by the team of de Boer, amongst others, is the effect of intra-patient variability in breast tissue due to differences in the composition and structure between patients \citep{de2016using, fanjul2018intra}. This may be especially problematic for the implemented SVM models, as the construction of a decision boundary becomes more challenging for heterogeneous classes as it may not align for all members of the 'healthy' class. One possible solution would be the usage of non-linear kernels, which map data to higher dimensions, but this would not be able to compensate for extensive heterogeneity within one class. Alternatively, the distinction of multiple tissue structures within the 'healthy' cluster may compensate for variation in tissue composition, as it would lead to more well-defined individual classes. Consequently, alterations in, for example, fat content could be recognised as such where previously, the heterogenous structure would have obscured the differences between healthy and cancerous tissue. The implementation of an algorithm capable of classifying multiple healthy tissue structures alongside tumours may, therefore, alleviate the concerns of intra-patient variation. However, future research should validate this through experimental implementation. 

\subsubsection{Skin cancer} \label{sec: Skin Cancer}

Skin cancer is one of the most common forms of cancer worldwide, with incidences increasing in recent decades, most likely due to environmental factors including increasing ultraviolet radiation \citep{furriel2024artificial, artosi2024epidemiological}. In this review, five previous studies considering the use of DRS for the identification of skin cancer were identified \citep{amouroux_classification_2009, murphy_toward_2005, kupriyanov_assessing_2023, kupriyanov_evaluating_2023, devpura_critical_2011}. Due to the ease of accessing the tissue, all five studies utilised in-vivo signals to classify healthy and pathological tissue, with four studies using human tissue and \citet{amouroux_classification_2009} recording mouse skin signals. All five studies collected spectra from multiple tissue classes with three studies conducting binary one vs. one or one vs. rest classification \citep{devpura_critical_2011, murphy_toward_2005, amouroux_classification_2009} and the remaining studies, conducted by the same team, implementing a multiclass classification of healthy tissue, compensatory hyperplasia, atypical hyperplasia and dysplasia \citep{kupriyanov_assessing_2023, kupriyanov_evaluating_2023}. The method of preprocessing and classification differed considerably between the studies. \citet{devpura_critical_2011} used Principal Component Analysis (PCA) to reduce the 226 features recorded from 350 to 850 nm to 4 principal components, classified by a KNN. The authors achieved a specificity of 94.8\% and a sensitivity of 87.7\% in differentiating healthy and malignant tissue from three collection sites. \citet{murphy_toward_2005} collected four types of tissue: common naevus, dysplastic naevus, in situ melanoma and invasive melanoma and conducted binary classification, comparing the different tissue types. The best classification result was achieved in the differentiation of common naevus from invasive melanoma (73\% specificity, 73\% sensitivity) and common naevus from dysplastic naevus (69\% specificity, 77\% sensitivity) using a Neural Network. \citet{amouroux_classification_2009} collected signals of four skin types: healthy tissue, compensatory hyperplasia, atypical hyperplasia and dysplasia of in-vivo mouse skin. The authors implemented a KNN for one vs. one classification and reported specificities ranging from 64\% to 82\% and sensitivities ranging between 46\% and 96\% depending on the task. The best performance was achieved for the differentiation of signals from normal tissue and dysplasia. Finally, the two studies by the team of Kupriyanov et al. \citep{kupriyanov_assessing_2023, kupriyanov_evaluating_2023} conducted multiclass classification of four different tissue types. The studies aimed to compare the success of DRS and autofluorescence for the classification task. DRS was found to outperform autofluorescence in both studies. \citet{kupriyanov_assessing_2023} presented the accuracy of five different classifiers for different source-detector separations, with the best classification accuracies achieved by Bagging Decision Trees with accuracies over 90\%. \citet{kupriyanov_evaluating_2023} conducted a similar study but implemented PCA before classification, achieving accuracies of 90.2\% through the use of an SVM classifier. Based on the analysed studies, it is clear that DRS can yield high classification results in applications to skin cancer. However, the range of accuracies is large, and studies varied in the reported methodologies and outcomes. Due to the ease of access, in-vivo data recording is straightforward, permitting the acquisition of representative samples. However, this introduces other confounds, such as varying measurement conditions and changes in source-detector distance. Additionally, the inter-personal differences in skin structure are large, adding additional noise to the data and making processing more difficult. Future research may wish to consider how to correct or account for these variations through, for example, the collection of larger, more diverse datasets or the construction of a transfer learning model. The latter would allow the combination of multiple datasets within an algorithm to stabilise performance and account for variations between patients. Overall, these results encourage the ability of DRS to identify skin cancer and differentiate it from other types of tissue. 

\subsubsection{Oral cancer} \label{sec: Oral Cancer}

Oral cancers, including cancers of the oral cavity, are collectively the 16th most common cancer worldwide, with more than 389,000 new diagnoses and over 188,000 deaths estimated worldwide in 2022 \citep{bray2024global}. Risk factors include alcohol and tobacco use as well as cigarette smoking. Five studies using DRS to identify oral cancer were retrieved in this review \citep{de_veld_autofluorescence_2005, jayanthi_diffuse_2011, brouwer_de_koning_toward_2018, skala_comparison_2007, a_v_kolpakov_decision_2023}. Three of the retrieved studies used in-vivo data \citep{de_veld_autofluorescence_2005, jayanthi_diffuse_2011, a_v_kolpakov_decision_2023}, likely owing to the ease of accessing the targeted region. The range of wavelengths considered varied between studies, with two recording from 400 - 700 mm \citep{de_veld_autofluorescence_2005, jayanthi_diffuse_2011}, one considering 400 - 1600 nm \citep{brouwer_de_koning_toward_2018}, another 325 - 1100 nm \citep{a_v_kolpakov_decision_2023} and the other analysing 350 - 600 nm \citep{skala_comparison_2007}. More recent studies tend to record larger wavelength ranges, likely due to technological advancements and the corresponding ease of signal collection. All studies utilised feature selection or extraction techniques before analysis to reduce the feature space and increase processing speed. The most popular method for feature extraction was PCA \citep{de_veld_autofluorescence_2005, jayanthi_diffuse_2011, skala_comparison_2007}. The remaining papers utilised downsampling with a factor of five, systematically reducing the feature space \citep{brouwer_de_koning_toward_2018} and the selection of significant wavelength ranges \citep{a_v_kolpakov_decision_2023}. All studies used binary classification with \citet{ brouwer_de_koning_toward_2018, skala_comparison_2007, a_v_kolpakov_decision_2023} differentiating healthy and pathological tissues, while \citet{de_veld_autofluorescence_2005, jayanthi_diffuse_2011} conducted multiple binary classifications to differentiate four different healthy and cancerous tissue classes. The implemented models included Linear Regression \citep{de_veld_autofluorescence_2005}, LDA \citep{jayanthi_diffuse_2011}, SVM \citep{skala_comparison_2007, brouwer_de_koning_toward_2018}, Random Forest, Logistic Regression and Gradient Boosting \citep{a_v_kolpakov_decision_2023}. All studies reported high accuracies with \citet{a_v_kolpakov_decision_2023} achieving 96\% accuracy using Logistic Regression. The ease of access to oral tissue permits the integration of DRS into the workflow and hence recommends the technique for the non-invasive identification of malignant tissue. The reviewed research demonstrates that DRS is capable of identifying cancerous tissue in both in and ex-vivo settings at high levels of accuracy. However, the studies remain limited by the small datasets employed. Moving forward, the collection of larger, more diverse datasets should be prioritised to validate findings and ensure the generalisability of the trained models. Factors such as smoking and tobacco use, skin colour and the effects of teeth whitening procedures should be taken into consideration as they may affect the reflective properties of tissue and hence introduce systematic bias. The implemented models are relatively small ML algorithms, such as SVM and LDA, which allow fast training and classification times. This recommends these systems for real-time surgical implementation, and future research may wish to focus on the assessment and optimisation of prediction speed. In the same vein of research, feature reduction through extraction or selection decreases processing times. However, different methodologies were proposed in the retrieved studies, with no consistent conclusion regarding the best approach. Future research may wish to elucidate whether a given feature selection or extraction technique consistently outperforms other approaches in this domain. Next to improvements in processing time, this may also permit the design of smaller recording devices if the necessary wavelength range is reduced, easing surgical integration. Overall, the high classification success supports DRS as a valid method for identifying cancerous oral tissue. 

\subsubsection{Brain cancer} \label{sec: Brain Cancer}

The incidence of brain cancer worldwide was recorded at over 347,000 new cases in 2019, with more than 246,000 recorded deaths from the illness \citep{ilic2023international}. This review identified three studies attempting to diagnose brain cancer using DRS \citep{li_situ_2023, lai_automated_2020, skyrman_diffuse_2022}. The study by \citet{li_situ_2023} conducted binary differentiation of four tissue types using a Gradient-Boosting Decision Tree. The authors recorded ex-vivo spectra of human brain tissue in the range of 500 - 800 nm. Based on these spectra, seven absorption and scattering parameters were extracted, which were utilised as input features to the ML algorithm. The authors achieved accuracies varying from 73\% to 94\%, with the best classification results recorded for the differentiation between normal white matter and high-grade glioma. \citet{lai_automated_2020} conducted a binary classification to differentiate white and grey matter in ex-vivo porcine tissue samples. The authors collected data from eight animals in the 676 - 954 nm range and implemented an SVM, achieving an AUC of 96\% (96\% sensitivity, 96\% specificity). The work by \citet{skyrman_diffuse_2022} implemented Random Forest for the multiclass classification of low and high-grade gliomas with normal white and grey matter. The authors recorded spectra of human ex-vivo tissue in the range of 450 - 1600 nm, and fit parameters were extracted for classification. Low-grade glioma was differentiated from normal white and grey matter with a specificity of 82.7\% and a sensitivity of 82\%. High-grade glioma was classified with a specificity of 43.2\% and a sensitivity of 93.3\%. The accuracies reported in these studies are encouraging the possibility of successful integration of ML and DRS for brain cancer diagnosis. However, several barriers to implementation remain. First, the high sensitivity of the surgical site makes the implementation of black-box machine-learning algorithms precarious. As such, the integration of explainable AI to ensure algorithm safety is especially vital in this context, as wrong classifications could have detrimental outcomes. Additionally, the acquisition of in-vivo data is challenging due to the inaccessibility of the site and the associated ethical constraints. However, given the sensitivity of the work, in-vivo training validation is crucial and should be prioritised in future research. 

\subsubsection{Cervical cancer} \label{sec: Cervical Cancer}

Cervical cancer is a major global health issue and is the fourth most common cancer in women worldwide. As the target of the World Health Organisation's elimination of cervical cancer program, the rates of cervical cancer have been decreasing, but the need for early diagnosis and treatment remains a priority \citep{yang2024cervical, caruso2024cervical}. During the systematic search, three studies implementing DRS in conjunction with ML for the identification of cancerous cervical tissue were identified \citep{liu_study_2021, shaikh_comparative_2017, marin_diffuse_2005}. All three studies use in-vivo data for the classification of cancerous and healthy tissue with \citep{liu_study_2021, shaikh_comparative_2017} conducting binary classification while \citep{marin_diffuse_2005} implemented multiclass classification to differentiate varying malignancy levels. The studies implemented feature engineering before classification with \citep{shaikh_comparative_2017, marin_diffuse_2005} applying PCA, while \citep{liu_study_2021} used characteristic curve shapes to identify eleven relevant features. The studies classified the tissue with consistently high accuracy rates. \citet{liu_study_2021} reported 95\% accuracy for the differentiation of healthy and cancerous tissue using an SVM. \citet{shaikh_comparative_2017} achieved a specificity of 95\% and a sensitivity of 85\% for the binary differentiation of cervical tumour and healthy cervical and vaginal tissue using an LDA. Finally, \citet{marin_diffuse_2005} differentiated three stages of cancer from healthy tissue with respective accuracies of 98\% for severe dysplasia, 94\% for moderate dysplasia, 93\% for mild dysplasia and 73\% for healthy tissue. The results are encouraging, especially in-vivo, and the next steps should focus on promoting algorithm integration. To this end, replication of results in a larger dataset should be targeted, assessing how individual differences may affect diagnostic accuracy. Furthermore, the changes associated with birth and menopause should be investigated to test whether the high accuracy rates are restricted to women of a certain age bracket. From a machine-learning point of view, algorithm speed and potential for real-time integration should be assessed. Researchers should optimise processing times through feature selection and efficient model choice. Furthermore, explainability measures should be integrated into the algorithm at this stage to conform with health regulations and prevent bias. Additionally, patient-wise data stratification is necessary for future research to prevent data leakage associated with individual differences.

\subsubsection{Liver cancer} \label{sec: Liver Cancer}

In 2022, liver cancer was the third most frequent cause of cancer deaths worldwide, also being the only one of the top five deadliest cancers to increase annually in occurrence as of 2019 \citep{siegel2019cancer, bray2024global}. Liver cancer occurs more frequently in developing countries with risk factors including hepatitis B and C, fatty liver disease, and various dietary exposures \citep{starley2010nonalcoholic, center2011international}. With only 15\% of patients being eligible for surgical removal, the prognosis of liver cancer is very poor, hinging on early diagnosis for the best outcomes \citep{anwanwan2020challenges}. As such, a way of identifying cancerous liver tissue in a surgical setting is paramount. This review identified three previous studies employing DRS in combination with ML to this end \citep{tanis_vivo_2016, reistad_distinguishing_2022, keller_diffuse_2018}. The work by \citet{tanis_vivo_2016} analyses in-vivo tissue signals of liver parenchyma, implementing a CART algorithm for the classification of tissues which have and have not been treated with chemotherapy. The authors collect reflectance spectra in the 400 - 800 nm range. The overall achieved classification results are a sensitivity of 95\% and a specificity of 92\%. This study makes the important contribution of in-vivo recordings to the diagnosis of liver cancer using DRS. However, one potential concern is the small number of seventeen specimens from which over 400 spectra are extracted. This may lead to data leakage between the training and test set due to shortcut learning of individual differences instead of the markers of cancerous and healthy tissue. Shortcut learning refers to an algorithm learning a simpler proxy pattern which functions well for classification in a given dataset but does not generalise to new data \citep{brown2023detecting}. As such, future research should endeavour to collect a larger number of source specimens and stratify the train-test split by specimen or patient to prevent data leakage. The studies by \citet{reistad_distinguishing_2022} and \citet{keller_diffuse_2018} both utilised ex-vivo tissue signals to differentiate between healthy and tumorous liver tissue. Both studies implemented SVM models to classify the tissue with \citet{reistad_distinguishing_2022} testing whether an extended wavelength range (450 - 1550 nm) in combination with PCA would yield improved results. The authors reported high accuracies of 100\% accuracy (100\% specificity, 99\% sensitivity, 100\% MCC) using two components extracted from the 450 - 1550 nm range. \citet{keller_diffuse_2018} investigated the use of different lighting for the best differentiation of tissues and found the use of selected LEDs to reach the highest accuracies, reporting results of 98.76\% accuracy (99.15\% specificity, 98.44\% sensitivity). These studies demonstrate the potential of extended wavelength DRS and adjusted lighting modalities for the improvement of classification accuracy of liver cancer. While both of these studies boast high accuracy rates, the equipment necessary for their implementation is more complex than that of standard DRS. This, in combination with the high incidence of liver cancer in developing countries, raises the question of whether the proposed methodology is a realistic possibility for the regions in need. As such, future research may wish to investigate whether more cost-efficient tools may be capable of achieving similar accuracy rates to promote implementation in developing countries. In addition, the studies of \citet{reistad_distinguishing_2022} and \citet{keller_diffuse_2018} both propose adaptions to standard DRS-measuring set-ups, which may improve accuracy levels. However, both alterations are conducted independently, and future research may wish to elucidate which combination of wavelength ranges and lighting modalities functions best, ideally within an in-vivo setting.

\subsubsection{Lung cancer} \label{sec: Lung Cancer}

In 2022, lung cancer was the most frequently diagnosed cancer worldwide, with almost 2.5 million new cases, constituting 12.4\% of all cancers diagnosed globally \citep{bray2024global}. This review identified two studies implementing DRS in combination with ML for the detection of cancerous lung tissue \citep{spliethoff_improved_2013, evers_diffuse_2012}. The work by \citet{spliethoff_improved_2013} tested whether the implementation of a DRS probe at the tip of a biopsy needle was feasible for the discrimination of healthy and cancerous tissue and the differentiation of two types of cancerous tissue. The authors recorded signals in the range of 400 - 1600 nm on ex-vivo human lung tissue. Physiological parameters were extracted from the recorded data as input to a CART classification algorithm. The differentiation of tumorous and healthy lung tissue achieved a specificity of 86\% and a sensitivity of 98\%. The classification of necrotic and vital tumours was accomplished with a specificity of 80\% and a sensitivity of 88\%. The second study by \citet{evers_diffuse_2012} also implemented a binary differentiation of normal and tumorous tissue using ex-vivo human tissue signals in the 500 - 1600 nm range. The authors used a Partial Least Squares Discriminant Analysis and a Decision Tree algorithm, with the latter achieving the highest performance, discriminating healthy and tumorous lung tissue with 89\% sensitivity, 79\% specificity and 84\% accuracy. Both studies concluded that DRS is capable of differentiating healthy and tumorous lung tissue in lung biopsy procedures, with the findings of \citet{spliethoff_improved_2013} showing that the differentiation of necrotic and vital tumour is best completed through the use of fluorescence spectroscopy. Several objectives for future research may be formulated based on these findings. First, a validation of the algorithmic performance on in-vivo tissue is necessitated to confirm whether reflective characteristics remain consistent and algorithmic performance is retained at an acceptable level. Second, the fusion of multiple optical techniques may be considered, given the superior performance of fluorescence spectroscopy for the differentiation of the two tumorous tissues. The combination of multiple optical techniques has been piloted successfully in other applications \citep{davey_analysis_2021, dao2023breast, hamdy2023monitoring} and integration for the classification of lung cancer may allow improved classification accuracy. Finally, while the implemented ML models achieve decent results, it should be investigated whether the implementation of more complex algorithms, such as deep neural networks, would improve results. Neural networks have inherent feature extraction capabilities, which may permit the automatic identification of relevant patterns and a resulting improvement in classification accuracy. Additionally, this would remove the need for preprocessing or feature engineering, potentially easing surgical workflow integration. However, to enable the integration of such algorithms into the workflow the collection of larger datasets should be prioritised as Neural Networks are data-hungry and large amounts of data are required to prevent overfitting.

\subsubsection{Oesophageal cancer} \label{sec: Gastrointestinal Cancer}

Gastrointestinal cancers are a major contributor to the global cancer rate, with 26.3\% of cancer cases and 35.4\% of cancer deaths being caused by gastrointestinal cancers in 2018 \citep{lu2021global}. Surgery targeting these cancers aims for complete resection of tumorous tissue with clear margins while preserving as much healthy tissue as possible. Incomplete resection with positive circumferential resection margins is associated with increased chances of local cancer recurrence and poorer long-term survival \citep{gkouzionis_real-time_2022}. This review identified two studies by the same team which aim to develop a real-time tracking and classification system that can differentiate healthy and cancerous stomach and oesophageal tissue using DRS in combination with ML \citep{gkouzionis_real-time_2022, nazarian_real-time_2022}. Both studies used ex-vivo data with \citet{gkouzionis_real-time_2022} recording spectra in the 400 - 1000 nm range and \citet{nazarian_real-time_2022} in the 475 - 725 nm range. A range of models was implemented for the binary classification task, with Extreme Gradient Boosting (XGB) performing best on average across the two studies. \citet{gkouzionis_real-time_2022} reported accuracies of 93.86\% for the differentiation of healthy and cancerous stomach tissue and 96.22\% for oesophageal tissue (93.63\% and 96.26\% respectively using light gradient boosting). \citet{nazarian_real-time_2022} reported analogous results, with XGB achieving classification accuracies of 93.8\% for stomach tissue and 96.2\% for oesophageal tissue. The high performance of XGB for this task is in line with expectations as the algorithm is capable of handling non-linear relationships while also providing robustness to noise and automatically performing feature selection. As such, the algorithm reduces the large feature space, guaranteeing efficient run times which promote real-time integrations in a surgical setting. Additionally, XGB is capable of providing feature importance scores, which allows straightforward implementation of explainability. Both studies utilised comparatively large datasets, given the medical context, emphasising the robustness of the findings. However, both studies utilised ex-vivo data, removing the influence of confounding factors like difficulty accessing the location and other fluids contaminating measurements. Future research should aim to validate findings in an in-vivo setting and use the explainability capabilities of XGB to test for potential bias in classification decisions. Overall, the team concludes that the high classification accuracies, in combination with the successful tracking system, promise the successful integration of DRS into the surgical workflow for the successful removal of cancerous tissue and the assurance of clean resection margins. 

\subsubsection{Other cancers} \label{sec: Other Cancers}

Three studies focusing on the implementation of DRS and ML in cancer research which do not belong to any of the above categories, were identified. Two studies are conducted by the team of Geldof et al. \citep{geldof_layer_2022, geldof_toward_2024} and the third by \citet{werahera_diffuse_2016}. \citet{geldof_layer_2022} implemented DRS for the prediction of tissue layer thickness and classification of the top two layers of an animal model for applications to oncological surgery. The authors used an SVM for both tasks and an additional Process Regression for the top-layer thickness prediction. Spectra were recorded in the range of 400 - 1600 nm, and a series of feature extraction methodologies were tested before classification and regression. The thickness prediction of the top layer was possible with an accuracy of 0.35 mm, and the classification of the first and second tissue layers achieved accuracies of 95\% and 99\%, respectively. While the initial findings regarding the ability of DRS to predict layer thickness are optimistic, the authors argue that future research needs to validate findings on tissue samples more akin to the real-world setting. As such, validation on excised, human ex-vivo tissue is paramount before testing in in-vivo settings. Within these new situations, tissue heterogeneity needs to be accounted for by models to ensure algorithm safety. The second study by the same team implemented DRS to differentiate healthy and tumorous tissue during soft tissue sarcoma surgery \citep{geldof_toward_2024}. The authors recorded signals of ex-vivo human tissue in the wavelength range of 400 - 1600 nm. Two classifications were completed: one binary, differentiating healthy and tumorous tissue, and one multiclass, comparing fat, muscle, skin, sarcoma, and well-differentiated liposarcomas. The binary classification achieved accuracies of 85\%, and the multiclass classification achieved accuracies of 90\%. Higher performance in the multiclass classification contradicts expectations as it is considered a more difficult ML task. One possible reason is that the additional classes have well-differentiated reflectance spectra, which were previously combined into healthy and tumorous classes, obscuring important patterns. One drawback of this research was a lack of patient-wise cross-validation due to the small sample size, which may lead to data leakage and artificially inflated performance measures. If a given patient's data is included in both the training and test set, the algorithm may learn patient-specific variations instead of clinically relevant patterns, and accuracies will thus not translate to new data. As such, future research should focus on expanding sample sizes to allow appropriate cross-validation methodologies. The final study by \citet{werahera_diffuse_2016} investigated the use of DRS for the differentiation of high and low-grade prostate cancer from benign tissue. The authors recorded ex-vivo human prostate tissue signals in the range of 500 - 700 nm. PCA was used for feature extraction, and the retention of different numbers of components was tested. The best results were achieved using three components and an SVM for classification, achieving specificity values as high as 91\% and sensitivities as high as 69\% depending on the task. Sensitivity measures a test's ability to correctly identify individuals with cancer. High sensitivity is desirable because it ensures that very few cases of the disease are missed. However, this often comes at the cost of specificity, which refers to the test’s ability to correctly identify healthy individuals. As a result, some healthy patients may be mistakenly diagnosed with cancer. While this is not ideal, the consequences of a false positive are generally less severe than missing an actual case of cancer. Therefore, high sensitivity suggests that the algorithm has the potential to be effective. The authors concluded that DRS may enable more accurate assessments of prostrate cancer and improvements in patient care. However, the dataset was highly unbalanced, affecting outcome measures. Additionally, the current results are based solely on ex-vivo samples, which may differ from in-vivo tissue in a range of reflective properties. As such, future research should aim to implement counter-balancing measures to deal with class imbalance before implementing in-vivo validation of the achieved results. 

\subsection{Other Applications} \label{sec: Other Applications}

In addition to the wide array of oncological deployments of DRS analysed above, thirty-one studies were identified that implemented DRS in combination with ML for different medical applications. The most common topic addressed was tissue classification with a range of aims, including orthopaedic and oral surgery refinement and the grading of tissue degradation in cases such as burn wounds. In addition, blood-grading and nerve identification studies are analysed in sections \ref{sec: Blood-Related Applications} and \ref{sec: Nerve Identification}, respectively. 

\subsubsection{Tissue classification} \label{sec: Tissue Classification}

Seventeen studies implementing DRS in combination with ML for classifying different tissues were identified. For clarity, the studies are divided by whether they classify different tissues (tissue differentiation) or different stages of the same tissue (tissue grading).

\paragraph{Tissue differentiation}

Fourteen studies implementing DRS in combination with ML for tissue differentiation were identified. Eleven of these studies used signals collected ex-vivo, with nine considering animal tissue \citep{zam_soft_2009, zam_tissue_2010, fanjul-velez_application_2020, sun_characterization_2022, li_frameworks_2023, li_wavelength_2022, stelzle_diffuse_2010, stelzle_optical_2011, gunaratne_wavelength_2020} and the remaining two using human samples \citep{kreis_diffuse_2019, gunaratne_machine_2019}. The remaining three studies conducted in-vivo used animal tissue \citep{davey_analysis_2021, dahlstrand_extended-wavelength_2019, stelzle_vivo_2012}. 
Three main teams working with the ex-vivo animal tissues were identified. The work by Li et al. \citep{li_frameworks_2023, li_wavelength_2022} aimed to differentiate joint tissues for integrating DRS into orthopaedic procedures. The authors conducted a multiclass and binary classification using Quadratic Discriminant Analysis and LDA, respectively. Both studies reached accuracies of 99\% or higher, demonstrating the feasibility of DRS for joint tissue discrimination. The main contribution of these works was the proposal and validation of feature selection frameworks. However, the findings require further validation in ex-vivo settings due to the method of dataset creation and the consequential risk of lacking generalisation. The second team of Zam et al. \citep{zam_soft_2009, zam_tissue_2010} implemented DRS for tissue differentiation with applications in oral and maxillary surgery. The two studies implemented binary differentiation of four classes, using PCA for preprocessing and feature extraction and LDA for classification. High accuracies between 90 - 100\% were achieved. Given the authors proposed real-time implementation of the developed algorithm, future research should aim to test and optimise processing and prediction times. While the small size and efficient processing of LDA recommends itself to this task, the use of PCA for preprocessing requires cubic runtime, which is computationally expensive and different methods of feature reduction should be tested in future research. Additionally, PCA is based on the entire wavelength spectrum, where feature selection techniques may allow the isolation of a smaller wavelength range and consequential smaller device design, which would permit easier medical integration. Furthermore, in-vivo validation is imperative to ensure the translation of the high accuracy levels achieved ex-vivo. The third team of Stelzle et al. \citep{stelzle_diffuse_2010, stelzle_optical_2011} used DRS in combination with ML for improved tissue differentiation and optical feedback during laser surgery. The authors differentiated five classes and achieved AUC scores as high as 98.9\%. Based on these findings, future research should attempt in-vivo replication. Additionally, the authors suggest that it should be investigated whether DRS can differentiate tissues altered by a surgical laser-induced carbonisation zone. Finally, the real-time application of these findings necessitates the exploration of run times and the optimisation of ML algorithms for fast prediction speed. The remaining animal ex-vivo studies reported relatively high classification accuracies in both the multiclass and binary case \citep{gunaratne_wavelength_2020, fanjul-velez_application_2020, sun_characterization_2022}. All three studies implemented feature selection before classification, and the best accuracy of 99\% was achieved by \citet{gunaratne_wavelength_2020} using an LDA to differentiate subchondral tissue from six other classes using just ten selected wavelengths. These findings emphasize the ability of DRS to differentiate tissue in the binary and multiclass case using ex-vivo animal tissue signals. 

The two studies considering ex-vivo human tissue implemented binary and multiclass classification, respectively \citep{kreis_diffuse_2019, gunaratne_machine_2019}. Both studies used feature extraction and achieved accuracies between 98 - 100\%. The authors concluded that DRS provides a cost-effective and accurate approach to differentiating tissues. In addition, \citet{kreis_diffuse_2019} demonstrated that using Bayesian decomposition for feature selection may increase accuracy. Here, Bayesian decomposition selected parameters by computing a Bayesian estimate and selecting those parameters with the highest likelihood of a low number of repetitions. One limitation of the work by \citet{gunaratne_machine_2019} is the considerable variation in group sizes. While the authors conducted undersampling to test whether this would decrease accuracy, this leads to a rampant reduction in sample size, which is not suitable for the deployment of machine learning models. As such, more sophisticated counterbalancing or evaluation methods should be implemented to validate the reported outcomes. One solution may be to oversample the minority class, to avoid the loss of data during preprocessing. While, oversampling can be problematic as it does not increase variance in the data, sophisticated methods such as SMOTE may be commendable as they can compensate for this problem  \citep{chawla2002smote}. Additionally, future studies should validate the reported results in an in-vivo setting and explore the potential of multimodal signal recording as proposed by \citet{kreis_diffuse_2019} to increase accuracy levels. 

The three identified in-vivo studies by \citet{davey_analysis_2021, dahlstrand_extended-wavelength_2019, stelzle_vivo_2012} aimed to differentiate tissue in the binary and multiclass case respectively. \citet{davey_analysis_2021} implemented a quadratic discriminant classifier to distinguish healthy and pathological tissue at different stages of development in a mouse model. The authors used PCA for feature extraction and achieved accuracies as high as 82.6\%. In addition to DRS, autofluorescence spectroscopy was implemented to serve as a comparison. The autofluorescence system outperformed DRS in three of the four tested conditions, achieving an average AUC of 0.85 compared to 0.79 reached by DRS. This research emphasizes the different strengths of the spectroscopic systems depending on the target tissue. The combination of the two modalities should be elucidated to test whether a hybrid system may compensate for the respective weaknesses of each method. \citet{dahlstrand_extended-wavelength_2019} used an SVM to differentiate five types of joint tissue, the signals of which were extracted from a pig model. Sensitivities between 96.4 - 99.0\% and specificities between 99.4 - 99.8\% were achieved using five principal components. The authors used an EWDRS system to test whether this yielded diagnostic advantages compared to smaller collection ranges. Interestingly, the extension from 450–900 nm to 450–1550 nm did not improve classification results, likely because no relevant chromophores were reflected in the extended range. However, the authors argue that this may change depending on probing depth, and future research may wish to elucidate this. \citet{stelzle_vivo_2012} implemented PCA to extract four features from the data and then used an LDA model to classify four tissue types, achieving an AUC of 75\%. The study is conducted on in-vivo tissue samples with a proposed application for the refinement of laser surgery. While these findings are promising, the authors highlight that during surgery, laser-induced changes, including exsiccation and carbonisation, may take place, which would alter tissue reflective properties, and the effects on classification performance remain to be tested. 

Interestingly, no studies attempting tissue differentiation using in-vivo human samples were retrieved. Given the high accuracy levels achieved in the present research, this should be considered as the next step for the validation of the developed systems. Potential applications for such research may include improved navigation in low-visibility surgery settings, including revision hip arthroplasty or refinements in laser surgery. Overall, these studies demonstrate the potential of DRS for tissue discrimination, and the high classification accuracies combined with feature extraction techniques are promising for implementing DRS and ML in a clinical setting. 

\paragraph{Tissue grading}

Three studies grading different developments of the same tissue were identified \citep{yeong_e_prediction_2005, zhu_early_2016, arista_romeu_diffuse_2018}. The work by \citet{yeong_e_prediction_2005} attempted to predict the healing time of human burn wounds using in-vivo signals. The authors used a Neural Network to predict whether tissue would heal in less or more than 14 days, achieving 86\% accuracy using the normalised spectra as input. \citet{zhu_early_2016} attempted to differentiate normal flaps from those with venous and arterial occlusion in a rat model using 470 - 700 nm wavelengths. The authors used Partial Least Squares components and an LDA for classification, achieving a maximum accuracy of 95.1\% after three hours of training. DRS is found to outperform autofluorescence in terms of sensitivity regarding differentiating the detection of progressive blood vessel occlusion and classifying arterial from venous occlusions. Both of these studies may suffer from intra-patient variability between different points of measurement due to changes in tissue structure. Future research may wish to investigate the implementation of a multimodal DRS and autofluorescence imaging system, which would allow the measurement of a larger tissue region to overcome this issue \citep{zhu_early_2016}. Finally, \citet{arista_romeu_diffuse_2018} used DRS to distinguish different stages of steatosis in a mouse model. The authors used a Linear Regression model and classified the spectra with an accuracy of 78.32\% after two weeks of a methionine–choline-deficient diet and with 87.88\% accuracy after eight weeks of this diet. However, the sample size in this study was relatively small, consisting of just 428 instances. Future research should aim to validate these findings in a larger dataset to assess whether the classification performances generalise to new data. Furthermore, the cross-validation did not consider potential data leakage due to sample variations, and the implementation of specimen-wise stratification should be considered in the future. The authors conclude that DRS demonstrates promise in evaluating the type and degree of steatosis. The wide range of tasks and high accuracy levels demonstrate the versatility of DRS and its suitability for the grading of tissue. 

\subsubsection{Blood-related applications} \label{sec: Blood-Related Applications}

Five studies applying DRS in combination with ML for blood-related applications were identified. The papers can be further divided by their proposed purpose with two papers aiming to diagnose blood-related diseases \citep{liu_blood_2018, lu_use_2023} and the remaining research characterising blood, based on species or constituents \citep{can_estimation_2018, banerjee_non-invasive_2023, li_identification_2018}. In the diagnostic aspect, \citep{liu_blood_2018} identify patients with blood hyperviscosity disease using in-vivo human samples and \citep{lu_use_2023} attempt to differentiate vein and artery spectra based on ex-vivo animal data. Both papers report reasonably high accuracies implementing feature reduction and classification using PCA and a Neural Network and Partial Least Squares and LDA, respectively; \citep{liu_blood_2018} reported 97\% accuracy for the diagnosis of hyperviscosity and \citep{lu_use_2023} achieved 92\% specificity and 100\% sensitivity for the differentiation of vein and artery spectra. The implementation of multimodal optical methods to increase classification accuracy may be of interest to future researchers. In addition, collecting more extensive datasets, especially in-vivo, is necessitated for model validation. The remaining studies used DRS to identify the species or constituents of blood with \citep{li_identification_2018} comparing the ability of DRS and infrared spectra to differentiate blood of different species, \citep{can_estimation_2018} estimating free haemoglobin levels in blood bags to detect storage lesions and \citep{banerjee_non-invasive_2023} attempting to estimate haemoglobin, bilirubin and oxygen saturation as a novel alternative to blood sampling. The work by \citet{li_identification_2018} demonstrates that the blood types of five species of animals, including humans, can be successfully differentiated using a Neural Network and PCA with reported accuracy values exceeding 97\% for all five classes. The remaining studies demonstrate that ML can estimate blood constituents with R$^2$ values as high as 97\% for the estimation of haemoglobin, 99\% for Bilirubin and 98\% for oxygen saturation \citep{ banerjee_non-invasive_2023, can_estimation_2018}. The best-performing model for this task is Artificial Neural Network. The identified research demonstrates the potential of DRS to support the diagnosis of blood-related diseases, emphasising that DRS can be used in combination with Neural Networks for the prediction of blood constituents with high levels of precision. However, Neural Networks are computationally expensive and non-transparent, making their medical integration difficult. Future research should consider developing smaller models to ease the computational load. One possibility would be using an SVM, which is advantageous due to its ability to detect non-linear relationships and differentiate similar classes. Furthermore, SVM is robust to noise, which makes it especially suitable for the processing of in-vivo DRS data, where other contaminants may obscure the signal. SVM has an easier explainability integration compared to deeper networks, which would furthermore allow the alignment with EU regulations \citep{EU_regulation_AI_act}. 

\subsubsection{Nerve identification} \label{sec: Nerve Identification}

Three studies using DRS for the identification of nerves with proposed applications to inter-surgical implementation for improved patient safety were retrieved \citep{schols_differentiation_2014, stelzle_diffuse_2010-1, langhout_vivo_2018}. The highest accuracies were achieved by \citet{stelzle_diffuse_2010-1} using an LDA with an average specificity of 96\% and a sensitivity of 94\%. However, it is worth noting that these results were based on ex-vivo spectra. The best results in-vivo were reported by \citet{schols_differentiation_2014}, using an SVM to differentiate nerve signals from two surrounding tissue classes with 95\% accuracy. The authors implemented domain-knowledge-based feature selection before classification. All three studies concluded that DRS provides a reliable method for differentiating nerves during surgical procedures to increase patient safety. \citet{stelzle_diffuse_2010-1} further argues that this may be a viable option for improving laser surgery. Given the proposed application, fast processing times for real-time implementation are essential for this task. SVM and LDA are both suitable for this, especially combined with the appropriate feature reduction to decrease the number of processed wavelengths. Selective relevant wavelength ranges through feature reduction is one possible research venue as it may improve processing times by reducing the recorded wavelength ranges and hence decreasing device size, easing surgical integration. Additionally, both SVM and LDA allow for the computation of feature weights and the consequential implementation of explainability, which can increase physician trust. Future research may hence wish to implement explainability and feature reduction to promote surgical integration of the developed algorithms. 

\subsubsection{Dental caries detection}
Dental caries, or tooth decay, is a chronic disease which forms over time through the interaction of acid-producing bacteria and fermentable carbohydrates. Individuals are susceptible to dental caries throughout their lifetime, but prevalent risk factors include poor oral hygiene and poverty. Dental caries cause severe oral pain and decay, and early identification is paramount to prevent spreading \citep{selwitz2007dental}. In this review, two studies addressing the identification of dental caries were retrieved \citep{charvat_diffuse_2020, a_prochazka_absorption_2023}. Both studies utilised in-vivo measurement with \citet{charvat_diffuse_2020} recording wavelengths from 400 - 1600 nm and \citet{a_prochazka_absorption_2023} measuring 400 - 1700 nm. Both studies conducted binary classification using SVM, Neural Networks, KNN and Naive Bayes classifiers. The highest performance was achieved by \citet{charvat_diffuse_2020} using only four features as input to a Neural Network and reaching 98.4\% accuracy. Both studies demonstrate high classification performance, especially given the in-vivo setting. The authors argue that future work should focus on developing more sophisticated methods, including deep learning methodologies, refining the task and improving accuracy. One problem with both studies is the relatively small sample size of 327 and 578 in \citet{a_prochazka_absorption_2023} and \citet{charvat_diffuse_2020} respectively. Pursuing the development of deep learning models should be treated with caution as these algorithms are frequently data-hungry and would require future researchers to amass larger datasets to prevent overfitting and ensure generalisability. In addition, Deep Neural Networks are computationally expensive and tend to be slower than smaller machine learning models such as SVMs. One possibility for the integration of deep learning would be the construction of a Transfer Learning model, as this could benefit from all existing datasets and hence would not require extensive additional data collection. An alternative possibility to maintain high accuracy values in an in-vivo setting may be exploring a range of feature selection methodologies. This would combat overfitting while ensuring efficient processing and targeting the highly correlated nature of DRS signals.

\section{Discussion} \label{sec:Discussion}

The identified studies show that supervised ML can successfully analyse DRS signals for classification and regression tasks in optics-based diagnostics. The retrieved studies reported consistently high accuracies in the binary and multiclass cases, highlighting the efficacy of ML models for DRS signal processing. Observed trends are discussed below, and recommendations for future research are provided based on these observations. 

\subsection{Tissue and Examination Type}
The majority of the analysed studies were conducted in the ex-vivo setting, where higher classification accuracies tend to be achieved due to better control of environmental influences and lower chances of tissue contamination. However, various applications, including clean resection margins and intra-surgical tumour identification, require in-vivo classification. Future research should validate the developed algorithms on data collected in this examination type to ensure generalisability to real-world data. In this review, lower accuracies were generally reported by in-vivo studies, emphasising that despite the importance of this application, further refinement is necessary to improve the collection and processing of these signals before deployment. In-vivo data collection is challenging due to a range of considerations. First, ethical and time constraints decrease dataset size, which makes the training of data-hungry machine learning models more challenging. Additionally, the presence of blood supply, tissue perfusion and heterogeneity as well as changes in probe pressure and angle create additional variance in the data, further complicating the classification of signals. As such, in-vivo classification is more challenging due to both data size and quality. However, in-vivo testing is a vital step before clinical adoption. As such, the collection of larger in-vivo datasets should be prioritised to assess the degree to which lower classification performance results from dataset size as opposed to the inherent difficulties associated with recording signals in the in-vivo setting. Alternatively, models capable of learning patterns across several datasets should be developed. Initial work by \citet{veluponnar_margin_2024}, who trained models on ex- and in-vivo spectra and deployed them on in-vivo signals showed that the transfer of ex-vivo knowledge to an in-vivo setting may be feasible. As such, future research may wish to examine the feasibility of implementing transfer learning models on ex-vivo data for classification in an in-vivo setting.

\subsection{Sample Size} \label{sec: Sample Size}

One overarching problem is the data-hungry nature of ML models juxtaposed with the relative scarcity of large datasets in the medical domain. While the collection of sufficiently large datasets is problematic in medicine in general, this is exacerbated by the rarity of many conditions. Illnesses associated with inherently smaller sample sizes, such as brain cancer, may be more difficult to access and are often subject to additional complications, such as variations in probe angle or the presence of obscuring fluids. Thus, there may be a correlation between smaller sample sizes and more complex classification tasks. Expanding data collection would help disentangle the impact of the unavoidable in-vivo challenges from the effects of limited sample sizes. While the reduced datasets are, hence, to a certain extent, a consequence of the design limitations of in-vivo signals, they can be more easily rectified and should thus be targeted going forward. One possible solution may be the implementation of ML algorithms which are adjusted to limited data or are capable of learning patterns from other datasets. Alternatively, future research may want to deploy condition-tailored data augmentation techniques, integrating domain knowledge for improved artificial data creation. 

\subsection{Extended Wavelength Diffuse Reflectance Spectroscopy}
A third trend is the recording of extended wavelength ranges for improvements in classification accuracy. These modifications generally extend the measurement range to include near-infrared and short-wave infrared wavelengths up to 2100 nm. However, it is worthwhile noting that most studies only extend to 1600 - 1800 nm, depending on the available equipment and optical properties of the targeted tissue. The majority of studies utilising EWDRS reported higher accuracies compared to research utilising DRS covering narrower wavelength ranges, emphasising the potential of EWDRS. However, while increasing the collected information, this methodology requires an additional spectrometer covering range beyond 1100 nm, which can potentially increase the instrumentation costs. From a computational point of view, EWDRS is advantageous as more information regarding sample properties is retrieved, increasing the potential for successful pattern identification and processing. However, in the same vein, the increased dimensionality may raise processing times, making real-time implementation more challenging. Additionally, the relevance of the information collected in the extended wavelength range cannot be guaranteed as reflectance depends on biomolecular components and microstructures, and depending on the target application, it may not provide useful information. As such, it may be recommended for pilot studies to assess whether EWDRS yields a sufficient increase in information to justify the increased cost, size and processing times. 

\subsection{Feature Reduction}
A fourth interesting trend is the wide range of implemented feature selection and extraction methods. Most of the retrieved research reduced the feature space before classification to decrease data dimensionality and improve processing times. The preprocessing methodologies include feature extraction, the extraction of statistical features, feature selection based on their relevance to the ML classifier and the use of biomarker concentrations and scattering properties extracted via fitting methods. Each of these methods yields certain merits. Feature extraction constructs novel latent features based on a given metric, such as variance or the relationship between predictors and the outcome variable. One common example is PCA, which identifies directions of maximum variance and hence accounts for the common problem of high co-variance in DRS data. However, it rests on the assumption that high variance represents relevant information, which is a false equivalent, as high variance may stem from environmental factors rather than relevant components and microstructures, and PCA may hence obscure rather than identify relevant spectral patterns. PLS is an alternative feature extraction technique, which creates latent features based on directions that maximize the covariance between predictors and the response variable. This is suitable for DRS, as it allows the processing of high-dimensional and highly collinear data. However, biological components in DRS exhibit broad absorption bands and often overlap. Consequentially, the selection of a single representative wavelength to account for this, may not be sufficient. Furthermore, the choice of the number of constructed latent variables is difficult and rather arbitrary, which may lead to over or underfitting. Statistical feature selection, extracting given aspects of the curve shape such as maxima and minima of the reflectance spectra, is mathematically simple and computationally cheap, but its domain compliance with broadband, smooth DRS signals is not assured. Data-driven feature selection based on feature importance computed by ML models is favourable as it works in tandem with the classification algorithm and may, hence, efficiently boost performance. It furthermore automatically considers feature dependencies, which may account for the inter-correlation of features in DRS data. However, it neglects domain knowledge and may in turn select insignificant wavelength ranges. This is a result of the ML model favouring features which maximise class separation. However, such features could be collection artifacts and may not be representative of the underlying classes. While the selected wavelengths may thus allow for accurate classification within the selected dataset, their importance may not generalise to new data \citep{kohavi1997wrappers}. Additionally, selected features are dependent on the classifier and may not remain the optimum if a different ML algorithm is deployed \citep{pudjihartono2022review}. Finally, the direct use of biomarker concentrations and scattering parameters extracted using forward Monte Carlo simulations and spectral fitting algorithms is the most domain-proximate method as it relies on the estimation of the underlying biochemical components. However, it is time-consuming, and therefore challenging to deploy in real time, and it may introduce some uncertainty, due to simplification of the geometry of the forward Monte Carlo model and the possible overfitting of the parameters extracted in the process \citep{nguyen2021machine}. While each method has merits, no single methodology was found to consistently perform best. Future research may wish to elucidate whether feature selection varies as a function of the data or domain or if a given method may consistently perform best when processing DRS signals. Additionally, the combination of feature reduction methods and model choice should be elucidated as certain models, including SVM, may require more stringent feature reduction than others, such as Neural Networks, which integrate feature engineering in the classification pipeline. In addition to reducing computational costs and increasing processing speed, feature selection is also a key consideration in device engineering. If equivalent accuracies can be achieved using a smaller wavelength subset, smaller devices could be designed, easing surgical integration and lowering cost.

\subsection{Bias Analysis}
Next to the bias within the studies discussed in Section \ref{sec: Bias Analysis}, publication bias in the retrieved works should be considered. Within this review, a wide array of applications of DRS and ML were analysed. However, it is important to consider that the frequency with which a given condition is diagnosed may not correspond to its severity. Given the geographical limitations, research is restricted to those illnesses with a sufficient local incidence to allow the retrieval of a representative number of spectra. Consequentially, rare conditions, which may benefit greatly from DRS and ML applications, may be neglected due to the difficulties associated with acquiring sufficient training data. With fewer incidences, the identification of patients and retrieval of signals becomes more challenging and researchers may thus choose to focus on more prominent conditions instead. The frequency at which a condition is studied may hence not be a consequence of its severity but rather of its prevalence and the resulting opportunity, or lack thereof, to study it. Additionally, the ease of access to a given tissue should be taken into consideration. While this impacts the proportion of in and ex-vivo spectra collected, it may also impact the number of studies conducted for a certain disease or condition. If signals are considered to be highly challenging to retrieve, researchers may focus on more easily accessible areas, affecting the proportions of the targeted conditions. One example is the identification of oral and brain cancer. Where the earlier tissue is easily accessible, the latter is comparatively more difficult to sample due to both location and ethical constraints. Consequentially, the trade-off between ease of signal collection, resulting sample size and classification should be taken into consideration during study design. It may be of interest to future research to test whether certain conditions have been favoured due to their ease of access rather than the accuracy they provide. 

\subsection{Machine Learning Compatibility}

The choice of a ML model for data analysis is critical as models yield varying advantages, depending on the data type and the proposed application, as discussed in Section \ref{subsec: Implemented  Models}. While no universal recommendation can be made for DRS data, in line with the 'no free lunch theory', several considerations should be taken into account. The first issue is the selected tissue's ease of access and the dataset's resulting size. If the number of spectra is limited, it is best to employ simpler machine learning models, such as LDAs or SVMs, as they require less training data and are less prone to overfitting, even in small datasets. Additionally, these models are more transparent, and it may be easier to implement explainability into their classification processes. Conversely, if the tissue under study is easily accessible and a large dataset can be collected across a wide range of wavelengths, a deep learning model, such as a Convolutional Neural Network, may be better suited as these algorithms can extract high-level features during their training phase and may hence achieve better classification levels in high-dimensional data, given that a sufficiently large dataset can be collected. However, the black-box nature and increased computational expense of these models should be considered, making them more difficult to implement in a surgical setting. A second consideration is the desired preprocessing. The high dimensionality and inter-correlation of DRS data make preprocessing paramount to the successful implementation of most machine learning classifiers. However, some algorithms, such as Convolutional Neural Networks, construct high-level features during the training process, rendering feature reduction methodologies moot. Preprocessing should hence be considered before model choice and may depend on the domain, the proposed deployment of the technology as well as the preferences of research groups. It is important to note that preprocessing and model choice are frequently considered in isolation, despite being strongly interlinked. Going forward, preprocessing and ML algorithms should be selected and optimised in tandem as their performances are strongly related. In addition to these considerations, model choice also strongly depends on the chosen task. This includes factors such as data complexity, class separation, and the number of classes. As such, future research should implement a wide range of models in tandem with preprocessing techniques to permit the successful selection of the most suitable model.  

\subsection{Future Research}

The analysis of the retrieved studies highlighted a series of trends in the research including increasing wavelength ranges and changes in tissue and examination types. Based on this analysis, this study proposes several considerations that future research may wish to address.  

\subsubsection{Patient Stratification} \label{subsec: patient stratification}

The most important consideration that is frequently lacking in current research is the stratification of patients. Most studies included in this review took multiple measures from a single patient or tissue specimen without separating these measures into training and test sets. As such, measures from the same patient or tissue piece may be considered in both datasets, leading to artificial inflation of accuracy measures due to intrapersonal differences and consequential data leakage or shortcut learning. Essentially, the algorithm learns the individual variation of a given patient in the training set and then recognises these in the test set, rather than extracting the condition-relevant patterns. This allows the algorithm to reach high classification accuracies by learning patient-specific patterns rather than illness-relevant features. As a result, the performance achieved in the dataset will not generalise to new data. To ensure future models' robustness and real-world applicability, researchers should enforce strict stratification, ensuring that no patient or specimen contributes to both training and test datasets. One approach to this may be patient-wise cross validation, where at each round of cross validation, a given number of patients is reserved for the testing set and excluded from the training data. This is an efficient way to avoid data leakage while maximising training data in small datasets. 

\subsubsection{Metric Reporting}

Future work should furthermore aim to improve the consistency of the reported dataset and outcome metrics. During the initial analysis of retrieved studies, exploratory metrics of sample size and number of wavelengths were computed. However, all results were purely estimate-based as the retrieved studies were largely inconsistent in their data reporting, with many only providing estimates of these numbers or failing to report them entirely. Another problem stems from inconsistency of reported classification metrics, with studies using a wide range of metrics as well as frequently reporting partial outcomes, making it impossible to conduct an assessment of algorithm performance trends. Going forward, researchers should take care to report, at minimum, the number of spectra, the number of wavelengths and the accuracy, sensitivity and specificity for each machine learning algorithm evaluated. This will ease the comparison of studies and the synthesis of cohesive results across research to more effectively identify gaps and inconsistencies and propose solutions for future works. 

\subsubsection{Reproducibility and Data Sharing}

The ability to reproduce research is a cornerstone of scientific research, as it allows the validation and extension of previous studies. This is especially important in the medical field, as a lack of vigorous testing could have severe adverse effects on vulnerable groups. As such, future research should aim to make both collected data and implemented code publicly available. This would not only benefit the research community by increasing data availability but also permit more efficient allocation of resources as it reduces the need for new data collection. However, one thing that needs to be taken into consideration when sharing data is the protection of patient privacy, especially in the medical field. As such, researchers should take care to anonymise data before publication. 

\subsubsection{Transfer Learning}

As repeatedly emphasized throughout this review, the limited data availability in the medical domain is a major challenge to the deployment of machine learning models. One possible solution to this is the construction of a transfer learning algorithm. These pre-trained models are capable of transferring patterns learned from related datasets to new problems, hence reducing the need for large-scale data collection. Given the ethical restrictions and availability limitations of medical data, this may improve algorithm performance in areas where data is limited. Based on the reviewed studies, several domains for the deployment of transfer-learning models are suggested. The first is the compensation for the limited data available in in-vivo settings by training models on larger ex-vivo datasets and then deploying them in-vivo. This is especially interesting as initial results are promising but the increased noise and changing measurement conditions associated with in-vivo data may prove an additional challenge for the algorithm. An alternative application is the deployment of transfer-learning for rare diseases. Many conditions may benefit from the implementation of DRS and ML, but due to their low incidence, the amount of collected data is insufficient for successful integration. Here, models could be trained on similar ailments and then fine-tuned on the given illness, reducing the need for large-scale data collection, which may not be feasible. Here, it would furthermore be of interest to identify conditions that are sufficiently similar to allow the transfer of learned patterns. As such, transfer learning holds considerable potential in the domain of DRS and future research should ascertain in which domains its integration is not only feasible but sensible. 

\subsubsection{Explainability}

A final important pathway for future research is the implementation of explainability for the classification of DRS signals. A recurring concern with the deployment of ML in the medical context is the "black-box" nature of the majority of models, whereby algorithms lack clear interpretability. Given the critical nature of medical tasks, the use of black-box models without auditability is problematic, as decisions may be biased, unreliable or result from shortcut learning. Without access to explanations regarding the decision-making process, it is impossible to ensure that the algorithm uses a correct, reliable pathway. As such, the implementation of explainable AI models is imperative to allow for both auditability and accountability, ensuring patient safety. Explainable AI encompasses a range of methodologies that improve model transparency while maintaining classification performance \citep{minh2022explainable}. In addition to preventing model bias, it has been shown to increase practitioner trust and is necessitated for medical implementations under the European Union guidelines \citep{rasheed2022explainable, EU_regulation_AI_act}. Consequently, there is a growing focus on integrating explainable AI techniques into medical ML applications, with recent reviews by \citet{allgaier2023does, loh2022application, van2022explainable} providing comprehensive insights into available explainable AI techniques, their benefits, and limitations. Despite this undeniable relevance of explainability in medical tasks, the vast majority of studies identified in this review do not implement any explainable AI methodologies, relying solely on performance metrics for model evaluation. Future research should consider both the implementation and evaluation of explainability techniques for DRS signals, which are more challenging to evaluate due to their highly correlated nature. While many approaches to explainability have been put forward, including techniques such as GradCAM and SHAP, their adaptation for DRS signals remains unexplored. This should be resolved in future research as implementing explainable AI for highly multi-collinear data such as DRS signals is challenging, as identified wavelengths may only be partially representative of important regions, and explanations may change considerably between model iterations.

\section{Conclusion} \label{sec: Conclusion}

The objective of this review was to identify the current research areas and applications of DRS and ML in the field of optically-guided diagnostics. Based on this, gaps in research and areas for improvement were identified and considerations for future research were provided. The findings confirm that DRS and ML hold significant promise in various clinical applications. However, much of the research remains limited to ex-vivo studies. Future work should prioritize in-vivo validation to mirror surgical conditions and test the validity of the measurement conditions. Furthermore, the successful integration of DRS with ML should be approached more holistically, with careful consideration of the clinical application, dataset size, data preprocessing, and ML model, all selected based on their inherent interconnections. Additionally, integrating explainability measures is crucial to mitigate algorithmic bias and ensure patient safety. Researchers must also prevent data leakage by stratifying training and testing data by patient or specimen or implementing patient-wise cross validation. In conclusion, while DRS and ML have the potential to enhance surgical accuracy and safety, refining these techniques with a strong emphasis on patient safety must be a key focus of future research. 

\section*{Funding and Acknowledgements}
This publication has emanated from research conducted with the financial support of Taighde Éireann - Research Ireland under Grant No. 18/CRT/6223 and 12/RC/2289-P2, which are co-funded under the European Regional Development Fund. For the purpose of Open Access, the author has applied a CC BY public copyright license to any Author Accepted Manuscript version arising from this submission.

This publication has emanated from research funded by Taighde Éireann – Research Ireland under the grant number 22/RP-2TF/10293. For the purpose of Open Access, the author has applied a CC BY public copyright licence to any Author Accepted Manuscript version arising from this submission.

\section{Availability of Data and Materials}
Data sharing does not apply to this article as no datasets were generated or analysed during the current study.

\section{Competing Interests}
The authors declare that they have no competing interests.

\bibliography{mybibfile}

\begin{thebibliography}{131}
\expandafter\ifx\csname natexlab\endcsname\relax\def\natexlab#1{#1}\fi
\providecommand{\url}[1]{\texttt{#1}}
\providecommand{\href}[2]{#2}
\providecommand{\path}[1]{#1}
\providecommand{\DOIprefix}{doi:}
\providecommand{\ArXivprefix}{arXiv:}
\providecommand{\URLprefix}{URL: }
\providecommand{\Pubmedprefix}{pmid:}
\providecommand{\doi}[1]{\href{http://dx.doi.org/#1}{\path{#1}}}
\providecommand{\Pubmed}[1]{\href{pmid:#1}{\path{#1}}}
\providecommand{\bibinfo}[2]{#2}
\ifx\xfnm\relax \def\xfnm[#1]{\unskip,\space#1}\fi
\bibitem[{Akter et~al.(2018)Akter, Hossain, Nishidate, Hazama \& Awazu}]{akter2018medical}
\bibinfo{author}{Akter, S.}, \bibinfo{author}{Hossain, M.~G.}, \bibinfo{author}{Nishidate, I.}, \bibinfo{author}{Hazama, H.}, \& \bibinfo{author}{Awazu, K.} (\bibinfo{year}{2018}).
\newblock \bibinfo{title}{Medical applications of reflectance spectroscopy in the diffusive and sub-diffusive regimes}.
\newblock {\it \bibinfo{journal}{Journal of Near Infrared Spectroscopy}\/},  {\it \bibinfo{volume}{26}\/}, \bibinfo{pages}{337--350}.
\bibitem[{Allgaier et~al.(2023)Allgaier, Mulansky, Draelos \& Pryss}]{allgaier2023does}
\bibinfo{author}{Allgaier, J.}, \bibinfo{author}{Mulansky, L.}, \bibinfo{author}{Draelos, R.~L.}, \& \bibinfo{author}{Pryss, R.} (\bibinfo{year}{2023}).
\newblock \bibinfo{title}{How does the model make predictions? a systematic literature review on the explainability power of machine learning in healthcare}.
\newblock {\it \bibinfo{journal}{Artificial Intelligence in Medicine}\/},  {\it \bibinfo{volume}{143}\/}, \bibinfo{pages}{102616}.
\bibitem[{Amouroux et~al.(2009)Amouroux, Díaz-Ayil, Blondel, Bourg-Heckly, Leroux \& Guillemin}]{amouroux_classification_2009}
\bibinfo{author}{Amouroux, M.}, \bibinfo{author}{Díaz-Ayil, G.}, \bibinfo{author}{Blondel, W. C. P.~M.}, \bibinfo{author}{Bourg-Heckly, G.}, \bibinfo{author}{Leroux, A.}, \& \bibinfo{author}{Guillemin, F.} (\bibinfo{year}{2009}).
\newblock \bibinfo{title}{Classification of ultraviolet irradiated mouse skin histological stages bybimodal spectroscopy: multiple excitation autofluorescence and diffusereflectance}.
\newblock {\it \bibinfo{journal}{J. Biomed. Opt.}\/},  {\it \bibinfo{volume}{14}\/}, \bibinfo{pages}{014011}.
\newblock \bibinfo{note}{Publisher: SPIE-Intl Soc Optical Eng}.
\bibitem[{Anwanwan et~al.(2020)Anwanwan, Singh, Singh, Saikam \& Singh}]{anwanwan2020challenges}
\bibinfo{author}{Anwanwan, D.}, \bibinfo{author}{Singh, S.~K.}, \bibinfo{author}{Singh, S.}, \bibinfo{author}{Saikam, V.}, \& \bibinfo{author}{Singh, R.} (\bibinfo{year}{2020}).
\newblock \bibinfo{title}{Challenges in liver cancer and possible treatment approaches}.
\newblock {\it \bibinfo{journal}{Biochimica et Biophysica Acta (BBA)-Reviews on Cancer}\/},  {\it \bibinfo{volume}{1873}\/}, \bibinfo{pages}{188314}.
\bibitem[{Arista~Romeu et~al.(2018)Arista~Romeu, Escobedo, Campos-Espinosa, Romero-Bello, Moreno-González, Fabila-Bustos, Reed, Isakina, Vázquez \& Guzmán}]{arista_romeu_diffuse_2018}
\bibinfo{author}{Arista~Romeu, E.~J.}, \bibinfo{author}{Escobedo, G.}, \bibinfo{author}{Campos-Espinosa, A.}, \bibinfo{author}{Romero-Bello, I.~I.}, \bibinfo{author}{Moreno-González, J.}, \bibinfo{author}{Fabila-Bustos, D.~A.}, \bibinfo{author}{Reed, A.~V.}, \bibinfo{author}{Isakina, S.~S.}, \bibinfo{author}{Vázquez, J. M. d. l.~R.}, \& \bibinfo{author}{Guzmán, C.} (\bibinfo{year}{2018}).
\newblock \bibinfo{title}{Diffuse reflectance spectroscopy accurately discriminates early andadvanced grades of fatty liver in mice}.
\newblock {\it \bibinfo{journal}{J. Biomed. Opt.}\/},  {\it \bibinfo{volume}{23}\/}, \bibinfo{pages}{1--8}.
\newblock \bibinfo{note}{Publisher: SPIE-Intl Soc Optical Eng}.
\bibitem[{Artosi et~al.(2024)Artosi, Costanza, Di~Prete, Garofalo, Lozzi, Dika, Cosio, Diluvio, Shumak, Lambiase et~al.}]{artosi2024epidemiological}
\bibinfo{author}{Artosi, F.}, \bibinfo{author}{Costanza, G.}, \bibinfo{author}{Di~Prete, M.}, \bibinfo{author}{Garofalo, V.}, \bibinfo{author}{Lozzi, F.}, \bibinfo{author}{Dika, E.}, \bibinfo{author}{Cosio, T.}, \bibinfo{author}{Diluvio, L.}, \bibinfo{author}{Shumak, R.~G.}, \bibinfo{author}{Lambiase, S.} et~al. (\bibinfo{year}{2024}).
\newblock \bibinfo{title}{Epidemiological and clinical analysis of exposure-related factors in non-melanoma skin cancer: A retrospective cohort study}.
\newblock {\it \bibinfo{journal}{Environmental Research}\/},  {\it \bibinfo{volume}{247}\/}, \bibinfo{pages}{118117}.
\bibitem[{Baltussen et~al.(2019{\natexlab{a}})Baltussen, Brouwer De~Koning, Sanders, Aalbers, Kok, Beets, Hendriks, Sterenborg, Kuhlmann \& Ruers}]{baltussen_tissue_2019}
\bibinfo{author}{Baltussen, E.}, \bibinfo{author}{Brouwer De~Koning, S.}, \bibinfo{author}{Sanders, J.}, \bibinfo{author}{Aalbers, A.}, \bibinfo{author}{Kok, N.}, \bibinfo{author}{Beets, G.}, \bibinfo{author}{Hendriks, B.}, \bibinfo{author}{Sterenborg, H.}, \bibinfo{author}{Kuhlmann, K.}, \& \bibinfo{author}{Ruers, T.} (\bibinfo{year}{2019}{\natexlab{a}}).
\newblock \bibinfo{title}{Tissue diagnosis during colorectal cancer surgery using optical sensing: {An} in vivo study}.
\newblock {\it \bibinfo{journal}{Journal of Translational Medicine}\/},  {\it \bibinfo{volume}{17}\/}.
\bibitem[{Baltussen et~al.(2020)Baltussen, Brouwer~de Koning, Sanders, Aalbers, Kok, Beets, Hendriks, Sterenborg, Kuhlmann \& Ruers}]{baltussen_using_2020}
\bibinfo{author}{Baltussen, E.}, \bibinfo{author}{Brouwer~de Koning, S.}, \bibinfo{author}{Sanders, J.}, \bibinfo{author}{Aalbers, A.}, \bibinfo{author}{Kok, N.}, \bibinfo{author}{Beets, G.}, \bibinfo{author}{Hendriks, B.}, \bibinfo{author}{Sterenborg, H.}, \bibinfo{author}{Kuhlmann, K.}, \& \bibinfo{author}{Ruers, T.} (\bibinfo{year}{2020}).
\newblock \bibinfo{title}{Using {Diffuse} {Reflectance} {Spectroscopy} to {Distinguish} {Tumor} {Tissue} {From} {Fibrosis} in {Rectal} {Cancer} {Patients} as a {Guide} to {Surgery}}.
\newblock {\it \bibinfo{journal}{Lasers in Surgery and Medicine}\/},  {\it \bibinfo{volume}{52}\/}, \bibinfo{pages}{604--611}.
\bibitem[{Baltussen et~al.(2017)Baltussen, Snaebjornsson, de~Koning, Sterenborg, Aalbers, Kok, Beets, Hendriks, Kuhlmann \& Ruers}]{baltussen_diffuse_2017}
\bibinfo{author}{Baltussen, E.}, \bibinfo{author}{Snaebjornsson, P.}, \bibinfo{author}{de~Koning, S.}, \bibinfo{author}{Sterenborg, H.}, \bibinfo{author}{Aalbers, A.}, \bibinfo{author}{Kok, N.}, \bibinfo{author}{Beets, G.}, \bibinfo{author}{Hendriks, B.}, \bibinfo{author}{Kuhlmann, K.}, \& \bibinfo{author}{Ruers, T.} (\bibinfo{year}{2017}).
\newblock \bibinfo{title}{Diffuse reflectance spectroscopy as a tool for real-time tissue assessment during colorectal cancer surgery}.
\newblock {\it \bibinfo{journal}{Journal of biomedical optics}\/},  {\it \bibinfo{volume}{22}\/}, \bibinfo{pages}{1--6}.
\bibitem[{Baltussen et~al.(2019{\natexlab{b}})Baltussen, Sterenborg, Ruers \& Dashtbozorg}]{baltussen_optimizing_2019}
\bibinfo{author}{Baltussen, E. J.~M.}, \bibinfo{author}{Sterenborg, H. J. C.~M.}, \bibinfo{author}{Ruers, T. J.~M.}, \& \bibinfo{author}{Dashtbozorg, B.} (\bibinfo{year}{2019}{\natexlab{b}}).
\newblock \bibinfo{title}{Optimizing algorithm development for tissue classification in colorectalcancer based on diffuse reflectance spectra}.
\newblock {\it \bibinfo{journal}{Biomed. Opt. Express}\/},  {\it \bibinfo{volume}{10}\/}, \bibinfo{pages}{6096--6113}.
\newblock \bibinfo{note}{Publisher: The Optical Society}.
\bibitem[{Banerjee et~al.(2023)Banerjee, Bhattacharyya, Ghosh, Singh, Adhikari, Mondal, Roy, Bajaj, Ghosh, Bhushan, Goswami, Ahmed, Moussa, Mondal, Mukhopadhyay, Bhattacharyya, Chattopadhyay, Ahmed, Mallick \& Pal}]{banerjee_non-invasive_2023}
\bibinfo{author}{Banerjee, A.}, \bibinfo{author}{Bhattacharyya, N.}, \bibinfo{author}{Ghosh, R.}, \bibinfo{author}{Singh, S.}, \bibinfo{author}{Adhikari, A.}, \bibinfo{author}{Mondal, S.}, \bibinfo{author}{Roy, L.}, \bibinfo{author}{Bajaj, A.}, \bibinfo{author}{Ghosh, N.}, \bibinfo{author}{Bhushan, A.}, \bibinfo{author}{Goswami, M.}, \bibinfo{author}{Ahmed, A. S.~A.}, \bibinfo{author}{Moussa, Z.}, \bibinfo{author}{Mondal, P.}, \bibinfo{author}{Mukhopadhyay, S.}, \bibinfo{author}{Bhattacharyya, D.}, \bibinfo{author}{Chattopadhyay, A.}, \bibinfo{author}{Ahmed, S.~A.}, \bibinfo{author}{Mallick, A.~K.}, \& \bibinfo{author}{Pal, S.~K.} (\bibinfo{year}{2023}).
\newblock \bibinfo{title}{Non-invasive estimation of hemoglobin, bilirubin and oxygen saturation ofneonates simultaneously using whole optical spectrum analysis at point ofcare}.
\newblock {\it \bibinfo{journal}{Sci. Rep.}\/},  {\it \bibinfo{volume}{13}\/}, \bibinfo{pages}{2370}.
\bibitem[{Begum et~al.(2021)Begum, Maiti, Chakravarty \& Das}]{begum2021diffuse}
\bibinfo{author}{Begum, N.}, \bibinfo{author}{Maiti, A.}, \bibinfo{author}{Chakravarty, D.}, \& \bibinfo{author}{Das, B.~S.} (\bibinfo{year}{2021}).
\newblock \bibinfo{title}{Diffuse reflectance spectroscopy based rapid coal rank estimation: A machine learning enabled framework}.
\newblock {\it \bibinfo{journal}{Spectrochimica Acta Part A: Molecular and Biomolecular Spectroscopy}\/},  {\it \bibinfo{volume}{263}\/}, \bibinfo{pages}{120150}.
\bibitem[{Bess et~al.(2019)Bess, Greening \& Muldoon}]{bess2019efficacy}
\bibinfo{author}{Bess, S.~N.}, \bibinfo{author}{Greening, G.~J.}, \& \bibinfo{author}{Muldoon, T.~J.} (\bibinfo{year}{2019}).
\newblock \bibinfo{title}{Efficacy and clinical monitoring strategies for immune checkpoint inhibitors and targeted cytokine immunotherapy for locally advanced and metastatic colorectal cancer}.
\newblock {\it \bibinfo{journal}{Cytokine \& growth factor reviews}\/},  {\it \bibinfo{volume}{49}\/}, \bibinfo{pages}{1--9}.
\bibitem[{Blondel et~al.(2021)Blondel, Delconte, Khairallah, Marchal, Gavoille \& Amouroux}]{blondel2021spatially}
\bibinfo{author}{Blondel, W.}, \bibinfo{author}{Delconte, A.}, \bibinfo{author}{Khairallah, G.}, \bibinfo{author}{Marchal, F.}, \bibinfo{author}{Gavoille, A.}, \& \bibinfo{author}{Amouroux, M.} (\bibinfo{year}{2021}).
\newblock \bibinfo{title}{Spatially-resolved multiply-excited autofluorescence and diffuse reflectance spectroscopy: Spectrolive medical device for skin in vivo optical biopsy}.
\newblock {\it \bibinfo{journal}{Electronics}\/},  {\it \bibinfo{volume}{10}\/}, \bibinfo{pages}{243}.
\bibitem[{de~Boer et~al.(2021)de~Boer, Kho, Van~de Vijver, Vranken~Peeters, van Duijnhoven, Hendriks, Sterenborg \& Ruers}]{de_boer_optical_2021}
\bibinfo{author}{de~Boer, L.~L.}, \bibinfo{author}{Kho, E.}, \bibinfo{author}{Van~de Vijver, K.~K.}, \bibinfo{author}{Vranken~Peeters, M.-J. T. F.~D.}, \bibinfo{author}{van Duijnhoven, F.}, \bibinfo{author}{Hendriks, B. H.~W.}, \bibinfo{author}{Sterenborg, H. J. C.~M.}, \& \bibinfo{author}{Ruers, T. J.~M.} (\bibinfo{year}{2021}).
\newblock \bibinfo{title}{Optical tissue measurements of invasive carcinoma and ductal carcinoma insitu for surgical guidance}.
\newblock {\it \bibinfo{journal}{Breast Cancer Res.}\/},  {\it \bibinfo{volume}{23}\/}, \bibinfo{pages}{59}.
\newblock \bibinfo{note}{Publisher: Springer Science and Business Media LLC}.
\bibitem[{Bona \& Andr{\'e}s(2007)}]{bona2007coal}
\bibinfo{author}{Bona, M.}, \& \bibinfo{author}{Andr{\'e}s, J.} (\bibinfo{year}{2007}).
\newblock \bibinfo{title}{Coal analysis by diffuse reflectance near-infrared spectroscopy: Hierarchical cluster and linear discriminant analysis}.
\newblock {\it \bibinfo{journal}{Talanta}\/},  {\it \bibinfo{volume}{72}\/}, \bibinfo{pages}{1423--1431}.
\bibitem[{Bray et~al.(2024)Bray, Laversanne, Sung, Ferlay, Siegel, Soerjomataram \& Jemal}]{bray2024global}
\bibinfo{author}{Bray, F.}, \bibinfo{author}{Laversanne, M.}, \bibinfo{author}{Sung, H.}, \bibinfo{author}{Ferlay, J.}, \bibinfo{author}{Siegel, R.~L.}, \bibinfo{author}{Soerjomataram, I.}, \& \bibinfo{author}{Jemal, A.} (\bibinfo{year}{2024}).
\newblock \bibinfo{title}{Global cancer statistics 2022: Globocan estimates of incidence and mortality worldwide for 36 cancers in 185 countries}.
\newblock {\it \bibinfo{journal}{CA: a cancer journal for clinicians}\/},  {\it \bibinfo{volume}{74}\/}, \bibinfo{pages}{229--263}.
\bibitem[{Brown et~al.(2023)Brown, Tomasev, Freyberg, Liu, Karthikesalingam \& Schrouff}]{brown2023detecting}
\bibinfo{author}{Brown, A.}, \bibinfo{author}{Tomasev, N.}, \bibinfo{author}{Freyberg, J.}, \bibinfo{author}{Liu, Y.}, \bibinfo{author}{Karthikesalingam, A.}, \& \bibinfo{author}{Schrouff, J.} (\bibinfo{year}{2023}).
\newblock \bibinfo{title}{Detecting shortcut learning for fair medical ai using shortcut testing}.
\newblock {\it \bibinfo{journal}{Nature communications}\/},  {\it \bibinfo{volume}{14}\/}, \bibinfo{pages}{4314}.
\bibitem[{Can \& Ülgen(2018)}]{can_estimation_2018}
\bibinfo{author}{Can, O.~M.}, \& \bibinfo{author}{Ülgen, Y.} (\bibinfo{year}{2018}).
\newblock \bibinfo{title}{Estimation of free hemoglobin concentrations in blood bags by diffusereflectance spectroscopy}.
\newblock {\it \bibinfo{journal}{J. Biomed. Opt.}\/},  {\it \bibinfo{volume}{23}\/}, \bibinfo{pages}{1--12}.
\newblock \bibinfo{note}{Publisher: SPIE-Intl Soc Optical Eng}.
\bibitem[{Caruso et~al.(2024)Caruso, Wagar, Hsu, Hoegl, Valzacchi, Fernandes, Cucinella, Aker, Jayraj, Mauro et~al.}]{caruso2024cervical}
\bibinfo{author}{Caruso, G.}, \bibinfo{author}{Wagar, M.~K.}, \bibinfo{author}{Hsu, H.-C.}, \bibinfo{author}{Hoegl, J.}, \bibinfo{author}{Valzacchi, G. M.~R.}, \bibinfo{author}{Fernandes, A.}, \bibinfo{author}{Cucinella, G.}, \bibinfo{author}{Aker, S.~S.}, \bibinfo{author}{Jayraj, A.~S.}, \bibinfo{author}{Mauro, J.} et~al. (\bibinfo{year}{2024}).
\newblock \bibinfo{title}{Cervical cancer: a new era}.
\newblock {\it \bibinfo{journal}{International Journal of Gynecologic Cancer}\/},  (pp. \bibinfo{pages}{ijgc--2024}).
\bibitem[{Center \& Jemal(2011)}]{center2011international}
\bibinfo{author}{Center, M.~M.}, \& \bibinfo{author}{Jemal, A.} (\bibinfo{year}{2011}).
\newblock \bibinfo{title}{International trends in liver cancer incidence rates}.
\newblock {\it \bibinfo{journal}{Cancer Epidemiology, Biomarkers \& Prevention}\/},  {\it \bibinfo{volume}{20}\/}, \bibinfo{pages}{2362--2368}.
\bibitem[{Charvát et~al.(2020)Charvát, Procházka, Fričl, Vyšata \& Himmlová}]{charvat_diffuse_2020}
\bibinfo{author}{Charvát, J.}, \bibinfo{author}{Procházka, A.}, \bibinfo{author}{Fričl, M.}, \bibinfo{author}{Vyšata, O.}, \& \bibinfo{author}{Himmlová, L.} (\bibinfo{year}{2020}).
\newblock \bibinfo{title}{Diffuse reflectance spectroscopy in dental caries detection and classification}.
\newblock {\it \bibinfo{journal}{Signal, Image and Video Processing}\/},  {\it \bibinfo{volume}{14}\/}, \bibinfo{pages}{1063--1070}.
\bibitem[{Chatterjee et~al.(2021)Chatterjee, Das, Ghosh, Banerjee, Bhattacharjee \& Banerjee}]{chatterjee2021performance}
\bibinfo{author}{Chatterjee, S.}, \bibinfo{author}{Das, A.~K.}, \bibinfo{author}{Ghosh, K.}, \bibinfo{author}{Banerjee, A.}, \bibinfo{author}{Bhattacharjee, M.}, \& \bibinfo{author}{Banerjee, S.} (\bibinfo{year}{2021}).
\newblock \bibinfo{title}{Performance improvement of artificial neural networks by addressing class overlapping problem}.
\newblock In {\it \bibinfo{booktitle}{Advances in Smart Communication Technology and Information Processing: OPTRONIX 2020}\/} (pp. \bibinfo{pages}{229--237}).
\newblock \bibinfo{organization}{Springer}.
\bibitem[{Chaudhry et~al.(2023)Chaudhry, Albinsson, Cinthio, Kröll, Malmsjö, Rydén, Sheikh, Reistad \& Zackrisson}]{chaudhry_breast_2023}
\bibinfo{author}{Chaudhry, N.}, \bibinfo{author}{Albinsson, J.}, \bibinfo{author}{Cinthio, M.}, \bibinfo{author}{Kröll, S.}, \bibinfo{author}{Malmsjö, M.}, \bibinfo{author}{Rydén, L.}, \bibinfo{author}{Sheikh, R.}, \bibinfo{author}{Reistad, N.}, \& \bibinfo{author}{Zackrisson, S.} (\bibinfo{year}{2023}).
\newblock \bibinfo{title}{Breast {Cancer} {Diagnosis} {Using} {Extended}-{Wavelength}–{Diffuse} {Reflectance} {Spectroscopy} ({EW}-{DRS})—{Proof} of {Concept} in {Ex} {Vivo} {Breast} {Specimens} {Using} {Machine} {Learning}}.
\newblock {\it \bibinfo{journal}{Diagnostics}\/},  {\it \bibinfo{volume}{13}\/}.
\bibitem[{Chawla et~al.(2002)Chawla, Bowyer, Hall \& Kegelmeyer}]{chawla2002smote}
\bibinfo{author}{Chawla, N.~V.}, \bibinfo{author}{Bowyer, K.~W.}, \bibinfo{author}{Hall, L.~O.}, \& \bibinfo{author}{Kegelmeyer, W.~P.} (\bibinfo{year}{2002}).
\newblock \bibinfo{title}{Smote: synthetic minority over-sampling technique}.
\newblock {\it \bibinfo{journal}{Journal of artificial intelligence research}\/},  {\it \bibinfo{volume}{16}\/}, \bibinfo{pages}{321--357}.
\bibitem[{Dahlstrand et~al.(2019)Dahlstrand, Sheikh, Ansson, Memarzadeh, Reistad \& Malmsjö}]{dahlstrand_extended-wavelength_2019}
\bibinfo{author}{Dahlstrand, U.}, \bibinfo{author}{Sheikh, R.}, \bibinfo{author}{Ansson, C.}, \bibinfo{author}{Memarzadeh, K.}, \bibinfo{author}{Reistad, N.}, \& \bibinfo{author}{Malmsjö, M.} (\bibinfo{year}{2019}).
\newblock \bibinfo{title}{Extended-wavelength diffuse reflectance spectroscopy with a machine-learning method for in vivo tissue classification}.
\newblock {\it \bibinfo{journal}{PLoS ONE}\/},  {\it \bibinfo{volume}{14}\/}.
\bibitem[{Dao et~al.(2023)Dao, Gohla, Williams, Lovrics, Badr, Fang, Farrell \& Farquharson}]{dao2023breast}
\bibinfo{author}{Dao, E.}, \bibinfo{author}{Gohla, G.}, \bibinfo{author}{Williams, P.}, \bibinfo{author}{Lovrics, P.}, \bibinfo{author}{Badr, F.}, \bibinfo{author}{Fang, Q.}, \bibinfo{author}{Farrell, T.}, \& \bibinfo{author}{Farquharson, M.} (\bibinfo{year}{2023}).
\newblock \bibinfo{title}{Breast tissue analysis using a clinically compatible combined time-resolved fluorescence and diffuse reflectance (trf-dr) system}.
\newblock {\it \bibinfo{journal}{Lasers in Surgery and Medicine}\/},  {\it \bibinfo{volume}{55}\/}, \bibinfo{pages}{769--783}.
\bibitem[{Dara \& Tumma(2018)}]{dara2018feature}
\bibinfo{author}{Dara, S.}, \& \bibinfo{author}{Tumma, P.} (\bibinfo{year}{2018}).
\newblock \bibinfo{title}{Feature extraction by using deep learning: A survey}.
\newblock In {\it \bibinfo{booktitle}{2018 Second international conference on electronics, communication and aerospace technology (ICECA)}\/} (pp. \bibinfo{pages}{1795--1801}).
\newblock \bibinfo{organization}{IEEE}.
\bibitem[{Davey et~al.(2021)Davey, Vasiljevski, O'Donohue, Fleming \& Schindeler}]{davey_analysis_2021}
\bibinfo{author}{Davey, C.~J.}, \bibinfo{author}{Vasiljevski, E.~R.}, \bibinfo{author}{O'Donohue, A.~K.}, \bibinfo{author}{Fleming, S.~C.}, \& \bibinfo{author}{Schindeler, A.} (\bibinfo{year}{2021}).
\newblock \bibinfo{title}{Analysis of muscle tissue in vivo using fiber-optic autofluorescence anddiffuse reflectance spectroscopy}.
\newblock {\it \bibinfo{journal}{J. Biomed. Opt.}\/},  {\it \bibinfo{volume}{26}\/}.
\newblock \bibinfo{note}{Publisher: SPIE-Intl Soc Optical Eng}.
\bibitem[{De~Boer et~al.(2018)De~Boer, Bydlon, Van~Duijnhoven, Vranken~Peeters, Loo, Winter-Warnars, Sanders, Sterenborg, Hendriks \& Ruers}]{de_boer_towards_2018}
\bibinfo{author}{De~Boer, L.}, \bibinfo{author}{Bydlon, T.}, \bibinfo{author}{Van~Duijnhoven, F.}, \bibinfo{author}{Vranken~Peeters, M.-J.}, \bibinfo{author}{Loo, C.}, \bibinfo{author}{Winter-Warnars, G.}, \bibinfo{author}{Sanders, J.}, \bibinfo{author}{Sterenborg, H.}, \bibinfo{author}{Hendriks, B.}, \& \bibinfo{author}{Ruers, T.} (\bibinfo{year}{2018}).
\newblock \bibinfo{title}{Towards the use of diffuse reflectance spectroscopy for real-time in vivo detection of breast cancer during surgery}.
\newblock {\it \bibinfo{journal}{Journal of Translational Medicine}\/},  {\it \bibinfo{volume}{16}\/}.
\bibitem[{De~Boer et~al.(2016)De~Boer, Hendriks, Van~Duijnhoven, Peeters-Baas, Van~de Vijver, Loo, J{\'o}{\'z}wiak, Sterenborg \& Ruers}]{de2016using}
\bibinfo{author}{De~Boer, L.~L.}, \bibinfo{author}{Hendriks, B.~H.}, \bibinfo{author}{Van~Duijnhoven, F.}, \bibinfo{author}{Peeters-Baas, M.-J. T.~V.}, \bibinfo{author}{Van~de Vijver, K.}, \bibinfo{author}{Loo, C.~E.}, \bibinfo{author}{J{\'o}{\'z}wiak, K.}, \bibinfo{author}{Sterenborg, H.~J.}, \& \bibinfo{author}{Ruers, T.~J.} (\bibinfo{year}{2016}).
\newblock \bibinfo{title}{Using drs during breast conserving surgery: identifying robust optical parameters and influence of inter-patient variation}.
\newblock {\it \bibinfo{journal}{Biomedical optics express}\/},  {\it \bibinfo{volume}{7}\/}, \bibinfo{pages}{5188--5200}.
\bibitem[{Devpura et~al.(2011)Devpura, Pattamadilok, Syed, Vemulapalli, Henderson, Rehse, Hamzavi, Lim \& Naik}]{devpura_critical_2011}
\bibinfo{author}{Devpura, S.}, \bibinfo{author}{Pattamadilok, B.}, \bibinfo{author}{Syed, Z.~U.}, \bibinfo{author}{Vemulapalli, P.}, \bibinfo{author}{Henderson, M.}, \bibinfo{author}{Rehse, S.~J.}, \bibinfo{author}{Hamzavi, I.}, \bibinfo{author}{Lim, H.~W.}, \& \bibinfo{author}{Naik, R.} (\bibinfo{year}{2011}).
\newblock \bibinfo{title}{Critical comparison of diffuse reflectance spectroscopy and colorimetry asdermatological diagnostic tools for acanthosis nigricans: a chemometricapproach}.
\newblock {\it \bibinfo{journal}{Biomed. Opt. Express}\/},  {\it \bibinfo{volume}{2}\/}, \bibinfo{pages}{1664--1673}.
\newblock \bibinfo{note}{Publisher: The Optical Society}.
\bibitem[{Dhanabal \& Chandramathi(2011)}]{dhanabal2011review}
\bibinfo{author}{Dhanabal, S.}, \& \bibinfo{author}{Chandramathi, S.} (\bibinfo{year}{2011}).
\newblock \bibinfo{title}{A review of various k-nearest neighbor query processing techniques}.
\newblock {\it \bibinfo{journal}{International Journal of Computer Applications}\/},  {\it \bibinfo{volume}{31}\/}, \bibinfo{pages}{14--22}.
\bibitem[{Duperron et~al.(2019)Duperron, Grygoryev, Nunan, Eason, Gunther, Burke, Manley \& O’brien}]{duperron2019diffuse}
\bibinfo{author}{Duperron, M.}, \bibinfo{author}{Grygoryev, K.}, \bibinfo{author}{Nunan, G.}, \bibinfo{author}{Eason, C.}, \bibinfo{author}{Gunther, J.}, \bibinfo{author}{Burke, R.}, \bibinfo{author}{Manley, K.}, \& \bibinfo{author}{O’brien, P.} (\bibinfo{year}{2019}).
\newblock \bibinfo{title}{Diffuse reflectance spectroscopy-enhanced drill for bone boundary detection}.
\newblock {\it \bibinfo{journal}{Biomedical optics express}\/},  {\it \bibinfo{volume}{10}\/}, \bibinfo{pages}{961--977}.
\bibitem[{Evers et~al.(2013)Evers, Nachabe, Vranken~Peeters, van~der Hage, Oldenburg, Rutgers, Lucassen, Hendriks, Wesseling \& Ruers}]{evers_diffuse_2013}
\bibinfo{author}{Evers, D.~J.}, \bibinfo{author}{Nachabe, R.}, \bibinfo{author}{Vranken~Peeters, M.-J.}, \bibinfo{author}{van~der Hage, J.~A.}, \bibinfo{author}{Oldenburg, H.~S.}, \bibinfo{author}{Rutgers, E.~J.}, \bibinfo{author}{Lucassen, G.~W.}, \bibinfo{author}{Hendriks, B. H.~W.}, \bibinfo{author}{Wesseling, J.}, \& \bibinfo{author}{Ruers, T. J.~M.} (\bibinfo{year}{2013}).
\newblock \bibinfo{title}{Diffuse reflectance spectroscopy: towards clinical application in breastcancer}.
\newblock {\it \bibinfo{journal}{Breast Cancer Res. Treat.}\/},  {\it \bibinfo{volume}{137}\/}, \bibinfo{pages}{155--165}.
\newblock \bibinfo{note}{Publisher: Springer Science and Business Media LLC}.
\bibitem[{Evers et~al.(2012)Evers, Nachabé, Klomp, van Sandick, Wouters, Lucassen, Hendriks, Wesseling \& Ruers}]{evers_diffuse_2012}
\bibinfo{author}{Evers, D.~J.}, \bibinfo{author}{Nachabé, R.}, \bibinfo{author}{Klomp, H.~M.}, \bibinfo{author}{van Sandick, J.~W.}, \bibinfo{author}{Wouters, M.~W.}, \bibinfo{author}{Lucassen, G.~W.}, \bibinfo{author}{Hendriks, B. H.~W.}, \bibinfo{author}{Wesseling, J.}, \& \bibinfo{author}{Ruers, T. J.~M.} (\bibinfo{year}{2012}).
\newblock \bibinfo{title}{Diffuse reflectance spectroscopy: a new guidance tool for improvement of biopsy procedures in lung malignancies}.
\newblock {\it \bibinfo{journal}{Clin. Lung Cancer}\/},  {\it \bibinfo{volume}{13}\/}, \bibinfo{pages}{424--431}.
\newblock \bibinfo{note}{Publisher: Elsevier BV}.
\bibitem[{Fanjul-V{\'e}lez et~al.(2018)Fanjul-V{\'e}lez, Ar{\'e}valo-D{\'\i}az \& Arce-Diego}]{fanjul2018intra}
\bibinfo{author}{Fanjul-V{\'e}lez, F.}, \bibinfo{author}{Ar{\'e}valo-D{\'\i}az, L.}, \& \bibinfo{author}{Arce-Diego, J.~L.} (\bibinfo{year}{2018}).
\newblock \bibinfo{title}{Intra-class variability in diffuse reflectance spectroscopy: application to porcine adipose tissue}.
\newblock {\it \bibinfo{journal}{Biomedical Optics Express}\/},  {\it \bibinfo{volume}{9}\/}, \bibinfo{pages}{2297--2303}.
\bibitem[{Fanjul-Vélez et~al.(2020)Fanjul-Vélez, Pampín-Suárez \& Arce-Diego}]{fanjul-velez_application_2020}
\bibinfo{author}{Fanjul-Vélez, F.}, \bibinfo{author}{Pampín-Suárez, S.}, \& \bibinfo{author}{Arce-Diego, J.~L.} (\bibinfo{year}{2020}).
\newblock \bibinfo{title}{Application of classification algorithms to diffuse reflectancespectroscopy measurements for ex vivo characterization of biologicaltissues}.
\newblock {\it \bibinfo{journal}{Entropy (Basel)}\/},  {\it \bibinfo{volume}{22}\/}, \bibinfo{pages}{736}.
\newblock \bibinfo{note}{Publisher: MDPI AG}.
\bibitem[{Furriel et~al.(2024)Furriel, Oliveira, Pr{\^o}a, Paiva, Loureiro, Calixto, Reis \& Giavina-Bianchi}]{furriel2024artificial}
\bibinfo{author}{Furriel, B.~C.}, \bibinfo{author}{Oliveira, B.~D.}, \bibinfo{author}{Pr{\^o}a, R.}, \bibinfo{author}{Paiva, J.~Q.}, \bibinfo{author}{Loureiro, R.~M.}, \bibinfo{author}{Calixto, W.~P.}, \bibinfo{author}{Reis, M.~R.}, \& \bibinfo{author}{Giavina-Bianchi, M.} (\bibinfo{year}{2024}).
\newblock \bibinfo{title}{Artificial intelligence for skin cancer detection and classification for clinical environment: a systematic review}.
\newblock {\it \bibinfo{journal}{Frontiers in Medicine}\/},  {\it \bibinfo{volume}{10}\/}, \bibinfo{pages}{1305954}.
\bibitem[{Garcia et~al.(2009)Garcia, Feldman, Gupta \& Srivastava}]{garcia2009completely}
\bibinfo{author}{Garcia, E.~K.}, \bibinfo{author}{Feldman, S.}, \bibinfo{author}{Gupta, M.~R.}, \& \bibinfo{author}{Srivastava, S.} (\bibinfo{year}{2009}).
\newblock \bibinfo{title}{Completely lazy learning}.
\newblock {\it \bibinfo{journal}{IEEE Transactions on Knowledge and Data Engineering}\/},  {\it \bibinfo{volume}{22}\/}, \bibinfo{pages}{1274--1285}.
\bibitem[{Geldof et~al.(2022)Geldof, Dashtbozorg, Hendriks, Sterenborg \& Ruers}]{geldof_layer_2022}
\bibinfo{author}{Geldof, F.}, \bibinfo{author}{Dashtbozorg, B.}, \bibinfo{author}{Hendriks, B.}, \bibinfo{author}{Sterenborg, H.}, \& \bibinfo{author}{Ruers, T.} (\bibinfo{year}{2022}).
\newblock \bibinfo{title}{Layer thickness prediction and tissue classification in two-layered tissue structures using diffuse reflectance spectroscopy}.
\newblock {\it \bibinfo{journal}{Scientific reports}\/},  {\it \bibinfo{volume}{12}\/}, \bibinfo{pages}{1698}.
\bibitem[{Geldof et~al.(2024)Geldof, Schrage, van Houdt, Sterenborg, Dashtbozorg \& Ruers}]{geldof_toward_2024}
\bibinfo{author}{Geldof, F.}, \bibinfo{author}{Schrage, Y.}, \bibinfo{author}{van Houdt, W.}, \bibinfo{author}{Sterenborg, H.}, \bibinfo{author}{Dashtbozorg, B.}, \& \bibinfo{author}{Ruers, T.} (\bibinfo{year}{2024}).
\newblock \bibinfo{title}{Toward the use of diffuse reflection spectroscopy for intra-operative tissue discrimination during sarcoma surgery}.
\newblock {\it \bibinfo{journal}{Journal of biomedical optics}\/},  {\it \bibinfo{volume}{29}\/}, \bibinfo{pages}{027001}.
\bibitem[{Geldof et~al.(2023)Geldof, Witteveen, Sterenborg, Ruers \& Dashtbozorg}]{geldof_diffuse_2023}
\bibinfo{author}{Geldof, F.}, \bibinfo{author}{Witteveen, M.}, \bibinfo{author}{Sterenborg, H.}, \bibinfo{author}{Ruers, T.}, \& \bibinfo{author}{Dashtbozorg, B.} (\bibinfo{year}{2023}).
\newblock \bibinfo{title}{Diffuse reflection spectroscopy at the fingertip: design and performance of a compact side-firing probe for tissue discrimination during colorectal cancer surgery}.
\newblock {\it \bibinfo{journal}{Biomedical Optics Express}\/},  {\it \bibinfo{volume}{14}\/}, \bibinfo{pages}{128--147}.
\bibitem[{Ghosh et~al.(2019)Ghosh, Dasgupta \& Swetapadma}]{ghosh2019study}
\bibinfo{author}{Ghosh, S.}, \bibinfo{author}{Dasgupta, A.}, \& \bibinfo{author}{Swetapadma, A.} (\bibinfo{year}{2019}).
\newblock \bibinfo{title}{A study on support vector machine based linear and non-linear pattern classification}.
\newblock In {\it \bibinfo{booktitle}{2019 International Conference on Intelligent Sustainable Systems (ICISS)}\/} (pp. \bibinfo{pages}{24--28}).
\newblock \bibinfo{organization}{IEEE}.
\bibitem[{Gkouzionis et~al.(2022)Gkouzionis, Nazarian, Kawka, Darzi, Patel, Peters \& Elson}]{gkouzionis_real-time_2022}
\bibinfo{author}{Gkouzionis, I.}, \bibinfo{author}{Nazarian, S.}, \bibinfo{author}{Kawka, M.}, \bibinfo{author}{Darzi, A.}, \bibinfo{author}{Patel, N.}, \bibinfo{author}{Peters, C.}, \& \bibinfo{author}{Elson, D.} (\bibinfo{year}{2022}).
\newblock \bibinfo{title}{Real-time tracking of a diffuse reflectance spectroscopy probe used to aid histological validation of margin assessment in upper gastrointestinal cancer resection surgery}.
\newblock {\it \bibinfo{journal}{Journal of biomedical optics}\/},  {\it \bibinfo{volume}{27}\/}.
\bibitem[{Gunaratne et~al.(2020)Gunaratne, Goncalves, Monteath, Sheh, Kapfer, Chipper, Robertson, Khan, Fick \& Ironside}]{gunaratne_wavelength_2020}
\bibinfo{author}{Gunaratne, R.}, \bibinfo{author}{Goncalves, J.}, \bibinfo{author}{Monteath, I.}, \bibinfo{author}{Sheh, R.}, \bibinfo{author}{Kapfer, M.}, \bibinfo{author}{Chipper, R.}, \bibinfo{author}{Robertson, B.}, \bibinfo{author}{Khan, R.}, \bibinfo{author}{Fick, D.}, \& \bibinfo{author}{Ironside, C.} (\bibinfo{year}{2020}).
\newblock \bibinfo{title}{Wavelength weightings in machine learning for ovine joint tissue differentiation using diffuse reflectance spectroscopy ({DRS})}.
\newblock {\it \bibinfo{journal}{Biomedical Optics Express}\/},  {\it \bibinfo{volume}{11}\/}, \bibinfo{pages}{5122--5131}.
\bibitem[{Gunaratne et~al.(2019)Gunaratne, Monteath, Goncalves, Sheh, Ironside, Kapfer, Chipper, Robertson, Khan \& Fick}]{gunaratne_machine_2019}
\bibinfo{author}{Gunaratne, R.}, \bibinfo{author}{Monteath, I.}, \bibinfo{author}{Goncalves, J.}, \bibinfo{author}{Sheh, R.}, \bibinfo{author}{Ironside, C.}, \bibinfo{author}{Kapfer, M.}, \bibinfo{author}{Chipper, R.}, \bibinfo{author}{Robertson, B.}, \bibinfo{author}{Khan, R.}, \& \bibinfo{author}{Fick, D.} (\bibinfo{year}{2019}).
\newblock \bibinfo{title}{Machine learning classification of human joint tissue from diffuse reflectance spectroscopy data}.
\newblock {\it \bibinfo{journal}{Biomedical Optics Express}\/},  {\it \bibinfo{volume}{10}\/}, \bibinfo{pages}{3889--3898}.
\bibitem[{Hamdy et~al.(2023)Hamdy, Abdel-Salam \& Abdel-Harith}]{hamdy2023monitoring}
\bibinfo{author}{Hamdy, O.}, \bibinfo{author}{Abdel-Salam, Z.}, \& \bibinfo{author}{Abdel-Harith, M.} (\bibinfo{year}{2023}).
\newblock \bibinfo{title}{Monitoring liver tissue photocoagulation using spatially-resolved diffuse reflectance and laser-induced fluorescence spectroscopy}.
\newblock In {\it \bibinfo{booktitle}{AIP Conference Proceedings}\/}.
\newblock \bibinfo{organization}{AIP Publishing} volume \bibinfo{volume}{2620}.
\bibitem[{Ilic \& Ilic(2023)}]{ilic2023international}
\bibinfo{author}{Ilic, I.}, \& \bibinfo{author}{Ilic, M.} (\bibinfo{year}{2023}).
\newblock \bibinfo{title}{International patterns and trends in the brain cancer incidence and mortality: An observational study based on the global burden of disease}.
\newblock {\it \bibinfo{journal}{Heliyon}\/},  {\it \bibinfo{volume}{9}\/}.
\bibitem[{Jayanthi et~al.(2011)Jayanthi, Nisha, Manju, Philip, Jeemon, Baiju, Beena \& Subhash}]{jayanthi_diffuse_2011}
\bibinfo{author}{Jayanthi, J.~L.}, \bibinfo{author}{Nisha, G.~U.}, \bibinfo{author}{Manju, S.}, \bibinfo{author}{Philip, E.~K.}, \bibinfo{author}{Jeemon, P.}, \bibinfo{author}{Baiju, K.~V.}, \bibinfo{author}{Beena, V.~T.}, \& \bibinfo{author}{Subhash, N.} (\bibinfo{year}{2011}).
\newblock \bibinfo{title}{Diffuse reflectance spectroscopy: diagnostic accuracy of a non-invasivescreening technique for early detection of malignant changes in the oralcavity}.
\newblock {\it \bibinfo{journal}{BMJ Open}\/},  {\it \bibinfo{volume}{1}\/}, \bibinfo{pages}{e000071--e000071}.
\newblock \bibinfo{note}{Publisher: BMJ}.
\bibitem[{Keller et~al.(2018)Keller, Bialecki, Wilhelm \& Vetter}]{keller_diffuse_2018}
\bibinfo{author}{Keller, A.}, \bibinfo{author}{Bialecki, P.}, \bibinfo{author}{Wilhelm, T.~J.}, \& \bibinfo{author}{Vetter, M.~K.} (\bibinfo{year}{2018}).
\newblock \bibinfo{title}{Diffuse reflectance spectroscopy of human liver tumor specimens - towardsa tissue differentiating optical biopsy needle using light emitting diodes}.
\newblock {\it \bibinfo{journal}{Biomed. Opt. Express}\/},  {\it \bibinfo{volume}{9}\/}, \bibinfo{pages}{1069}.
\newblock \bibinfo{note}{Publisher: The Optical Society}.
\bibitem[{Kim et~al.(2020)Kim, Wales \& Yang}]{kim2020optical}
\bibinfo{author}{Kim, J.~A.}, \bibinfo{author}{Wales, D.~J.}, \& \bibinfo{author}{Yang, G.-Z.} (\bibinfo{year}{2020}).
\newblock \bibinfo{title}{Optical spectroscopy for in vivo medical diagnosis—a review of the state of the art and future perspectives}.
\newblock {\it \bibinfo{journal}{Progress in Biomedical Engineering}\/},  {\it \bibinfo{volume}{2}\/}, \bibinfo{pages}{042001}.
\bibitem[{Kohavi \& John(1997)}]{kohavi1997wrappers}
\bibinfo{author}{Kohavi, R.}, \& \bibinfo{author}{John, G.~H.} (\bibinfo{year}{1997}).
\newblock \bibinfo{title}{Wrappers for feature subset selection}.
\newblock {\it \bibinfo{journal}{Artificial intelligence}\/},  {\it \bibinfo{volume}{97}\/}, \bibinfo{pages}{273--324}.
\bibitem[{Kolpakov et~al.(2023)Kolpakov, Yanushevich, Mamatsashvili, Sokolova, Abramova, Parshkov, Moshkova \& Zolotnitsky}]{a_v_kolpakov_decision_2023}
\bibinfo{author}{Kolpakov, A.}, \bibinfo{author}{Yanushevich, O.~O.}, \bibinfo{author}{Mamatsashvili, V.~G.}, \bibinfo{author}{Sokolova, D.~Y.}, \bibinfo{author}{Abramova, M.~Y.}, \bibinfo{author}{Parshkov, V.~V.}, \bibinfo{author}{Moshkova, A.~A.}, \& \bibinfo{author}{Zolotnitsky, I.~V.} (\bibinfo{year}{2023}).
\newblock \bibinfo{title}{Decision {Support} {System} for {Oncoscreening} of the {Oral} {Mucosa} and {Red} {Border} of the {Lips} {Using} {Diffuse} {Reflectance} {Spectroscopy} of the {Visible} qnd {Near}-{IR} {Range} in {Vivo}}.
\newblock (pp. \bibinfo{pages}{198--201}).
\bibitem[{Brouwer~de Koning et~al.(2018)Brouwer~de Koning, Baltussen, Karakullukcu, Dashtbozorg, Smit, Dirven, Hendriks, Sterenborg \& Ruers}]{brouwer_de_koning_toward_2018}
\bibinfo{author}{Brouwer~de Koning, S.}, \bibinfo{author}{Baltussen, E.}, \bibinfo{author}{Karakullukcu, M.}, \bibinfo{author}{Dashtbozorg, B.}, \bibinfo{author}{Smit, L.}, \bibinfo{author}{Dirven, R.}, \bibinfo{author}{Hendriks, B.}, \bibinfo{author}{Sterenborg, H.}, \& \bibinfo{author}{Ruers, T.} (\bibinfo{year}{2018}).
\newblock \bibinfo{title}{Toward complete oral cavity cancer resection using a handheld diffuse reflectance spectroscopy probe}.
\newblock {\it \bibinfo{journal}{Journal of biomedical optics}\/},  {\it \bibinfo{volume}{23}\/}, \bibinfo{pages}{1--8}.
\bibitem[{Kononenko(2001)}]{kononenko2001machine}
\bibinfo{author}{Kononenko, I.} (\bibinfo{year}{2001}).
\newblock \bibinfo{title}{Machine learning for medical diagnosis: history, state of the art and perspective}.
\newblock {\it \bibinfo{journal}{Artificial Intelligence in medicine}\/},  {\it \bibinfo{volume}{23}\/}, \bibinfo{pages}{89--109}.
\bibitem[{Kreiß et~al.(2019)Kreiß, Hohmann, Klämpfl, Schürmann, Dehghani, Schmidt, Friedrich \& Büchler}]{kreis_diffuse_2019}
\bibinfo{author}{Kreiß, L.}, \bibinfo{author}{Hohmann, M.}, \bibinfo{author}{Klämpfl, F.}, \bibinfo{author}{Schürmann, S.}, \bibinfo{author}{Dehghani, F.}, \bibinfo{author}{Schmidt, M.}, \bibinfo{author}{Friedrich, O.}, \& \bibinfo{author}{Büchler, L.} (\bibinfo{year}{2019}).
\newblock \bibinfo{title}{Diffuse reflectance spectroscopy and {Raman} spectroscopy for label-free molecular characterization and automated detection of human cartilage and subchondral bone}.
\newblock {\it \bibinfo{journal}{Sensors and Actuators, B: Chemical}\/},  {\it \bibinfo{volume}{301}\/}.
\bibitem[{Kretzschmar et~al.(2005)Kretzschmar, Karayiannis \& Eggimann}]{kretzschmar2005feedforward}
\bibinfo{author}{Kretzschmar, R.}, \bibinfo{author}{Karayiannis, N.~B.}, \& \bibinfo{author}{Eggimann, F.} (\bibinfo{year}{2005}).
\newblock \bibinfo{title}{Feedforward neural network models for handling class overlap and class imbalance}.
\newblock {\it \bibinfo{journal}{International journal of neural systems}\/},  {\it \bibinfo{volume}{15}\/}, \bibinfo{pages}{323--338}.
\bibitem[{Kupriyanov et~al.(2023{\natexlab{a}})Kupriyanov, Blondel, Daul, Amouroux \& Kistenev}]{kupriyanov_assessing_2023}
\bibinfo{author}{Kupriyanov, V.}, \bibinfo{author}{Blondel, W.}, \bibinfo{author}{Daul, C.}, \bibinfo{author}{Amouroux, M.}, \& \bibinfo{author}{Kistenev, Y.} (\bibinfo{year}{2023}{\natexlab{a}}).
\newblock \bibinfo{title}{Assessing the effectiveness of multimodal data fusion methods for in-vivo diagnosis of pre-cancerous skin conditions in a preclinical model}.
\newblock {\it \bibinfo{journal}{Progress in Biomedical Optics and Imaging - Proceedings of SPIE}\/},  {\it \bibinfo{volume}{12373}\/}.
\bibitem[{Kupriyanov et~al.(2023{\natexlab{b}})Kupriyanov, Blondel, Daul, Amouroux \& Kistenev}]{kupriyanov_evaluating_2023}
\bibinfo{author}{Kupriyanov, V.}, \bibinfo{author}{Blondel, W.}, \bibinfo{author}{Daul, C.}, \bibinfo{author}{Amouroux, M.}, \& \bibinfo{author}{Kistenev, Y.} (\bibinfo{year}{2023}{\natexlab{b}}).
\newblock \bibinfo{title}{Evaluating the efficacy of different data processing methods in diagnosing precancerous skin conditions in a preclinical in vivo model using bimodal spectroscopy}.
\newblock volume \bibinfo{volume}{12378}.
\bibitem[{Lai et~al.(2020)Lai, Skyrman, Shan, Paulussen, Manni, Swamy, Babic, Edstrom, Persson, Burstrom, Elmi-Terander, Hendriks \& De~With}]{lai_automated_2020}
\bibinfo{author}{Lai, M.}, \bibinfo{author}{Skyrman, S.}, \bibinfo{author}{Shan, C.}, \bibinfo{author}{Paulussen, E.}, \bibinfo{author}{Manni, F.}, \bibinfo{author}{Swamy, A.}, \bibinfo{author}{Babic, D.}, \bibinfo{author}{Edstrom, E.}, \bibinfo{author}{Persson, O.}, \bibinfo{author}{Burstrom, G.}, \bibinfo{author}{Elmi-Terander, A.}, \bibinfo{author}{Hendriks, B.}, \& \bibinfo{author}{De~With, P.} (\bibinfo{year}{2020}).
\newblock \bibinfo{title}{Automated classification of brain tissue: {Comparison} between hyperspectral imaging and diffuse reflectance spectroscopy}.
\newblock volume \bibinfo{volume}{11315}.
\bibitem[{Langhout et~al.(2018)Langhout, Kuhlmann, Schreuder, Bydlon, Smeele, van~den Brekel, Sterenborg, Hendriks \& Ruers}]{langhout_vivo_2018}
\bibinfo{author}{Langhout, G.~C.}, \bibinfo{author}{Kuhlmann, K. F.~D.}, \bibinfo{author}{Schreuder, P.}, \bibinfo{author}{Bydlon, T.}, \bibinfo{author}{Smeele, L.~E.}, \bibinfo{author}{van~den Brekel, M. W.~M.}, \bibinfo{author}{Sterenborg, H. J. C.~M.}, \bibinfo{author}{Hendriks, B. H.~W.}, \& \bibinfo{author}{Ruers, T. J.~M.} (\bibinfo{year}{2018}).
\newblock \bibinfo{title}{In vivo nerve identification in head and neck surgery using diffusereflectance spectroscopy}.
\newblock {\it \bibinfo{journal}{Laryngoscope Investig. Otolaryngol.}\/},  {\it \bibinfo{volume}{3}\/}, \bibinfo{pages}{349--355}.
\newblock \bibinfo{note}{Publisher: Wiley}.
\bibitem[{Lawrence et~al.(1997)Lawrence, Giles \& Tsoi}]{lawrence1997lessons}
\bibinfo{author}{Lawrence, S.}, \bibinfo{author}{Giles, C.~L.}, \& \bibinfo{author}{Tsoi, A.~C.} (\bibinfo{year}{1997}).
\newblock \bibinfo{title}{Lessons in neural network training: Overfitting may be harder than expected}.
\newblock In {\it \bibinfo{booktitle}{Aaai/iaai}\/} (pp. \bibinfo{pages}{540--545}).
\bibitem[{Li et~al.(2022)Li, Fisher, Komolibus, Grygoryev, Burke \& Andersson-Engels}]{li_wavelength_2022}
\bibinfo{author}{Li, C.}, \bibinfo{author}{Fisher, C.}, \bibinfo{author}{Komolibus, K.}, \bibinfo{author}{Grygoryev, K.}, \bibinfo{author}{Burke, R.}, \& \bibinfo{author}{Andersson-Engels, S.} (\bibinfo{year}{2022}).
\newblock \bibinfo{title}{Wavelength selection using diffuse reflectance spectra and machine learning algorithms for tissue differentiation in orthopedic surgery}.
\bibitem[{Li et~al.(2023{\natexlab{a}})Li, Fisher, Komolibus, Grygoryev, Lu, Burke, Visentin \& Andersson-Engels}]{li_frameworks_2023}
\bibinfo{author}{Li, C.~L.}, \bibinfo{author}{Fisher, C.~J.}, \bibinfo{author}{Komolibus, K.}, \bibinfo{author}{Grygoryev, K.}, \bibinfo{author}{Lu, H.}, \bibinfo{author}{Burke, R.}, \bibinfo{author}{Visentin, A.}, \& \bibinfo{author}{Andersson-Engels, S.} (\bibinfo{year}{2023}{\natexlab{a}}).
\newblock \bibinfo{title}{Frameworks of wavelength selection in diffuse reflectance spectroscopy fortissue differentiation in orthopedic surgery}.
\newblock {\it \bibinfo{journal}{J. Biomed. Opt.}\/},  {\it \bibinfo{volume}{28}\/}, \bibinfo{pages}{121207}.
\bibitem[{Li et~al.(2018)Li, Sun, Xiang, Lin, Qin \& Li}]{li_identification_2018}
\bibinfo{author}{Li, H.}, \bibinfo{author}{Sun, M.}, \bibinfo{author}{Xiang, Z.}, \bibinfo{author}{Lin, L.}, \bibinfo{author}{Qin, C.}, \& \bibinfo{author}{Li, Y.} (\bibinfo{year}{2018}).
\newblock \bibinfo{title}{Identification of blood species based on diffuse reflectance andtransmission joint spectra with machine learning method}.
\newblock {\it \bibinfo{journal}{Infrared Phys. Technol.}\/},  {\it \bibinfo{volume}{88}\/}, \bibinfo{pages}{200--205}.
\newblock \bibinfo{note}{Publisher: Elsevier BV}.
\bibitem[{Li et~al.(2023{\natexlab{b}})Li, Wu, Feng, Zhao, Jin, Qiu, Gu \& Chen}]{li_situ_2023}
\bibinfo{author}{Li, K.}, \bibinfo{author}{Wu, Q.}, \bibinfo{author}{Feng, S.}, \bibinfo{author}{Zhao, H.}, \bibinfo{author}{Jin, W.}, \bibinfo{author}{Qiu, H.}, \bibinfo{author}{Gu, Y.}, \& \bibinfo{author}{Chen, D.} (\bibinfo{year}{2023}{\natexlab{b}}).
\newblock \bibinfo{title}{In situ detection of human glioma based on tissue optical properties using diffuse reflectance spectroscopy}.
\newblock {\it \bibinfo{journal}{Journal of biophotonics}\/},  {\it \bibinfo{volume}{16}\/}, \bibinfo{pages}{e202300195}.
\bibitem[{Liu et~al.(2018)Liu, Zhao, Lu, Li, Wu \& Zhang}]{liu_blood_2018}
\bibinfo{author}{Liu, M.}, \bibinfo{author}{Zhao, J.}, \bibinfo{author}{Lu, X.}, \bibinfo{author}{Li, G.}, \bibinfo{author}{Wu, T.}, \& \bibinfo{author}{Zhang, L.} (\bibinfo{year}{2018}).
\newblock \bibinfo{title}{Blood hyperviscosity identification with reflective spectroscopy of tongue tip based on principal component analysis combining artificial neural network.}
\newblock {\it \bibinfo{journal}{BioMedical Engineering OnLine}\/},  {\it \bibinfo{volume}{17}\/}, \bibinfo{pages}{N.PAG--N.PAG}.
\newblock \bibinfo{note}{Publisher: BioMed Central}.
\bibitem[{Liu et~al.(2021)Liu, Jin, Li, Xue, Li, Qian, Li \& Yan}]{liu_study_2021}
\bibinfo{author}{Liu, W.}, \bibinfo{author}{Jin, X.}, \bibinfo{author}{Li, J.}, \bibinfo{author}{Xue, Y.}, \bibinfo{author}{Li, Y.}, \bibinfo{author}{Qian, Z.}, \bibinfo{author}{Li, W.}, \& \bibinfo{author}{Yan, X.} (\bibinfo{year}{2021}).
\newblock \bibinfo{title}{Study of cervical precancerous lesions detection by spectroscopy and support vector machine.}
\newblock {\it \bibinfo{journal}{Minimally Invasive Therapy \& Allied Technologies}\/},  {\it \bibinfo{volume}{30}\/}, \bibinfo{pages}{208--214}.
\newblock \bibinfo{note}{Place: Philadelphia, Pennsylvania Publisher: Taylor \& Francis Ltd}.
\bibitem[{Loh et~al.(2022)Loh, Ooi, Seoni, Barua, Molinari \& Acharya}]{loh2022application}
\bibinfo{author}{Loh, H.~W.}, \bibinfo{author}{Ooi, C.~P.}, \bibinfo{author}{Seoni, S.}, \bibinfo{author}{Barua, P.~D.}, \bibinfo{author}{Molinari, F.}, \& \bibinfo{author}{Acharya, U.~R.} (\bibinfo{year}{2022}).
\newblock \bibinfo{title}{Application of explainable artificial intelligence for healthcare: A systematic review of the last decade (2011--2022)}.
\newblock {\it \bibinfo{journal}{Computer Methods and Programs in Biomedicine}\/},  {\it \bibinfo{volume}{226}\/}, \bibinfo{pages}{107161}.
\bibitem[{Lu et~al.(2023)Lu, O'Dowling, Caplice, Burke \& Andersson-Engels}]{lu_use_2023}
\bibinfo{author}{Lu, H.}, \bibinfo{author}{O'Dowling, C.}, \bibinfo{author}{Caplice, N.}, \bibinfo{author}{Burke, R.}, \& \bibinfo{author}{Andersson-Engels, S.} (\bibinfo{year}{2023}).
\newblock \bibinfo{title}{Use of derivative diffuse reflectance spectroscopy in {CAM} assay combinedwith multivariate analysis as an approach to detect features of vulnerableplaques}.
\newblock \bibinfo{publisher}{SPIE}.
\bibitem[{Lu et~al.(2021)Lu, Mullins, Schafmayer, Zei{\ss}ig \& Linnebacher}]{lu2021global}
\bibinfo{author}{Lu, L.}, \bibinfo{author}{Mullins, C.~S.}, \bibinfo{author}{Schafmayer, C.}, \bibinfo{author}{Zei{\ss}ig, S.}, \& \bibinfo{author}{Linnebacher, M.} (\bibinfo{year}{2021}).
\newblock \bibinfo{title}{A global assessment of recent trends in gastrointestinal cancer and lifestyle-associated risk factors}.
\newblock {\it \bibinfo{journal}{Cancer Communications}\/},  {\it \bibinfo{volume}{41}\/}, \bibinfo{pages}{1137--1151}.
\bibitem[{Majumder et~al.(2008)Majumder, Keller, Boulos, Kelley \& Mahadevan-Jansen}]{majumder_comparison_2008}
\bibinfo{author}{Majumder, S.~K.}, \bibinfo{author}{Keller, M.~D.}, \bibinfo{author}{Boulos, F.~I.}, \bibinfo{author}{Kelley, M.~C.}, \& \bibinfo{author}{Mahadevan-Jansen, A.} (\bibinfo{year}{2008}).
\newblock \bibinfo{title}{Comparison of autofluorescence, diffuse reflectance, and {Ramanspectroscopy} for breast tissue discrimination}.
\newblock {\it \bibinfo{journal}{J. Biomed. Opt.}\/},  {\it \bibinfo{volume}{13}\/}, \bibinfo{pages}{054009}.
\newblock \bibinfo{note}{Publisher: SPIE-Intl Soc Optical Eng}.
\bibitem[{Marín et~al.(2005)Marín, Milbourne, Rhodes, Ehlen, Miller, Benedet, Richards-Kortum \& Follen}]{marin_diffuse_2005}
\bibinfo{author}{Marín, N.~M.}, \bibinfo{author}{Milbourne, A.}, \bibinfo{author}{Rhodes, H.}, \bibinfo{author}{Ehlen, T.}, \bibinfo{author}{Miller, D.}, \bibinfo{author}{Benedet, L.}, \bibinfo{author}{Richards-Kortum, R.}, \& \bibinfo{author}{Follen, M.} (\bibinfo{year}{2005}).
\newblock \bibinfo{title}{Diffuse reflectance patterns in cervical spectroscopy}.
\newblock {\it \bibinfo{journal}{Gynecol. Oncol.}\/},  {\it \bibinfo{volume}{99}\/}, \bibinfo{pages}{S116--20}.
\newblock \bibinfo{note}{Publisher: Elsevier BV}.
\bibitem[{Mehta et~al.(2023)Mehta, Thapa, Singh, Joshi, Sarangi, Mishra \& Srivastava}]{mehta2023multimodal}
\bibinfo{author}{Mehta, D.~S.}, \bibinfo{author}{Thapa, P.}, \bibinfo{author}{Singh, V.}, \bibinfo{author}{Joshi, H.}, \bibinfo{author}{Sarangi, D.~J.}, \bibinfo{author}{Mishra, D.}, \& \bibinfo{author}{Srivastava, A.} (\bibinfo{year}{2023}).
\newblock \bibinfo{title}{Multimodal and multispectral diagnostic devices for oral and breast cancer screening in low resource settings}.
\newblock {\it \bibinfo{journal}{Current Opinion in Biomedical Engineering}\/},  (p. \bibinfo{pages}{100485}).
\bibitem[{Meza~Ramirez et~al.(2021)Meza~Ramirez, Greenop, Ashton \& Rehman}]{meza2021applications}
\bibinfo{author}{Meza~Ramirez, C.~A.}, \bibinfo{author}{Greenop, M.}, \bibinfo{author}{Ashton, L.}, \& \bibinfo{author}{Rehman, I.~U.} (\bibinfo{year}{2021}).
\newblock \bibinfo{title}{Applications of machine learning in spectroscopy}.
\newblock {\it \bibinfo{journal}{Applied Spectroscopy Reviews}\/},  {\it \bibinfo{volume}{56}\/}, \bibinfo{pages}{733--763}.
\bibitem[{Minh et~al.(2022)Minh, Wang, Li \& Nguyen}]{minh2022explainable}
\bibinfo{author}{Minh, D.}, \bibinfo{author}{Wang, H.~X.}, \bibinfo{author}{Li, Y.~F.}, \& \bibinfo{author}{Nguyen, T.~N.} (\bibinfo{year}{2022}).
\newblock \bibinfo{title}{Explainable artificial intelligence: a comprehensive review}.
\newblock {\it \bibinfo{journal}{Artificial Intelligence Review}\/},  (pp. \bibinfo{pages}{1--66}).
\bibitem[{Murphy et~al.(2005)Murphy, Webster, Turlach, Quirk, Clay, Heenan \& Sampson}]{murphy_toward_2005}
\bibinfo{author}{Murphy, B.}, \bibinfo{author}{Webster, R.}, \bibinfo{author}{Turlach, B.}, \bibinfo{author}{Quirk, C.}, \bibinfo{author}{Clay, C.}, \bibinfo{author}{Heenan, P.}, \& \bibinfo{author}{Sampson, D.} (\bibinfo{year}{2005}).
\newblock \bibinfo{title}{Toward the discrimination of early melanoma from common and dysplastic nevus using fiber optic diffuse reflectance spectroscopy.}
\newblock {\it \bibinfo{journal}{Journal of biomedical optics}\/},  {\it \bibinfo{volume}{10}\/}, \bibinfo{pages}{064020}.
\bibitem[{Nachabé et~al.(2011)Nachabé, Evers, Hendriks, Lucassen, van~der Voort, Rutgers, Peeters, Van~der Hage, Oldenburg, Wesseling \& Ruers}]{nachabe_diagnosis_2011}
\bibinfo{author}{Nachabé, R.}, \bibinfo{author}{Evers, D.~J.}, \bibinfo{author}{Hendriks, B. H.~W.}, \bibinfo{author}{Lucassen, G.~W.}, \bibinfo{author}{van~der Voort, M.}, \bibinfo{author}{Rutgers, E.~J.}, \bibinfo{author}{Peeters, M.-J.~V.}, \bibinfo{author}{Van~der Hage, J.~A.}, \bibinfo{author}{Oldenburg, H.~S.}, \bibinfo{author}{Wesseling, J.}, \& \bibinfo{author}{Ruers, T. J.~M.} (\bibinfo{year}{2011}).
\newblock \bibinfo{title}{Diagnosis of breast cancer using diffuse optical spectroscopy from 500 to1600 nm: comparison of classification methods}.
\newblock {\it \bibinfo{journal}{J. Biomed. Opt.}\/},  {\it \bibinfo{volume}{16}\/}, \bibinfo{pages}{087010}.
\newblock \bibinfo{note}{Publisher: SPIE-Intl Soc Optical Eng}.
\bibitem[{Nazarian et~al.(2022)Nazarian, Gkouzionis, Kawka, Jamroziak, Lloyd, Darzi, Patel, Elson \& Peters}]{nazarian_real-time_2022}
\bibinfo{author}{Nazarian, S.}, \bibinfo{author}{Gkouzionis, I.}, \bibinfo{author}{Kawka, M.}, \bibinfo{author}{Jamroziak, M.}, \bibinfo{author}{Lloyd, J.}, \bibinfo{author}{Darzi, A.}, \bibinfo{author}{Patel, N.}, \bibinfo{author}{Elson, D.}, \& \bibinfo{author}{Peters, C.} (\bibinfo{year}{2022}).
\newblock \bibinfo{title}{Real-time {Tracking} and {Classification} of {Tumor} and {Nontumor} {Tissue} in {Upper} {Gastrointestinal} {Cancers} {Using} {Diffuse} {Reflectance} {Spectroscopy} for {Resection} {Margin} {Assessment}}.
\newblock {\it \bibinfo{journal}{JAMA Surgery}\/},  {\it \bibinfo{volume}{157}\/}, \bibinfo{pages}{E223899}.
\bibitem[{Nazarian et~al.(2024)Nazarian, Gkouzionis, Murphy, Darzi, Patel, Peters \& Elson}]{nazarian_real-time_2024}
\bibinfo{author}{Nazarian, S.}, \bibinfo{author}{Gkouzionis, I.}, \bibinfo{author}{Murphy, J.}, \bibinfo{author}{Darzi, A.}, \bibinfo{author}{Patel, N.}, \bibinfo{author}{Peters, C.}, \& \bibinfo{author}{Elson, D.} (\bibinfo{year}{2024}).
\newblock \bibinfo{title}{Real-time classification of tumour and non-tumour tissue in colorectal cancer using diffuse reflectance spectroscopy and neural networks to aid margin assessment}.
\newblock {\it \bibinfo{journal}{International journal of surgery (London, England)}\/}, .
\bibitem[{Nguyen et~al.(2021)Nguyen, Zhang, Wang, De~La Garza Evia~Linan, Markey \& Tunnell}]{nguyen2021machine}
\bibinfo{author}{Nguyen, M.~H.}, \bibinfo{author}{Zhang, Y.}, \bibinfo{author}{Wang, F.}, \bibinfo{author}{De~La Garza Evia~Linan, J.}, \bibinfo{author}{Markey, M.~K.}, \& \bibinfo{author}{Tunnell, J.~W.} (\bibinfo{year}{2021}).
\newblock \bibinfo{title}{Machine learning to extract physiological parameters from multispectral diffuse reflectance spectroscopy}.
\newblock {\it \bibinfo{journal}{Journal of Biomedical Optics}\/},  {\it \bibinfo{volume}{26}\/}, \bibinfo{pages}{052912--052912}.
\bibitem[{Nogueira et~al.(2024)Nogueira, Maryam, Amissah, Killeen, O'Riordain \& Andersson-Engels}]{nogueira_intestinal_2024}
\bibinfo{author}{Nogueira, M.}, \bibinfo{author}{Maryam, S.}, \bibinfo{author}{Amissah, M.}, \bibinfo{author}{Killeen, S.}, \bibinfo{author}{O'Riordain, M.}, \& \bibinfo{author}{Andersson-Engels, S.} (\bibinfo{year}{2024}).
\newblock \bibinfo{title}{Intestinal anastomosis automation through diffuse reflectance spectroscopy: {Towards} real time guidance during robotic surgery}.
\newblock volume \bibinfo{volume}{12853}.
\bibitem[{Nogueira et~al.(2021{\natexlab{a}})Nogueira, Raju, Gunther, Maryam, Amissah, Lu, Killeen, O'Riordain \& Andersson-Engels}]{nogueira_tissue_2021}
\bibinfo{author}{Nogueira, M.}, \bibinfo{author}{Raju, M.}, \bibinfo{author}{Gunther, J.}, \bibinfo{author}{Maryam, S.}, \bibinfo{author}{Amissah, M.}, \bibinfo{author}{Lu, H.}, \bibinfo{author}{Killeen, S.}, \bibinfo{author}{O'Riordain, M.}, \& \bibinfo{author}{Andersson-Engels, S.} (\bibinfo{year}{2021}{\natexlab{a}}).
\newblock \bibinfo{title}{Tissue biomolecular and microstructure profiles in optical colorectal cancer delineation}.
\newblock {\it \bibinfo{journal}{Journal of Physics D: Applied Physics}\/},  {\it \bibinfo{volume}{54}\/}.
\bibitem[{Nogueira et~al.(2021{\natexlab{b}})Nogueira, Maryam, Amissah, Lu, Lynch, Killeen, O’Riordain \& Andersson-Engels}]{nogueira_evaluation_2021}
\bibinfo{author}{Nogueira, M.~S.}, \bibinfo{author}{Maryam, S.}, \bibinfo{author}{Amissah, M.}, \bibinfo{author}{Lu, H.}, \bibinfo{author}{Lynch, N.}, \bibinfo{author}{Killeen, S.}, \bibinfo{author}{O’Riordain, M.}, \& \bibinfo{author}{Andersson-Engels, S.} (\bibinfo{year}{2021}{\natexlab{b}}).
\newblock \bibinfo{title}{Evaluation of wavelength ranges and tissue depth probed by diffuse reflectance spectroscopy for colorectal cancer detection}.
\newblock {\it \bibinfo{journal}{Scientific Reports}\/},  {\it \bibinfo{volume}{11}\/}.
\newblock \bibinfo{note}{Publisher: Springer Science and Business Media LLC}.
\bibitem[{Nogueira et~al.(2021{\natexlab{c}})Nogueira, Maryam, Amissah, Lynch, Killeen, O'Riordain \& Andersson-Engels}]{nogueira_benefit_2021}
\bibinfo{author}{Nogueira, M.~S.}, \bibinfo{author}{Maryam, S.}, \bibinfo{author}{Amissah, M.}, \bibinfo{author}{Lynch, N.}, \bibinfo{author}{Killeen, S.}, \bibinfo{author}{O'Riordain, M.}, \& \bibinfo{author}{Andersson-Engels, S.} (\bibinfo{year}{2021}{\natexlab{c}}).
\newblock \bibinfo{title}{Benefit of extending near-infrared wavelength range of diffuse reflectancespectroscopy for colorectal cancer detection using machine learning}.
\newblock \bibinfo{publisher}{SPIE}.
\bibitem[{Page et~al.(2021)Page, McKenzie, Bossuyt, Boutron, Hoffmann, Mulrow, Shamseer, Tetzlaff, Akl, Brennan et~al.}]{page2021prisma}
\bibinfo{author}{Page, M.~J.}, \bibinfo{author}{McKenzie, J.~E.}, \bibinfo{author}{Bossuyt, P.~M.}, \bibinfo{author}{Boutron, I.}, \bibinfo{author}{Hoffmann, T.~C.}, \bibinfo{author}{Mulrow, C.~D.}, \bibinfo{author}{Shamseer, L.}, \bibinfo{author}{Tetzlaff, J.~M.}, \bibinfo{author}{Akl, E.~A.}, \bibinfo{author}{Brennan, S.~E.} et~al. (\bibinfo{year}{2021}).
\newblock \bibinfo{title}{The prisma 2020 statement: an updated guideline for reporting systematic reviews}.
\newblock {\it \bibinfo{journal}{bmj}\/},  {\it \bibinfo{volume}{372}\/}.
\bibitem[{Procházkam et~al.(2023)Procházkam, Martynek \& Charvát}]{a_prochazka_absorption_2023}
\bibinfo{author}{Procházkam, A.}, \bibinfo{author}{Martynek, D.}, \& \bibinfo{author}{Charvát, J.} (\bibinfo{year}{2023}).
\newblock \bibinfo{title}{Absorption {Spectroscopy} in {Dental} {Tissue} {Analysis}}.
\newblock {\it \bibinfo{journal}{IEEE Access}\/},  {\it \bibinfo{volume}{11}\/}, \bibinfo{pages}{17569--17575}.
\bibitem[{Pudjihartono et~al.(2022)Pudjihartono, Fadason, Kempa-Liehr \& O'Sullivan}]{pudjihartono2022review}
\bibinfo{author}{Pudjihartono, N.}, \bibinfo{author}{Fadason, T.}, \bibinfo{author}{Kempa-Liehr, A.~W.}, \& \bibinfo{author}{O'Sullivan, J.~M.} (\bibinfo{year}{2022}).
\newblock \bibinfo{title}{A review of feature selection methods for machine learning-based disease risk prediction}.
\newblock {\it \bibinfo{journal}{Frontiers in Bioinformatics}\/},  {\it \bibinfo{volume}{2}\/}, \bibinfo{pages}{927312}.
\bibitem[{Rasheed et~al.(2022)Rasheed, Qayyum, Ghaly, Al-Fuqaha, Razi \& Qadir}]{rasheed2022explainable}
\bibinfo{author}{Rasheed, K.}, \bibinfo{author}{Qayyum, A.}, \bibinfo{author}{Ghaly, M.}, \bibinfo{author}{Al-Fuqaha, A.}, \bibinfo{author}{Razi, A.}, \& \bibinfo{author}{Qadir, J.} (\bibinfo{year}{2022}).
\newblock \bibinfo{title}{Explainable, trustworthy, and ethical machine learning for healthcare: A survey}.
\newblock {\it \bibinfo{journal}{Computers in Biology and Medicine}\/},  (p. \bibinfo{pages}{106043}).
\bibitem[{Reistad \& Sturesson(2022)}]{reistad_distinguishing_2022}
\bibinfo{author}{Reistad, N.}, \& \bibinfo{author}{Sturesson, C.} (\bibinfo{year}{2022}).
\newblock \bibinfo{title}{Distinguishing tumor from healthy tissue in human liver ex vivo using machine learning and multivariate analysis of diffuse reflectance spectra}.
\newblock {\it \bibinfo{journal}{Journal of biophotonics}\/},  (p. \bibinfo{pages}{e202200140}).
\bibitem[{Rossel \& Behrens(2010)}]{rossel2010using}
\bibinfo{author}{Rossel, R.~V.}, \& \bibinfo{author}{Behrens, T.} (\bibinfo{year}{2010}).
\newblock \bibinfo{title}{Using data mining to model and interpret soil diffuse reflectance spectra}.
\newblock {\it \bibinfo{journal}{Geoderma}\/},  {\it \bibinfo{volume}{158}\/}, \bibinfo{pages}{46--54}.
\bibitem[{Saito~Nogueira et~al.(2023)Saito~Nogueira, Maryam, Amissah, Killeen, O'Riordain \& Andersson-Engels}]{saito_nogueira_diffuse_2023}
\bibinfo{author}{Saito~Nogueira, M.}, \bibinfo{author}{Maryam, S.}, \bibinfo{author}{Amissah, M.}, \bibinfo{author}{Killeen, S.}, \bibinfo{author}{O'Riordain, M.}, \& \bibinfo{author}{Andersson-Engels, S.} (\bibinfo{year}{2023}).
\newblock \bibinfo{title}{Diffuse reflectance spectroscopy for colorectal cancer surgical guidance: towards real-time tissue characterization and new biomarkers}.
\newblock {\it \bibinfo{journal}{The Analyst}\/},  {\it \bibinfo{volume}{149}\/}, \bibinfo{pages}{88--99}.
\bibitem[{Saito~Nogueira et~al.(2022{\natexlab{a}})Saito~Nogueira, Maryam, Amissah, Lu, Lynch, Killeen, O'Riordain \& Andersson-Engels}]{saito_nogueira_improving_2022}
\bibinfo{author}{Saito~Nogueira, M.}, \bibinfo{author}{Maryam, S.}, \bibinfo{author}{Amissah, M.}, \bibinfo{author}{Lu, H.}, \bibinfo{author}{Lynch, N.}, \bibinfo{author}{Killeen, S.}, \bibinfo{author}{O'Riordain, M.}, \& \bibinfo{author}{Andersson-Engels, S.} (\bibinfo{year}{2022}{\natexlab{a}}).
\newblock \bibinfo{title}{Improving colorectal cancer detection by extending the near-infrared wavelength range and tissue probed depth of diffuse reflectance spectroscopy: {A} support vector machine approach}.
\newblock volume \bibinfo{volume}{11954}.
\bibitem[{Saito~Nogueira et~al.(2022{\natexlab{b}})Saito~Nogueira, Maryam, Amissah, McGuire, Spillane, Killeen, Andersson-Engels \& O'Riordain}]{saito_nogueira_insights_2022}
\bibinfo{author}{Saito~Nogueira, M.}, \bibinfo{author}{Maryam, S.}, \bibinfo{author}{Amissah, M.}, \bibinfo{author}{McGuire, A.}, \bibinfo{author}{Spillane, C.}, \bibinfo{author}{Killeen, S.}, \bibinfo{author}{Andersson-Engels, S.}, \& \bibinfo{author}{O'Riordain, M.} (\bibinfo{year}{2022}{\natexlab{b}}).
\newblock \bibinfo{title}{Insights into biochemical sources and diffuse reflectance spectralfeatures for colorectal cancer detection and localization}.
\newblock {\it \bibinfo{journal}{Cancers (Basel)}\/},  {\it \bibinfo{volume}{14}\/}, \bibinfo{pages}{5715}.
\newblock \bibinfo{note}{Publisher: MDPI AG}.
\bibitem[{Schols et~al.(2014)Schols, ter Laan, Stassen, Bouvy, Amelink, Wieringa \& Alic}]{schols_differentiation_2014}
\bibinfo{author}{Schols, R.~M.}, \bibinfo{author}{ter Laan, M.}, \bibinfo{author}{Stassen, L. P.~S.}, \bibinfo{author}{Bouvy, N.~D.}, \bibinfo{author}{Amelink, A.}, \bibinfo{author}{Wieringa, F.~P.}, \& \bibinfo{author}{Alic, L.} (\bibinfo{year}{2014}).
\newblock \bibinfo{title}{Differentiation between nerve and adipose tissue using wide-band(350-1,830 nm) in vivo diffuse reflectance spectroscopy}.
\newblock {\it \bibinfo{journal}{Lasers Surg. Med.}\/},  {\it \bibinfo{volume}{46}\/}, \bibinfo{pages}{538--545}.
\newblock \bibinfo{note}{Publisher: Wiley}.
\bibitem[{Selwitz et~al.(2007)Selwitz, Ismail \& Pitts}]{selwitz2007dental}
\bibinfo{author}{Selwitz, R.~H.}, \bibinfo{author}{Ismail, A.~I.}, \& \bibinfo{author}{Pitts, N.~B.} (\bibinfo{year}{2007}).
\newblock \bibinfo{title}{Dental caries}.
\newblock {\it \bibinfo{journal}{The Lancet}\/},  {\it \bibinfo{volume}{369}\/}, \bibinfo{pages}{51--59}.
\bibitem[{Shaik et~al.(2023)Shaik, Tao, Higgins, Li, Gururajan, Zhou \& Acharya}]{shaik2023remote}
\bibinfo{author}{Shaik, T.}, \bibinfo{author}{Tao, X.}, \bibinfo{author}{Higgins, N.}, \bibinfo{author}{Li, L.}, \bibinfo{author}{Gururajan, R.}, \bibinfo{author}{Zhou, X.}, \& \bibinfo{author}{Acharya, U.~R.} (\bibinfo{year}{2023}).
\newblock \bibinfo{title}{Remote patient monitoring using artificial intelligence: Current state, applications, and challenges}.
\newblock {\it \bibinfo{journal}{Wiley Interdisciplinary Reviews: Data Mining and Knowledge Discovery}\/},  {\it \bibinfo{volume}{13}\/}, \bibinfo{pages}{e1485}.
\bibitem[{Shaikh et~al.(2017)Shaikh, Prabitha, Dora, Chopra, Maheshwari, Deodhar, Rekhi, Sukumar, Krishna \& Subhash}]{shaikh_comparative_2017}
\bibinfo{author}{Shaikh, R.}, \bibinfo{author}{Prabitha, V.~G.}, \bibinfo{author}{Dora, T.~K.}, \bibinfo{author}{Chopra, S.}, \bibinfo{author}{Maheshwari, A.}, \bibinfo{author}{Deodhar, K.}, \bibinfo{author}{Rekhi, B.}, \bibinfo{author}{Sukumar, N.}, \bibinfo{author}{Krishna, C.~M.}, \& \bibinfo{author}{Subhash, N.} (\bibinfo{year}{2017}).
\newblock \bibinfo{title}{A comparative evaluation of diffuse reflectance and {Raman} spectroscopy inthe detection of cervical cancer}.
\newblock {\it \bibinfo{journal}{J. Biophotonics}\/},  {\it \bibinfo{volume}{10}\/}, \bibinfo{pages}{242--252}.
\newblock \bibinfo{note}{Publisher: Wiley}.
\bibitem[{Sidey-Gibbons \& Sidey-Gibbons(2019)}]{sidey2019machine}
\bibinfo{author}{Sidey-Gibbons, J.~A.}, \& \bibinfo{author}{Sidey-Gibbons, C.~J.} (\bibinfo{year}{2019}).
\newblock \bibinfo{title}{Machine learning in medicine: a practical introduction}.
\newblock {\it \bibinfo{journal}{BMC medical research methodology}\/},  {\it \bibinfo{volume}{19}\/}, \bibinfo{pages}{1--18}.
\bibitem[{Siegel et~al.(2024)Siegel, Giaquinto \& Jemal}]{siegel2024cancer}
\bibinfo{author}{Siegel, R.~L.}, \bibinfo{author}{Giaquinto, A.~N.}, \& \bibinfo{author}{Jemal, A.} (\bibinfo{year}{2024}).
\newblock \bibinfo{title}{Cancer statistics, 2024.}
\newblock {\it \bibinfo{journal}{CA: a cancer journal for clinicians}\/},  {\it \bibinfo{volume}{74}\/}.
\bibitem[{Siegel et~al.(2019)Siegel, Miller \& Jemal}]{siegel2019cancer}
\bibinfo{author}{Siegel, R.~L.}, \bibinfo{author}{Miller, K.~D.}, \& \bibinfo{author}{Jemal, A.} (\bibinfo{year}{2019}).
\newblock \bibinfo{title}{Cancer statistics, 2019}.
\newblock {\it \bibinfo{journal}{CA: a cancer journal for clinicians}\/},  {\it \bibinfo{volume}{69}\/}, \bibinfo{pages}{7--34}.
\bibitem[{Skala et~al.(2007)Skala, Palmer, Vrotsos, Gendron-Fitzpatrick \& Ramanujam}]{skala_comparison_2007}
\bibinfo{author}{Skala, M.~C.}, \bibinfo{author}{Palmer, G.~M.}, \bibinfo{author}{Vrotsos, K.~M.}, \bibinfo{author}{Gendron-Fitzpatrick, A.}, \& \bibinfo{author}{Ramanujam, N.} (\bibinfo{year}{2007}).
\newblock \bibinfo{title}{Comparison of a physical model and principal component analysis for thediagnosis of epithelial neoplasias in vivo using diffuse reflectancespectroscopy}.
\newblock {\it \bibinfo{journal}{Opt. Express}\/},  {\it \bibinfo{volume}{15}\/}, \bibinfo{pages}{7863--7875}.
\newblock \bibinfo{note}{Publisher: The Optical Society}.
\bibitem[{Skyrman et~al.(2022)Skyrman, Burström, Lai, Manni, Hendriks, Frostell, Edström, Persson \& Elmi-Terander}]{skyrman_diffuse_2022}
\bibinfo{author}{Skyrman, S.}, \bibinfo{author}{Burström, G.}, \bibinfo{author}{Lai, M.}, \bibinfo{author}{Manni, F.}, \bibinfo{author}{Hendriks, B.}, \bibinfo{author}{Frostell, A.}, \bibinfo{author}{Edström, E.}, \bibinfo{author}{Persson, O.}, \& \bibinfo{author}{Elmi-Terander, A.} (\bibinfo{year}{2022}).
\newblock \bibinfo{title}{Diffuse reflectance spectroscopy sensor to differentiate between glial tumor and healthy brain tissue: {A} proof-of-concept study}.
\newblock {\it \bibinfo{journal}{Biomedical Optics Express}\/},  {\it \bibinfo{volume}{13}\/}, \bibinfo{pages}{6470--6483}.
\bibitem[{Soares et~al.(2013)Soares, Barman, Dingari, Volynskaya, Liu, Klein, Plecha, Dasari \& Fitzmaurice}]{soares_diagnostic_2013}
\bibinfo{author}{Soares, J.~S.}, \bibinfo{author}{Barman, I.}, \bibinfo{author}{Dingari, N.~C.}, \bibinfo{author}{Volynskaya, Z.}, \bibinfo{author}{Liu, W.}, \bibinfo{author}{Klein, N.}, \bibinfo{author}{Plecha, D.}, \bibinfo{author}{Dasari, R.~R.}, \& \bibinfo{author}{Fitzmaurice, M.} (\bibinfo{year}{2013}).
\newblock \bibinfo{title}{Diagnostic power of diffuse reflectance spectroscopy for targeteddetection of breast lesions with microcalcifications}.
\newblock {\it \bibinfo{journal}{Proc. Natl. Acad. Sci. U. S. A.}\/},  {\it \bibinfo{volume}{110}\/}, \bibinfo{pages}{471--476}.
\newblock \bibinfo{note}{Publisher: Proceedings of the National Academy of Sciences}.
\bibitem[{Song et~al.(2022)Song, Ren \& He}]{song2022privacy}
\bibinfo{author}{Song, Z.}, \bibinfo{author}{Ren, Y.}, \& \bibinfo{author}{He, G.} (\bibinfo{year}{2022}).
\newblock \bibinfo{title}{Privacy-preserving knn classification algorithm for smart grid}.
\newblock {\it \bibinfo{journal}{Security and Communication Networks}\/},  {\it \bibinfo{volume}{2022}\/}, \bibinfo{pages}{7333175}.
\bibitem[{Spliethoff et~al.(2013)Spliethoff, Evers, Klomp, van Sandick, Wouters, Nachabe, Lucassen, Hendriks, Wesseling \& Ruers}]{spliethoff_improved_2013}
\bibinfo{author}{Spliethoff, J.~W.}, \bibinfo{author}{Evers, D.~J.}, \bibinfo{author}{Klomp, H.~M.}, \bibinfo{author}{van Sandick, J.~W.}, \bibinfo{author}{Wouters, M.~W.}, \bibinfo{author}{Nachabe, R.}, \bibinfo{author}{Lucassen, G.~W.}, \bibinfo{author}{Hendriks, B. H.~W.}, \bibinfo{author}{Wesseling, J.}, \& \bibinfo{author}{Ruers, T. J.~M.} (\bibinfo{year}{2013}).
\newblock \bibinfo{title}{Improved identification of peripheral lung tumors by using diffusereflectance and fluorescence spectroscopy}.
\newblock {\it \bibinfo{journal}{Lung Cancer}\/},  {\it \bibinfo{volume}{80}\/}, \bibinfo{pages}{165--171}.
\newblock \bibinfo{note}{Publisher: Elsevier BV}.
\bibitem[{Starley et~al.(2010)Starley, Calcagno \& Harrison}]{starley2010nonalcoholic}
\bibinfo{author}{Starley, B.~Q.}, \bibinfo{author}{Calcagno, C.~J.}, \& \bibinfo{author}{Harrison, S.~A.} (\bibinfo{year}{2010}).
\newblock \bibinfo{title}{Nonalcoholic fatty liver disease and hepatocellular carcinoma: a weighty connection}.
\newblock {\it \bibinfo{journal}{Hepatology}\/},  {\it \bibinfo{volume}{51}\/}, \bibinfo{pages}{1820--1832}.
\bibitem[{Stelzle et~al.(2012)Stelzle, Adler, Zam, Tangermann-Gerk, Knipfer, Douplik, Schmidt \& Nkenke}]{stelzle_vivo_2012}
\bibinfo{author}{Stelzle, F.}, \bibinfo{author}{Adler, W.}, \bibinfo{author}{Zam, A.}, \bibinfo{author}{Tangermann-Gerk, K.}, \bibinfo{author}{Knipfer, C.}, \bibinfo{author}{Douplik, A.}, \bibinfo{author}{Schmidt, M.}, \& \bibinfo{author}{Nkenke, E.} (\bibinfo{year}{2012}).
\newblock \bibinfo{title}{In vivo optical tissue differentiation by diffuse reflectancespectroscopy: preliminary results for tissue-specific laser surgery}.
\newblock {\it \bibinfo{journal}{Surg. Innov.}\/},  {\it \bibinfo{volume}{19}\/}, \bibinfo{pages}{385--393}.
\newblock \bibinfo{note}{Publisher: SAGE Publications}.
\bibitem[{Stelzle et~al.(2010{\natexlab{a}})Stelzle, Tangermann-Gerk, Adler, Zam, Schmidt, Douplik \& Nkenke}]{stelzle_diffuse_2010}
\bibinfo{author}{Stelzle, F.}, \bibinfo{author}{Tangermann-Gerk, K.}, \bibinfo{author}{Adler, W.}, \bibinfo{author}{Zam, A.}, \bibinfo{author}{Schmidt, M.}, \bibinfo{author}{Douplik, A.}, \& \bibinfo{author}{Nkenke, E.} (\bibinfo{year}{2010}{\natexlab{a}}).
\newblock \bibinfo{title}{Diffuse reflectance spectroscopy for optical soft tissue differentiationas remote feedback control for tissue-specific laser surgery}.
\newblock {\it \bibinfo{journal}{Lasers Surg. Med.}\/},  {\it \bibinfo{volume}{42}\/}, \bibinfo{pages}{319--325}.
\newblock \bibinfo{note}{Publisher: Wiley}.
\bibitem[{Stelzle et~al.(2010{\natexlab{b}})Stelzle, Zam, Adler, Douplik, Tangermann-Gerk, Nkenke, Neukam \& Schmidt}]{stelzle_diffuse_2010-1}
\bibinfo{author}{Stelzle, F.}, \bibinfo{author}{Zam, A.}, \bibinfo{author}{Adler, W.}, \bibinfo{author}{Douplik, A.}, \bibinfo{author}{Tangermann-Gerk, K.}, \bibinfo{author}{Nkenke, E.}, \bibinfo{author}{Neukam, F.~W.}, \& \bibinfo{author}{Schmidt, M.} (\bibinfo{year}{2010}{\natexlab{b}}).
\newblock \bibinfo{title}{Diffuse reflectance spectroscopy for optical nerve identification}.
\newblock {\it \bibinfo{journal}{Phys. Procedia}\/},  {\it \bibinfo{volume}{5}\/}, \bibinfo{pages}{647--654}.
\newblock \bibinfo{note}{Publisher: Elsevier BV}.
\bibitem[{Stelzle et~al.(2011)Stelzle, Zam, Adler, Tangermann-Gerk, Douplik, Nkenke \& Schmidt}]{stelzle_optical_2011}
\bibinfo{author}{Stelzle, F.}, \bibinfo{author}{Zam, A.}, \bibinfo{author}{Adler, W.}, \bibinfo{author}{Tangermann-Gerk, K.}, \bibinfo{author}{Douplik, A.}, \bibinfo{author}{Nkenke, E.}, \& \bibinfo{author}{Schmidt, M.} (\bibinfo{year}{2011}).
\newblock \bibinfo{title}{Optical nerve detection by diffuse reflectance spectroscopy for feedbackcontrolled oral and maxillofacial laser surgery}.
\newblock {\it \bibinfo{journal}{J. Transl. Med.}\/},  {\it \bibinfo{volume}{9}\/}, \bibinfo{pages}{20}.
\newblock \bibinfo{note}{Publisher: Springer Nature}.
\bibitem[{Sun \& Patil(2022)}]{sun_characterization_2022}
\bibinfo{author}{Sun, Y.}, \& \bibinfo{author}{Patil, C.~A.} (\bibinfo{year}{2022}).
\newblock \bibinfo{title}{Characterization of extended-wavelength diffuse reflectance spectroscopyfor identification of neurovascular bundles using {Monte} {Carlo} models,optical phantoms, and ex vivo animal tissues}.
\newblock \bibinfo{publisher}{SPIE}.
\bibitem[{Tanis et~al.(2016)Tanis, Evers, Spliethoff, Pully, Kuhlmann, van Coevorden, Hendriks, Sanders, Prevoo \& Ruers}]{tanis_vivo_2016}
\bibinfo{author}{Tanis, E.}, \bibinfo{author}{Evers, D.~J.}, \bibinfo{author}{Spliethoff, J.~W.}, \bibinfo{author}{Pully, V.~V.}, \bibinfo{author}{Kuhlmann, K.}, \bibinfo{author}{van Coevorden, F.}, \bibinfo{author}{Hendriks, B. H.~W.}, \bibinfo{author}{Sanders, J.}, \bibinfo{author}{Prevoo, W.}, \& \bibinfo{author}{Ruers, T. J.~M.} (\bibinfo{year}{2016}).
\newblock \bibinfo{title}{In vivo tumor identification of colorectal liver metastases with diffusereflectance and fluorescence spectroscopy}.
\newblock {\it \bibinfo{journal}{Lasers Surg. Med.}\/},  {\it \bibinfo{volume}{48}\/}, \bibinfo{pages}{820--827}.
\newblock \bibinfo{note}{Publisher: Wiley}.
\bibitem[{Union()}]{EU_regulation_AI_act}
\bibinfo{editor}{Union, E.} (Ed.) ().
\newblock {\it \bibinfo{title}{Regulation (EU) 2024/1689 of the European parliament and of the Council of 13 June 2024 laying down harmonised rules on artificial intelligence and amending Regulations (EC) No 300/2008, (EU) No 167/2013, (EU) No 168/2013, (EU) 2018/858, (EU) 2018/1139 and (EU) 2019/2144 and Directives 2014/90/EU, shorttitle (EU) 2016/797 and (EU) 2020/1828 (Artificial Intelligence Act]}\/}.
\newblock Number \bibinfo{number}{2024/1689} in \bibinfo{series}{L}.
\newblock \bibinfo{publisher}{EU}.
\bibitem[{Valkenborg et~al.(2023)Valkenborg, Rousseau, Geubbelmans \& Burzykowski}]{valkenborg2023support}
\bibinfo{author}{Valkenborg, D.}, \bibinfo{author}{Rousseau, A.-J.}, \bibinfo{author}{Geubbelmans, M.}, \& \bibinfo{author}{Burzykowski, T.} (\bibinfo{year}{2023}).
\newblock \bibinfo{title}{Support vector machines}.
\newblock {\it \bibinfo{journal}{American Journal of Orthodontics and Dentofacial Orthopedics}\/},  {\it \bibinfo{volume}{164}\/}, \bibinfo{pages}{754--757}.
\bibitem[{Vamathevan et~al.(2019)Vamathevan, Clark, Czodrowski, Dunham, Ferran, Lee, Li, Madabhushi, Shah, Spitzer et~al.}]{vamathevan2019applications}
\bibinfo{author}{Vamathevan, J.}, \bibinfo{author}{Clark, D.}, \bibinfo{author}{Czodrowski, P.}, \bibinfo{author}{Dunham, I.}, \bibinfo{author}{Ferran, E.}, \bibinfo{author}{Lee, G.}, \bibinfo{author}{Li, B.}, \bibinfo{author}{Madabhushi, A.}, \bibinfo{author}{Shah, P.}, \bibinfo{author}{Spitzer, M.} et~al. (\bibinfo{year}{2019}).
\newblock \bibinfo{title}{Applications of machine learning in drug discovery and development}.
\newblock {\it \bibinfo{journal}{Nature reviews Drug discovery}\/},  {\it \bibinfo{volume}{18}\/}, \bibinfo{pages}{463--477}.
\bibitem[{de~Veld et~al.(2005)de~Veld, Skurichina, Witjes, Duin, Sterenborg \& Roodenburg}]{de_veld_autofluorescence_2005}
\bibinfo{author}{de~Veld, D. C.~G.}, \bibinfo{author}{Skurichina, M.}, \bibinfo{author}{Witjes, M. J.~H.}, \bibinfo{author}{Duin, R. P.~W.}, \bibinfo{author}{Sterenborg, H. J. C.~M.}, \& \bibinfo{author}{Roodenburg, J. L.~N.} (\bibinfo{year}{2005}).
\newblock \bibinfo{title}{Autofluorescence and diffuse reflectance spectroscopy for oral oncology}.
\newblock {\it \bibinfo{journal}{Lasers Surg. Med.}\/},  {\it \bibinfo{volume}{36}\/}, \bibinfo{pages}{356--364}.
\newblock \bibinfo{note}{Publisher: Wiley}.
\bibitem[{Van~der Velden et~al.(2022)Van~der Velden, Kuijf, Gilhuijs \& Viergever}]{van2022explainable}
\bibinfo{author}{Van~der Velden, B.~H.}, \bibinfo{author}{Kuijf, H.~J.}, \bibinfo{author}{Gilhuijs, K.~G.}, \& \bibinfo{author}{Viergever, M.~A.} (\bibinfo{year}{2022}).
\newblock \bibinfo{title}{Explainable artificial intelligence (xai) in deep learning-based medical image analysis}.
\newblock {\it \bibinfo{journal}{Medical Image Analysis}\/},  {\it \bibinfo{volume}{79}\/}, \bibinfo{pages}{102470}.
\bibitem[{Veluponnar et~al.(2024)Veluponnar, de~Boer, Dashtbozorg, Jong, Geldof, Guimaraes, Sterenborg, Vrancken-Peeters, van Duijnhoven \& Ruers}]{veluponnar_margin_2024}
\bibinfo{author}{Veluponnar, D.}, \bibinfo{author}{de~Boer, L.~L.}, \bibinfo{author}{Dashtbozorg, B.}, \bibinfo{author}{Jong, L.-J.~S.}, \bibinfo{author}{Geldof, F.}, \bibinfo{author}{Guimaraes, M. D.~S.}, \bibinfo{author}{Sterenborg, H. J. C.~M.}, \bibinfo{author}{Vrancken-Peeters, M.-J. T. F.~D.}, \bibinfo{author}{van Duijnhoven, F.}, \& \bibinfo{author}{Ruers, T.} (\bibinfo{year}{2024}).
\newblock \bibinfo{title}{Margin assessment during breast conserving surgery using diffusereflectance spectroscopy}.
\newblock {\it \bibinfo{journal}{J. Biomed. Opt.}\/},  {\it \bibinfo{volume}{29}\/}.
\newblock \bibinfo{note}{Publisher: SPIE-Intl Soc Optical Eng}.
\bibitem[{Veluponnar et~al.(2023)Veluponnar, Dashtbozorg, Jong, Geldof, Da~Silva~Guimaraes, Vrancken~Peeters, Van~Duijnhoven, Sterenborg, Ruers \& De~Boer}]{veluponnar_diffuse_2023}
\bibinfo{author}{Veluponnar, D.}, \bibinfo{author}{Dashtbozorg, B.}, \bibinfo{author}{Jong, L.-J.}, \bibinfo{author}{Geldof, F.}, \bibinfo{author}{Da~Silva~Guimaraes, M.}, \bibinfo{author}{Vrancken~Peeters, M.-J.}, \bibinfo{author}{Van~Duijnhoven, F.}, \bibinfo{author}{Sterenborg, H.}, \bibinfo{author}{Ruers, T.}, \& \bibinfo{author}{De~Boer, L.} (\bibinfo{year}{2023}).
\newblock \bibinfo{title}{Diffuse reflectance spectroscopy for accurate margin assessment in breast-conserving surgeries: importance of an optimal number of fibers}.
\newblock {\it \bibinfo{journal}{Biomedical Optics Express}\/},  {\it \bibinfo{volume}{14}\/}, \bibinfo{pages}{4017--4036}.
\bibitem[{Vishwanath et~al.(2010)Vishwanath, Chang, Klein, Deng, Chang, Phelps \& Ramanujam}]{vishwanath2010portable}
\bibinfo{author}{Vishwanath, K.}, \bibinfo{author}{Chang, K.}, \bibinfo{author}{Klein, D.}, \bibinfo{author}{Deng, Y.~F.}, \bibinfo{author}{Chang, V.}, \bibinfo{author}{Phelps, J.~E.}, \& \bibinfo{author}{Ramanujam, N.} (\bibinfo{year}{2010}).
\newblock \bibinfo{title}{Portable, fiber-based, diffuse reflection spectroscopy (drs) systems for estimating tissue optical properties}.
\newblock {\it \bibinfo{journal}{Applied spectroscopy}\/},  {\it \bibinfo{volume}{65}\/}, \bibinfo{pages}{206--215}.
\bibitem[{Werahera et~al.(2016)Werahera, Jasion, David~Crawford, Lucia, van Bokhoven, Sullivan, Kim, Maroni, David~Port, Daily \& Rosa}]{werahera_diffuse_2016}
\bibinfo{author}{Werahera, P.~N.}, \bibinfo{author}{Jasion, E.~A.}, \bibinfo{author}{David~Crawford, E.}, \bibinfo{author}{Lucia, M.~S.}, \bibinfo{author}{van Bokhoven, A.}, \bibinfo{author}{Sullivan, H.~T.}, \bibinfo{author}{Kim, F.~J.}, \bibinfo{author}{Maroni, P.~D.}, \bibinfo{author}{David~Port, J.}, \bibinfo{author}{Daily, J.~W.}, \& \bibinfo{author}{Rosa, F. G.~L.} (\bibinfo{year}{2016}).
\newblock \bibinfo{title}{Diffuse reflectance spectroscopy can differentiate high grade and lowgrade prostatic carcinoma}.
\newblock \bibinfo{publisher}{IEEE}.
\bibitem[{Yang et~al.(2024)Yang, Wang, Liu, Chang, Chang \& Lee}]{yang2024cervical}
\bibinfo{author}{Yang, S.-T.}, \bibinfo{author}{Wang, P.-H.}, \bibinfo{author}{Liu, H.-H.}, \bibinfo{author}{Chang, C.-W.}, \bibinfo{author}{Chang, W.-H.}, \& \bibinfo{author}{Lee, W.-L.} (\bibinfo{year}{2024}).
\newblock \bibinfo{title}{Cervical cancer: Part ii the landscape of treatment for persistent, recurrent and metastatic diseases (i)}.
\newblock {\it \bibinfo{journal}{Taiwanese Journal of Obstetrics and Gynecology}\/},  {\it \bibinfo{volume}{63}\/}, \bibinfo{pages}{637--650}.
\bibitem[{{Yeong E} et~al.(2005){Yeong E}, {Hsiao T}, {Chiang HK} \& {Lin C}}]{yeong_e_prediction_2005}
\bibinfo{author}{{Yeong E}}, \bibinfo{author}{{Hsiao T}}, \bibinfo{author}{{Chiang HK}}, \& \bibinfo{author}{{Lin C}} (\bibinfo{year}{2005}).
\newblock \bibinfo{title}{Prediction of burn healing time using artificial neural networks and reflectance spectrometer.}
\newblock {\it \bibinfo{journal}{Burns (03054179)}\/},  {\it \bibinfo{volume}{31}\/}, \bibinfo{pages}{415--420}.
\newblock \bibinfo{note}{Place: Philadelphia, Pennsylvania Publisher: Elsevier B.V.}
\bibitem[{Zam et~al.(2009)Zam, Stelzle, Nkenke, Tangermann-Gerk, Schmidt, Adler \& Douplik}]{zam_soft_2009}
\bibinfo{author}{Zam, A.}, \bibinfo{author}{Stelzle, F.}, \bibinfo{author}{Nkenke, E.}, \bibinfo{author}{Tangermann-Gerk, K.}, \bibinfo{author}{Schmidt, M.}, \bibinfo{author}{Adler, W.}, \& \bibinfo{author}{Douplik, A.} (\bibinfo{year}{2009}).
\newblock \bibinfo{title}{Soft tissue differentiation by diffuse reflectance spectroscopy}.
\newblock \bibinfo{publisher}{SPIE}.
\bibitem[{Zam et~al.(2010)Zam, Stelzle, Tangermann-Gerk, Adler, Nkenke, Schmidt \& Douplik}]{zam_tissue_2010}
\bibinfo{author}{Zam, A.}, \bibinfo{author}{Stelzle, F.}, \bibinfo{author}{Tangermann-Gerk, K.}, \bibinfo{author}{Adler, W.}, \bibinfo{author}{Nkenke, E.}, \bibinfo{author}{Schmidt, M.}, \& \bibinfo{author}{Douplik, A.} (\bibinfo{year}{2010}).
\newblock \bibinfo{title}{Tissue differentiation by diffuse reflectance spectroscopy for automatedoral and maxillofacial laser surgery: ex vivo pilot study}.
\newblock \bibinfo{publisher}{SPIE}.
\bibitem[{Zhang et~al.(2022)Zhang, Zhang, Zhao, Chen, Ke, Xu \& He}]{zhang2022brief}
\bibinfo{author}{Zhang, D.}, \bibinfo{author}{Zhang, H.}, \bibinfo{author}{Zhao, Y.}, \bibinfo{author}{Chen, Y.}, \bibinfo{author}{Ke, C.}, \bibinfo{author}{Xu, T.}, \& \bibinfo{author}{He, Y.} (\bibinfo{year}{2022}).
\newblock \bibinfo{title}{A brief review of new data analysis methods of laser-induced breakdown spectroscopy: machine learning}.
\newblock {\it \bibinfo{journal}{Applied Spectroscopy Reviews}\/},  {\it \bibinfo{volume}{57}\/}, \bibinfo{pages}{89--111}.
\bibitem[{Zhu et~al.(2016)Zhu, Chen, Chui, Tan \& Liu}]{zhu_early_2016}
\bibinfo{author}{Zhu, C.}, \bibinfo{author}{Chen, S.}, \bibinfo{author}{Chui, C. H.-K.}, \bibinfo{author}{Tan, B.-K.}, \& \bibinfo{author}{Liu, Q.} (\bibinfo{year}{2016}).
\newblock \bibinfo{title}{Early detection and differentiation of venous and arterial occlusion inskin flaps using visible diffuse reflectance spectroscopy andautofluorescence spectroscopy}.
\newblock {\it \bibinfo{journal}{Biomed. Opt. Express}\/},  {\it \bibinfo{volume}{7}\/}, \bibinfo{pages}{570--580}.
\newblock \bibinfo{note}{Publisher: The Optical Society}.
\bibitem[{Zhu et~al.(2006)Zhu, Palmer, Breslin, Harter \& Ramanujam}]{zhu_diagnosis_2006}
\bibinfo{author}{Zhu, C.}, \bibinfo{author}{Palmer, G.~M.}, \bibinfo{author}{Breslin, T.~M.}, \bibinfo{author}{Harter, J.}, \& \bibinfo{author}{Ramanujam, N.} (\bibinfo{year}{2006}).
\newblock \bibinfo{title}{Diagnosis of breast cancer using diffuse reflectance spectroscopy:{Comparison} of a {Monte} {Carlo} versus partial least squares analysis basedfeature extraction technique}.
\newblock {\it \bibinfo{journal}{Lasers Surg. Med.}\/},  {\it \bibinfo{volume}{38}\/}, \bibinfo{pages}{714--724}.
\newblock \bibinfo{note}{Publisher: Wiley}.
\bibitem[{Zhu et~al.(2005)Zhu, Palmer, Breslin, Xu \& Ramanujam}]{zhu_use_2005}
\bibinfo{author}{Zhu, C.}, \bibinfo{author}{Palmer, G.~M.}, \bibinfo{author}{Breslin, T.~M.}, \bibinfo{author}{Xu, F.}, \& \bibinfo{author}{Ramanujam, N.} (\bibinfo{year}{2005}).
\newblock \bibinfo{title}{Use of a multiseparation fiber optic probe for the optical diagnosis ofbreast cancer}.
\newblock {\it \bibinfo{journal}{J. Biomed. Opt.}\/},  {\it \bibinfo{volume}{10}\/}, \bibinfo{pages}{024032}.
\newblock \bibinfo{note}{Publisher: SPIE-Intl Soc Optical Eng}.

\end{thebibliography}

\end{document}